\documentclass[12pt,a4paper, 
titlepage, oneside, openright, epsfig]{scrbook}
\usepackage[dvips]{graphicx}
\usepackage{epsfig}
\usepackage{latexsym}
\usepackage{fancyhdr}
\def\Tr{\mbox{Tr}\,}
\def\re{\mbox{Re}\,}
\def\im{\mbox{Im}\,}

\def\D{{\mathcal{D}}}
\def\half{\frac{1}{2}}

\def\d{\partial}
\def\be{\begin{eqnarray}}
\def\ee{\end{eqnarray}}

\def\d{\partial}
\def\m{\mu}
\def\n{\nu}
\def\l{\lambda}
\def\r{\rho}
\def\half{\frac{1}{2}}
\def\M{\mathcal}
\def\e{\epsilon}

\def\slash#1{\, /\kern-0.6em{#1}}

\begin{document}
\begin{center}

\thispagestyle{empty}
{\Huge \bf Some Non Perturbative 
\vskip 0.3cm
Aspects of Gauge theories}
\vskip 0.3cm

\vskip 2.0 true cm
Thesis Submitted for the degree of\\
{\bf Doctor of Philosophy (Science)}\\
of \\
{\bf UNIVERSITY OF JADAVPUR}\\ KOLKATA
\vskip 3.0 true cm
2010 \\
{\large \bf Chandrasekhar Chatterjee}\\
S. N. Bose National Centre for Basic Sciences\\
JD Block, Sector III \\
Salt Lake City \\
Kolkata  700098\\
India
\end{center}
\frontmatter
\newpage
\thispagestyle{empty}
\vskip 6.0cm
\begin{center}
{\Huge \bf  Dedicated}
\vskip 0.5 cm
{\Huge \bf to }
\vskip 0.5cm
{ \Huge \bf   My Parents}
\end{center}
\newpage
\begin{center}
{\large \bf ACKNOWLEDGMENTS}
\end{center}
 
As a sense of fulfilment at the completion of this phase of academic endeavour,
 I wish to express my gratitude to all those who made this thesis possible.

It has been my privilege to work under the able guidance of my revered thesis 
advisor, Dr.Amitabha Lahiri. His insights into various problems and insistence
 on clarity have been most useful and inspiring. I express my deep sense of 
 gratitude to him for his patience, persistence and his prompt and 
 sincere help whenever I needed it most.

\noindent I extend my sincere thanks to Dr. Samir Kumar Paul for many fruitful, 
enthusiastic and illuminating discussions, academic or otherwise. He has 
always motivated me to do my best.

\noindent I also want acknowledge with sincere gratitude 
 Prof Jayanta Bhattacharya for all the academic and non academic help  I received from him.
  I am grateful to him for always being 
there to help me in all matters. 
  
 \noindent I acknowledge Dr. Manu Mathur,
Prof. Rabin Banerjee,  Dr. Biswajit Chakraborty, Prof. Binayak Dutta Roy,
Prof. Abhijit Mookerjee, Dr Ranjan Choudhury, Prof. Alak Kumar Majumdar
  for the academic and non academic help that 
I got from them during my working period.

\noindent I am grateful to Prof. S. Dattagupta, 
ex. Director of Satyendra Nath Bose National Centre for
Basic Sciences, for giving me the opportunity to do research here.

\noindent I am grateful to Prof. A. K. Raychaudhuri, 
the Director of Satyendra Nath Bose National Centre for
Basic Sciences, for extending my fellowship 
 to complete my thesis. 
  
\noindent I thank all the  administrative staff of SNBNCBS 
for helping me in many ways.  In particular, I am thankful to  the office of Dean (AP), the Library staff
and   all  account section staff for  providing me  excellent assistance.    

\noindent It is my pleasure to thank Ashish, Saikat, 
Santosh , Arya, Arnab , Tomagna, Bipul, Subrata, Ayan,
Sagar, Anjanda, Kunal, Saurav, Mitali, Ankushda, Aftabda and my all other friends 
for their cooperation 
and help. 
I have had a very nice time with  
 Debmalya, Rudranil, Indrakshi, Amartya, Nirupam, 
Atanu Kumar, Anshuman, Arghya and many other junior students over all these years.

Finally and most importantly, I express my whole hearted gratitude 
to my parents Mr. Basudeb Chatterjee \& Pranati Chatterjee, my uncles Sukdeb Chatterjee \& Debdeb Chatterjee,
my brother Rajsekhar, my wife Sima,   and all other family members.
 It is the love and unflinching support of my full family that enabled me to pursue 
this line of study which finally culminated in this thesis.

\vskip 2.0cm
\qquad\qquad \qquad\qquad\qquad \qquad\qquad\qquad \qquad\qquad\qquad\qquad Chandrasekhar Chatterjee

\noindent  
\tableofcontents
\mainmatter
\chapter{Introduction and Overview}
Strong interactions are adequately described at high energies by
quantum chromodynamics (QCD). At low energies, the QCD coupling is
large and color (QCD charge) is confined, but a precise description of how that
happens is as yet unknown. 
The hadron spectrum found in nature consists of 
color-singlet combinations of color non singlet
objects: the quarks and gluons. Unlike atomic physics, 
 where we can separate electrons from atoms,
it is not possible to separate quarks from hadrons. 
So there is no color-charge version of ionization in hadronic physics.
This problem is often referred to
as `color confinement'. The discovery of renormalization
 and asymptotic freedom \cite{Gross:1973ju, Politzer:1973fx} of non-Abelian gauge theory  established  SU(3) gauge theory as the theory
 which  describes the  dynamics of quarks and gluons. 
 The quark charge density  $ \rho^a_{quark} $ is the source of 
 static color electric field, as required by  
 Gauss law
 \begin{eqnarray}
\nabla\cdot \vec{E^a} = \rho^a_{quark} - g f^{abc} A^b_k E^c_k.
\end{eqnarray}

The last term containing the structure constant of the gauge group $ f^{abc} $
 and the gauge field $ A^a_k $ reflects the non-vanishing color electric
charge of the gluons. However, gluons are in the adjoint representation of 
SU(3) whereas the quarks are in the fundamental $ 3$ representation of SU(3). 
 So the color electric field of an isolated quark could only end on another isolated
 quark or else extend out to infinity. \\ To extract a single
 quark from a hadron we can scatter hadrons with high energy
 photons. As the struck quark begins to move away
  from the other quarks in the hadron, it brings along the color
electric field also. 
However, the system becomes unstable if the energy stored in the color
electric field becomes large enough and it  creates a light quark-antiquark pair.
The final states will be highly excited two color-singlet hadrons and they decay
into lighter hadrons. So at the end there  is no free quark
or a color-ionized hadron but only a shower of ordinary hadrons. 
 
\begin{figure}[t]
\makebox[\textwidth]{
\epsfig{file=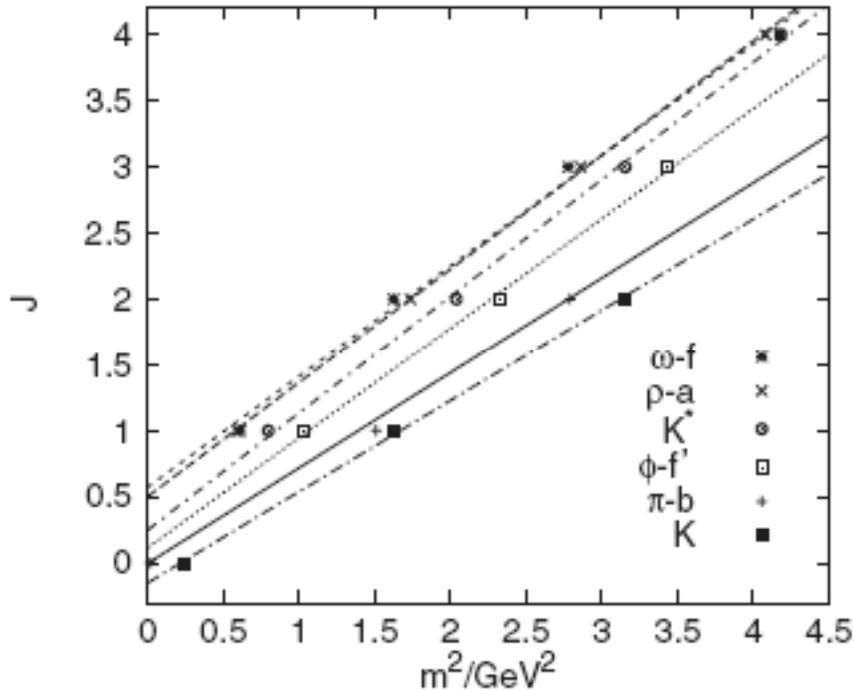}}
\caption{Regge trajectories \,(taken from \cite{Bali:2000gf})} \label{white}
\end{figure}

The hadron scattering process alone cannot fully determine the form of the
 confining potential. More knowledge about the  potential comes from what are known as `Regge trajectories'.
Experimentally it is found that all mesons  and baryons have many excited states (resonances) and 
when the spins of mesons (and baryons) are plotted against their squared masses,  the mesons and
baryons of given flavor quantum numbers seem to lie on nearly parallel straight lines, known
as  Regge trajectories.
  Regge trajectories are given by the equation 
 $ J = \alpha(s) $, where $ J $ is the angular momentum and 
 $ s = M^2 $ (the square of energy in the center of mass frame).
 Resonance occurs for some values of $ s $ for which $ \alpha(s) $
 is a non negative integer (mesons) or a half integer (baryons).

   For a fixed $ s $ the largest $ J $ is called the ` leading trajectory'.
 Experimentally it is seen that the leading trajectories are almost
 linear in $ s $:
 \begin{eqnarray}
\alpha(s) = \alpha(0) + \sigma' s.
\end{eqnarray} 
 There are also other trajectories for which
 \begin{eqnarray}
\alpha(s) = \alpha(0) - n + \sigma' s
\end{eqnarray}
where $ n $ is an integer.
$ \alpha(0) $ depends on the quantum numbers such as
strangeness and baryon number. The value of
the Regge slope $ \sigma' $ is  approximately 1 $GeV^{-2}$  \cite{hooftstringnotes}
and its value is universal.

A simple model satisfies Regge trajectories. 
Let us suppose that  a meson is constructed by a flux tube, 
with a quark and an anti quark attached to the ends of the tube. 
Then all lines of force of the color field are confined inside the tube.
We shall ignore the contribution to the energy by the quarks.
We can imagine a rotating  flux tube as a rotating straight line whose
end points are moving at a speed of light.
Suppose the mass per unit length of the flux tube is $ \sigma $ 
and the length of the tube is $ 2R $. Let us assume that the 
flux tube is rotating around a axis which  
perpendicularly bisects the straight fluxtube of length $ 2R $.
 If we take the quarks to be mass less, the endpoints move essentially at the speed of light  $ c = 1 $. 
Then we can write the mass of the rotating flux tube\,(spinning stick) 
 in the center of mass frame as
\begin{eqnarray}
m = Energy &=& 2 \int^R_0 \frac{\sigma dr}{\sqrt{1 - v^2(r)}}\\
&=& 2 \int^R_0 \frac{\sigma dr}{\sqrt{1 - r^2/R^2}}\\
&=& \pi\sigma R,
\end{eqnarray}
where $ v(r) $ is the speed at distance $ r $.
The angular momentum $ J $ will be
\begin{eqnarray}
J  &=& 2 \int^R_0 \frac{\sigma v(r) dr}{\sqrt{1 - v^2(r)}}\\
&=& \frac{2}{R} \int^R_0 \frac{\sigma r dr}{\sqrt{1 - r^2/R^2}}\\
&=& \half \pi\sigma R^2.
\end{eqnarray}
So, we can calculate
\begin{eqnarray}
\frac{J}{E^2} &=& \frac{1}{2\pi\sigma}.\\
J &=& \sigma' E^2,
\end{eqnarray}
here $ \sigma' $ is the Regge slope.  
 From experimental data we can estimate the value of $ \sigma $.
\begin{eqnarray}
\sigma' &=& \frac{1}{2\pi\sigma} = 0.9\, GeV^{-2}\\
\sigma &\approx & 0.18 \,GeV^2 
\end{eqnarray}

We can make the model more realistic by taking the flux tube as a 
string, instead of taking it as a rigid stick. This model 
can explain the existence of other trajectories also.

 The idea of constructing flux tubes to explain quark confinement  was first given by Nambu and Mandelstam  
  \cite{Mandelstam:1974pi, Nambu:1975ba}. Following
 them  it is now generally
thought that the QCD vacuum behaves like a dual superconductor,
created by condensation of magnetic monopoles, in which confinement
is analogous to a dual Meissner effect \cite{hooft:1976, hooft:1999, 'tHooft:1999au}.
 A meson state is then formed by attaching a quark-anti-quark pair to the ends of a flux string analogous
to the Abrikosov-Nielsen-Olesen  (ANO) vortex string of Abelian gauge
theory~\cite{Abrikosov:1956sx, Nielsen:1973cs}. As a consequence, the
energy of the pair increases linearly with their separation and quarks are
confined in hadrons. Calculation with explicit  models of this type \cite{Baker:1984dv, Maedan:1988yi}
have been compared both with experimental data and with Monte Carlo simulation
of QCD \cite{Baker:1997bg}.

However,  this model has its limitations.
One difference with reality is  that the static  Abrikosov-Nielsen-Olesen vortex  string carries
  magnetic field but static QCD flux strings must carry only electric field.
 The construction of flux tubes in field theory are formed via spontaneous symmetry breaking
 by scalar fields. However, it is not very clear 
 whether or how the symmetry is broken at low energy in QCD.   
 We can build a model for  confinement by confining magnetic particles by magnetic
 flux tubes and try to describe the system by dual variable to compare 
the system with the real system. 
Flux strings in the Weinberg-Salam theory was
suggested by Nambu~\cite{Nambu:1977ag}, in which a pair of magnetic
monopoles are bound by a flux string of Z condensate. Another construction 
of flux tubes in the Weinberg-Salam theory were also given in \cite{Vachaspati:1992fi}.
A different construction of flux strings, involving two adjoint scalar fields in
an SU(2) gauge theory, has been discussed in~\cite{Nielsen:1973cs,
  deVega:1976rt}. Recently there has been a resurgence of interest
in such constructions~\cite{Auzzi:2003fs, Hanany:2004ea,
  Shifman:2002yi, 'tHooft:1999au}.

In this thesis I construct flux strings and write the action in
terms of string variables as a dual gauge theory. I will show that,
in these dual gauge theories monopoles are attached at the end of the 
flux strings. In chapter $ 2 $ we shall give a brief description 
of magnetic monopoles in electromagnetism and Proca massive electrodynamics. Here I will 
discuss the  quantization of charge in Proca massive electromagnetic theory   by quantizing angular
 momentum in the presence of monopoles. I will also discuss  CPT symmetries
 in the presence of magnetic monopoles in this chapter.

  In chapter $ 3 $  I will first  give a review of flux string
 configuration in Abelian Higgs model. In the presence of these strings,
  I first dualize the scalar field to
find the strings interacting via an antisymmetric tensor
potential~\cite{Kalb:1974yc, Davis:1988rw, Lund:1976ze}, while the Abelian gauge
field is dualized ~\cite{Mathur:1991ip, Lee:1993ty} to a `magnetic' photon~\cite{Baker:1991bc}.
 Next I introduce fermionic magnetic monopoles into
the theory and minimally couple these to the magnetic photon.
Parity conservation of Maxwell equation suggests that the monopole current
may be an axial current. However, the axial current produces an anomaly and I
cancel the anomaly by postulating additional species of fermionic
monopoles. Then I dualize the resulting theory again, to find a
theory of magnetic flux tubes interacting with a massive Abelian
vector gauge field. The tubes are sealed at the ends by fermions,
thus providing a toy model for quark confinement. 
 

 For non-Abelian gauge theories, the construction is a little different, as the 
 theories themselves contain magnetic monopole solutions. 
 In chapters 4  and 5 I consider configurations corresponding 
 to a pair of 't Hooft - Polyakov monopole
  \cite{'tHooft:1974qc, Polyakov:1974ek, Prasad:1975kr}\,and anti-monopole attached to the end
 of a flux tube in SU(2) gauge theory. This corresponds to Nambu's picture of confinement
 but one in which the confined monopoles are the magnetic monopoles (topological objects)
 in the theory.

In Chapter $ 4 $ I  review  spontaneous
symmetry breaking and magnetic monopoles. 
Then I  describe flux tube solutions in non Abelian gauge theory
by two stage symmetry breaking. This can be done for
SU(2) gauge theory by starting with two scalr fields.
The first scalar field breaks the SU(2) symmetry to U(1).
Breaking  this U(1) produces a flux tube.
The U(1) can be broken in two ways by taking the second 
 scalar field  in one of two different representations of SU(2).
 One way is to take it  in the adjoint representation of SU(2), 
  and other is to have it in the fundamental representation of
SU(2). The idea of two-scale symmetry breaking in SU(2), the first to
produce monopoles and the second to produce strings, has appeared
in~\cite{Hindmarsh:1985xc}. Later this idea was used in a
supersymmetric setting in~\cite{Shifman:2004dr,  Kneipp:2003ue, Auzzi:2003em,
  Eto:2006dx}.

 In Chapter $ 5 $ we start with SU(2) gauge theory with two 
scalar fields. One of them, call it $\phi_1\,,$ acquires a vacuum
expectation value (vev) $\vec \xi_1$ which is a vector in internal
space, and breaks the symmetry group down to U(1). The 't
Hooft-Polyakov monopoles are associated with this breaking.
 The other scalar field also has a non-vanishing vev. 
 This second field is in the adjoint representation and it 
is free to wind around $\vec \xi_1$ in the internal
space. This winding is mapped to a circle in space, giving rise to
the vortex string.  We will construct two effective low energy
Lagrangians with monopole and strings, one for two adjoint scalars and the other
for one adjoint and one fundamental scalar. We shall see that these two
Lagrangians are the same except for the values of two parameters,
  the coupling constant and mass of the photon.
 We then dualize the fields as in~
 \cite{Davis:1988rw, Mathur:1991ip, Lee:1993ty, Akhmedov:1995mw, Chatterjee:2006iq}
  to write the action in terms of string
variables and here we shall  show  the attachment of monopoles
at the end of flux tubes.  The idea of flux matching, following
Nambu~\cite{Nambu:1977ag} also appears in this thesis. 





\chapter{Magnetic monopoles in  Electrodynamics}


\section{Duality and magnetic monopoles}
The equations which describe the  electromagnetic field 
 with sources $ \rho(x), \vec J(x) $ are
\be 
\label{maxwell1}
\nabla\cdot \vec E &=& \rho \\
\label{maxwell2}
\nabla\times \vec B &=&  \frac{\partial\vec E }{\partial t } + \vec J\\
\label{maxwell3}
\nabla\cdot \vec B &=& 0\\
\label{maxwell4}
\nabla\times \vec E &=& - \frac{\partial\vec B }{\partial t }.
\ee
 
 We can define vector  potential from Eq. (\ref{maxwell3}) as
\begin{eqnarray}
\label{vectorpotential1}
\vec B = \nabla \times \vec A, \qquad B^i = \epsilon^{ijk} \d_j A^k.
\end{eqnarray}
 This is a unification procedure, because 
 by defining the  vector potential we can describe both
 electric and magnetic fields. From Helmholtz theorem
 we know that any vector field is uniquely specified by its divergence and curl. 
However, Eq. (\ref{vectorpotential1}) is only the curl part of
the  field $ \vec{A} $. So we can fix $ \nabla\cdot\vec A $
by hand to uniquely specify the field $ \vec A $. This is a
process that we call gauge fixing in electrodynamics.

The Maxwell equations without any sources are invariant under the transformation 
\be
\label{dualityEB1}
\vec E \rightarrow \vec B\, , \qquad
\vec B \rightarrow  - \vec E.
\ee
This invariance is called the duality symmetry of the Maxwell equations. This 
 duality symmetry breaks down when we add an electric charge current 
 density  into the equations. To restore the 
 symmetry we have to add a magnetic charge density and a magnetic current density
 into the equations. If we add both  magnetic  and electric currents into the 
  Maxwell equations then the equations look like
\begin{eqnarray}
\label{maxl_monopole1}
\nabla\cdot \vec E &=& \rho_e \\
\label{maxl_monopole2}
\nabla\times \vec B &=& \vec j_e + \partial_t \vec E \\
\label{maxl_monopole3}
\nabla\cdot \vec B &=& \rho_m\\
\label{maxl_monopole4}
\nabla\times \vec E &=& - \vec j_m - \partial_t \vec B.
\end{eqnarray} 
It follows from Eq. (\ref{dualityEB1}) and the  Maxwell equations 
with electric and magnetic currents that the duality  transformation for the currents are
\be
\rho_e \rightarrow \rho_m,\qquad \rho_m \rightarrow - \rho_e\\
\vec j_e \rightarrow \vec j_m,\qquad\vec j_m \rightarrow - \vec j_e.
 \ee

One can generalize the transformations by introducing a
 parameter $\xi$, and the transformations
can be written as
\be
\vec E' = \vec E \,cos \xi + \vec B\, sin \xi&,& \,
\vec B' = - \vec E\, sin\xi + \vec B\, cos\xi \\
\rho'_e = \rho_e\, cos\xi + \rho_m \, sin\xi &,&\,
\rho'_m = - \rho_e\, sin\xi + \rho_m \, cos\xi \\
\vec j'_e = \vec j_e\, cos\xi + \vec j_m \, sin\xi&,& \,
 \vec j'_m = -\vec j_e\, sin\xi + \vec j_m \, cos\xi.
\ee
It is then  a matter of convention to say that a particle has magnetic charge or 
electric charge, because it fully depends on the value of $ \xi $ that we choose. 
The question can be asked whether all particles have the same ratio of magnetic 
to electric charge. If they are the same,  we can choose the angle $ \xi $ in the above equations
 so that $\rho_m = 0,\, \vec j_m = 0$. We then have the Maxwell equations as they are usually known. 
If we choose the electric and magnetic charges of an electron as $q_e = -e, q_m = 0$,
 then it is known  \,\cite{Jackson}\,that for a proton, $q_e = +e$ (with the 
 error limit $| q_e(electron) + q_e(proton)|/e \simeq 10^{-20}$) and 
$|q_m(nucleon)| < 2 \times 10^{-24}$. 
This  limit on the magnetic charge of a proton or neutron 
follows directly from knowing that the average magnetic field at the surface of 
the earth is not more than $10^{-4}$ T. To a very high degree of 
precision we can conclude that the particles of ordinary matter possess only electric charge or, 
equivalently, they all have the same ratio of magnetic to electric charge. For 
 unstable particles the question of magnetic charge is more open, but there exists no 
positive evidence.

\section{Dynamical system with magnetic charge}
 Following the equations (\ref{maxl_monopole1})-(\ref{maxl_monopole4}), we
 can write down the Maxwell equations with electric and magnetic currents ($  j^e_\m ,j^m_\m $),
\be
\label{maxwell_monopole}
\partial_\mu F^{\mu\nu} = j_e^\nu\,,\, \partial_\mu {*F}^{\mu\nu} = - j_m^\nu.
\ee
Here $*F^{\mu\nu} = \half \e^{\m\n\r\l}F_{\r\l}  $ and 
\begin{eqnarray}
F^{i0} = E^i,\qquad 
F_{ij} = \e_{ijk}B^k.
\end{eqnarray}

If the currents result from point particle sources then we can  write
\be
j_e^\m = \sum_i\, e_i \int {dx_i}^{\m}\delta^4(x-x_i),\\
j_m^\m = \sum_i\, g_i \int {dx_i}^{\m}\delta^4(x-x_i),
\ee
where the integral over $x_i$ is taken along the world line of the $ i $-th particle whose
electric and magnetic charges are $e_i$ and $g_i$, respectively.
The Lorentz force law can be generalized for a particle carrying magnetic
as well as electric charge,
\be
\label{lorentzforce1}
m\frac {d^2x^\m}{d\tau^2} = (eF^{\m\n} + g {*F}^{\m\n})\frac {dx_\n}{d\tau} .
\ee
Here $ \tau $ parametrizes the world line of the particle.
  The dynamics of classical particle with electric and magnetic charge
  can be completely  described
by the above equations (\ref{maxwell_monopole}) and (\ref{lorentzforce1}).
The field due to a static monopole is
\be
\label{monopolefield}
\vec B = \frac g{4\pi r^2} \hat r. 
\ee
 The dynamics of an electrically charged particle of charge $ e $ in a static monopole field is governed
  by  the generalized Lorentz force Eq. (\ref{lorentzforce1}),
\be
\label{lorentzforce2}
m\frac {d^2 \vec r}{dt^2} &=& e \frac {d\vec r}{dt} \times \vec B. 
\ee
Though the force in (\ref{lorentzforce2}) is not a central 
force (i.e. not directed towards the origin), angular momentum is conserved
 because of the spherically symmetric nature of  magnetic field of Eq. (\ref{monopolefield}).
 The rate of change of orbital 
angular momentum  of the particle can be written as
\begin{eqnarray}
\frac {d}{dt}(\vec r \times m \frac {d\vec r}{dt}) &=& \vec r \times m \frac {d^2\vec r}{dt^2}\\
&=& e \vec r \times (\frac {d\vec r}{dt} \times \vec B)\\
&=&\frac {eg}{4\pi r^3}\vec r \times (\frac {d\vec r}{dt} \times \vec r)\\
&=& \frac {d}{dt}\left(\frac {eg}{4\pi } \hat r\right).
\end{eqnarray}
Thus it is possible to define a conserved quantity 
\begin{equation}
\label{totalangularmomentum}
\vec{\tilde{J}} = \vec r \times m \frac {d\vec r}{dt} - \frac {eg}{4\pi } \hat r.
\end{equation} 
 The second term in  above Eq. (\ref{totalangularmomentum})
is actually the contribution from the electro-magnetic field.

To see this, we  calculate the angular momentum of the electromagnetic field
 for a system with one electric charge at some point $ \vec{r}_e $  and one magnetic charge at the origin,
\begin{eqnarray}
\label{abcd}
\vec J_{em} &=& \int d^3x \,\vec r \times (\vec E \times \vec B)\\
&=& \frac g{4\pi} \int d^3x \,\vec r \times (\vec E \times \frac {\vec r}{r^3})\\
&=& \frac g{4\pi} \int d^3x\, \frac {1}{r} (\vec E - (\vec E \cdot \hat r )\hat r).
\end{eqnarray}
Using the identity $\displaystyle{{\vec E}\over r} = 
\displaystyle{{\hat{r}(\hat{r}\cdot {\vec E})}\over r} + ({\vec E}\cdot \nabla)
\hat{r}$, we get
\begin{eqnarray}
\vec{J}_{em} &=& \frac g{4\pi} \int d^3x\,({\vec E}\cdot \nabla)\hat r \\
 &=& - \frac g{4\pi} \int d^3x\,\hat r \,\nabla \cdot {\vec E} + 
 \frac g{4\pi} \int  \hat r  \,\vec E \cdot \vec ds'.
\end{eqnarray}
where the second integral is over a surface $ S' $ at infinity and $ \vec{ds'} $ is directed along the outward 
normal to that surface. With $ \vec E $ for a point charge  this surface integral reduces to 
$ \frac{e}{4\pi}\int\hat r d\Omega $  because the integral (\ref{abcd}) for this system is invariant under
a shift of origin.  Since $ \hat r $ is radially directed,   it has zero angular average.  
Thus the second term vanishes.

The second term is a surface term at spatial boundary and all the components of the field 
${\vec E} $ go to zero on the spatial boundary. So $ J_{em} $ becomes
\begin{eqnarray}
\label{angular_momentum}
\vec{J}_{em} = - \frac g{4\pi} \int d^3x\,\hat r \,\nabla \cdot {\vec E}\,
= - \frac g{4\pi} \int d^3x\,\hat r \, \rho(\vec r).
\end{eqnarray}
  Since there is a static electric charge $ e $ at $ \vec r_e $, the 
 charge density $ \rho(r) = e\,\delta^3(\vec r - \vec r_e) $.
 Using this the above Eq. (\ref{angular_momentum}) becomes
 \begin{eqnarray}
\vec{J}_{em} = - \frac g{4\pi} \int d^3x\,\hat r \, \delta^3(\vec r - \vec r_e),\,
= - \frac {ge}{4\pi} \,\hat r_e.
\end{eqnarray}
This result \cite{J.J.Thomson, M.N.Saha} exactly matches the second 
 term of Eq. (\ref{totalangularmomentum})\, with $ \hat r = \hat r_e $.
So the total angular momentum of particle and electromagnetic field is
\begin{eqnarray}
\vec{\tilde{J}} = \vec{J} + \vec{J}_{em} =
 m_e \vec{r} \times \frac{d\vec r}{d t} - \frac{ge}{4\pi} \hat{r_e}.
\end{eqnarray}

The component of the total angular momentum  in the direction of the electric charge is
\begin{equation}
\vec{\tilde{J}}\cdot \hat r_e = - \frac{eg}{4\pi}. 
\end{equation}
  $\vec{\tilde{J}}$ is a constant of motion, so the particle will 
rotate around $ - \vec{\tilde{J}}$ with the angle\\
  $ \cos^{-1}( {eg} / {4\pi |\vec {\tilde {J}|}}) $
   ( between position vector $ \vec r_e$ and $ - \vec{\tilde{J}}$) .
 Since the angular momentum is quantized, we can choose the $ z $ axis along $ \vec r_e $
 and find
  \begin{equation}
  \label{dirac_quantization}
J_z = \frac{eg}{4\pi} = \frac {n \hbar}2.
\end{equation} 
Eq. (\ref{dirac_quantization}) is called the Dirac quantization condition. It follows that if
 there is at least one magnetic monopole in the universe,
 electric charge is quantized in multiples of some fundamental unit of charge.
\section{Monopole gauge field and Dirac string}

  Dirac \cite{ Dirac:1931kp, Dirac : 1948}  was the first to take
   magnetic monopoles seriously and tried to establish a theory
  for magnetic monopoles.
       The main problem was that if there are isolated magnetic charges, 
    the vector potential cannot be smooth and differentiable everywhere. 
    In particular, if we consider the radial magnetic field of Eq. (\ref{monopolefield})
     for any closed 
    surface $\vec S $ containing the origin, then
\begin{eqnarray}
\label{mono_charge}
 g = \int_S {\vec B}\cdot d\vec S.
\end{eqnarray}
However, if $ {\vec B} = \nabla \times \vec A,$ the integral (\ref{mono_charge}) would have to vanish. 
 Thus $ \vec A $ cannot be smooth and differentiable everywhere on S. 
       Let us consider the field due to 
          an infinitely long and thin solenoid placed along the 
          negative $z$ axis with  one end at the origin (with total flux
           strength $g$) . Its magnetic field would be:
\begin{eqnarray}
  {\vec B}_{sol} = {g\over {4\pi r^2}}\hat r + g \theta(- z)\delta(x)\delta(y)\hat z ,
\end{eqnarray}
 where $ \hat z$ is a unit vector in the $z$ direction and 
 $ \theta(\xi) = 0$ if $\xi < 0,\theta(\xi) = 1$ if $\xi>0$. 
     This magnetic field differs from $\vec B$ only by the
      singular magnetic flux along the solenoid but it is
       clearly source free, i.e, $ \nabla \cdot {\vec B}_{sol}$
        vanishes even at the origin. Thus we can define a
        vector potential  $ \vec A $ everywhere to write $ \vec B_{sol} = \nabla\times\vec A $,
        i.e.
\begin{eqnarray}
 \displaystyle{g\over{4\pi} }\displaystyle{\hat { r}\over r^2} = 
 \nabla\times{\vec A} - g \theta(- z)\delta(x)\delta(y)\hat z .
\end{eqnarray}
 The line occupied by the solenoid is called the Dirac string.  
   We should think of the field $ \vec B $ as being represented
    not just by $\vec A $,  but by $\vec A$ together with a line on which it is singular. 
   Given our choice of the position of the negative $z$ axis we can 
   calculate an explicit form for $ \vec A$ by exploiting axial symmetry. 
   The magnetic field due to the
   monopole contains only the radial component in spherical polar coordinates.
   So from symmetry we can choose the vector potential as 
   $ \vec A = A(r, \theta)\hat \varphi $, where ($ r, \theta, \varphi $) are
   the spherical polar coordinates.
    The magnetic flux through a circle $C$ (corresponding to fixed values 
    of $ r $ and $\theta$, and $\varphi$ ranging over the values 0 to $2\pi$) 
    is given by solid angle subtended by C at the origin
     multiplied by $\displaystyle{g\over{4\pi}}$ namely 
     $\displaystyle{{1\over2}g(1 - \cos\theta)}$. 
     But we can also write the flux using Stokes' theorem as
\begin{eqnarray}
{1\over2}g(1 - \cos\theta) = \int_{S} {\vec B}\cdot{\vec dS}\,=
\oint_C \vec{A}\cdot\vec{dl} = 2\pi\, A(r,\theta) \,r\, \sin \theta, 
\end{eqnarray}
here $ S $ is the surface enclosed by the circle $ C $. 
It follows that we can write the vector potential as
\begin{eqnarray}
\label{monopolepotential1}
{\vec A(\vec r)} = {g\over{4\pi}}{(1 - \cos\theta)\over {\sin\theta}}\hat{\varphi}.
\end{eqnarray}

 This vector potential (\ref{monopolepotential1}) shows the anticipated singularity 
on the negative $z$ axis $ \theta = \pi $. There are other ways of defining  this vector potential.
Define a two form $ F $ in three dimension
\begin{eqnarray}
\label{F0}
F = F_{ij}\,dx^i\wedge dx^j.
\end{eqnarray}
Now if we go from Cartesian coordinates $ x^i $ to other coordinates
 $ \xi^\alpha $, we have
\begin{eqnarray}
\label{F2}
F = F_{ij} \frac {\d x^i}{\d \xi^\alpha} 
\frac {\d x^j}{\d \xi^\beta}\,d\xi^\alpha\wedge d\xi^\beta.
\end{eqnarray} 
Now suppose $ \xi^\alpha = (r, \theta,\varphi) $.

  We can write  Eq. (\ref{F2}) as
\begin{eqnarray}
\label{F4}
F = \epsilon^{ijk}  B^k \left[ \frac{dx^i}{dr} \frac{dx^j}{d\theta} dr\wedge d\theta 
+ \frac{dx^i}{d\theta} \frac{dx^j}{d\varphi} d\theta\wedge d\varphi 
+ \frac{dx^i}{d\varphi} \frac{dx^j}{dr} d\varphi\wedge dr  \right].
\end{eqnarray}
Let us write
$ x^1 =  r\sin\theta \cos\varphi $, $ x^2=  r\sin\theta \sin\varphi $,
$ x^3 = r \cos\theta $ for spherical polar coordinate. Calculating all the derivatives 
we can write Eq. (\ref{F4}) as 
\begin{eqnarray}
\label{F3}
F = Q_m  \sin \theta \, d\theta\wedge d\varphi,
\end{eqnarray}  
with the corresponding vector potential 
\begin{equation}
\label{monopotantialform}
{ A} = - Q_m \cos\theta { d}\varphi.
\end{equation}
Though the term $- Q_m \cos\theta  $ is a smooth function, the vector 
potential is not smooth everywhere. To see the singularity let us write down 
vector potential in Cartesian coordinate. 
Using 
\begin{equation}
\cos\theta = \frac zr ,\qquad \tan\varphi = \frac yx,
\end{equation}
we can write  Eq. (\ref{monopotantialform}) as
\begin{eqnarray}
\label{monopotantialform2}
{A} = - Q_m\left[-\frac {zy}{r(x^2+y^2)} 
{ d}x + \frac{zx}{r(x^2+y^2)} { d}y\right].
\end{eqnarray}

 Now we want to check the behavior near the  z-axis, i.e in a region where 
 $ \displaystyle{\frac {|z|}r \approx 1 }$.
So for the positive  $ z $-axis, Eq. (\ref{monopotantialform2}) becomes
\begin{eqnarray}
\label{monopotantialform3}
{ A}_{\frac {z}r \approx 1} &=& - Q_m\left[-\frac {y}{(x^2+y^2)} { d}x + 
\frac{x}{(x^2+y^2)} { d}y\right].
 \end{eqnarray}
The above expression (\ref{monopotantialform3}) is singular
on the positive $ z $-axis and in the same way we can show that it is also
singular on the negative $ z $-axis. So the expression (\ref{monopotantialform}) 
cannot be treated as a gauge potential for the monopole. If we 
calculate the magnetic field for this vector potential, we will get the
magnetic field of a monopole along with a  singular line magnetic field 
along the whole $ z $-axis.

   The gauge potential (\ref{monopotantialform2})
 near the $ z$-axis takes the form as
\begin{eqnarray}
{ A}_{\frac {|z|}r \approx 1} &=& -Q_m \,{ d}\varphi,\, 
\mbox{for $ z > 0 $ }\\
&=& \,\,\,\,\,Q_m \,{ d}\varphi,\, 
\mbox{for $ z < 0 $} .
\end{eqnarray}
 If we add $ \pm Q_m { d}\varphi $ with the expression
 (\ref{monopotantialform3})   then the gauge
 potential is singular only on one of the $ z $-axis. So the vector potential
 can be written as
 \begin{eqnarray}
 \label{A_1}
{A}_1 &=& -Q_m \left( \cos \theta - 1\right){ d}\varphi,\,
\mbox{singular along $ \theta = \pi $},\\
\label{A_2}
{A}_2 &=& -Q_m \left( \cos \theta + 1\right){ d}\varphi,\,
\mbox{singular along $ \theta = 0 $}.
\end{eqnarray}
The vector potential $ {A_1} $ gives the magnetic field of a monopole field at the 
origin with a singular 
field line (Dirac string) along the negative $ z $-axis. For the 
vector potential $ { A}_2 $ the singular string will appear on the  positive $ z $-axis.
So to get only the monopole field we have to subtract the
singular  field due to the string  from the curl of the vector potentials $ { A}_1 $ and $ { A}_2 $.

     Alternatively, we can remove all singular strings by the 
Wu-Yang construction \cite{Wu:1976qk}. In this, space is covered 
by two coordinate patches $ R_a $ and $ R_b $. Using spherical coordinates 
($ r, \theta, \varphi $)  with the monopole at the origin we choose $ R_a $ and $ R_b $ as 
\begin{eqnarray}
R_a: \qquad 0 \leq \theta < \half \pi + \delta ,\qquad 0 < r , \qquad  0\leq \varphi < 2\pi\\
R_b: \qquad \half \pi - \delta \leq \theta < \pi  ,\qquad 0 < r , \qquad  0\leq \varphi < 2\pi.
\end{eqnarray}
Then the vector potential $ A $ is defined  to be $ A_1 $ in $ R_a $ and $ A_2 $ in the
patch $ R_b $. 
In the intersection of two patches the two gauge potentials
are related by gauge transformation 
\begin{eqnarray}
A_1 = A_2 + \frac{i}{e} G_{ab}dG^{-1}_{ab},
\end{eqnarray}
with $ G_{ab} = \displaystyle{e^{2ieQ_m\varphi}} $. 
Then in order to 
make the gauge transformation single valued we must require
the Dirac quantization condition $ eQ_m = \frac{n}{2} $. 
  
    Now we will discuss the type of the singularity of magnetic field
    due to the vector potential $ A_2 $.
     The expression (\ref{A_2}) is singular at the
  origin $ r=0 $ and on the positive $ z $-axis.
  The singularity at the origin reflects the singularity of
  the  monopole field. The singularity on the positive
  $ z $-axis constitute the string, which was not present in the pure
  monopole field. To control the singularity,
  we can regularize the gauge potential  \cite{felsagar},
  \begin{eqnarray}
  \label{Aepsilon}
{ A} \rightarrow { A_\epsilon} = 
Q_m \left[\frac y{R(R-z)}, -\frac x{R(R-z)}, 0\right]\\
\mbox{with}\, R = \sqrt{r^2 + \epsilon^2}\nonumber.
\end{eqnarray}
Then
\begin{eqnarray}
\frac{\d}{\d x} \left[\frac{1}{R(R-z)} \right] &=& - \frac{x(2R-z)}{R^3(R-z)^2}\\
\frac{\d}{\d y} \left[\frac{1}{R(R-z)} \right] &=& - \frac{y(2R-z)}{R^3(R-z)^2}\\
\frac{\d}{\d z}\left[\frac{1}{R(R-z)}\right] &=& \frac{1}{R^3}.
\end{eqnarray}
The magnetic field that we get from the vector potential (\ref{Aepsilon}) is 
\begin{eqnarray}
\vec B_\epsilon = Q_m\left[ \frac{\vec r}{R^3} - 
\frac{\epsilon^2(2R-z)}{R^3(R-z)^2} \hat k \right].
\end{eqnarray}
 The first term is  the monopole term. Let us call the second term  $ \vec B^S_\e $, it will
 produce the magnetic field of a singular string as we take $ \epsilon \rightarrow 0 $,  
 \begin{eqnarray}
 \label{Bstring}
 \vec B^{S}_\epsilon = - Q_m \frac{2\e ^2}{R^3(R-z)}\hat k 
 - Q_m \frac{\e^2 z}{R^3(R-z)^2} \hat k .
 \end{eqnarray}
 It is easy to check that $ \vec B_\e^S $ is zero everywhere other than on the positive
 $ z $-axis as $ \epsilon \rightarrow 0 $ .  The first term of  Eq. (\ref{Bstring}) gives no
 contribution to the singular part on the positive $ z $-axis
 while on the positive $ z $-axis the second terms becomes,
 \begin{eqnarray}
- Q_m \frac{1}{\e^2}.
\end{eqnarray}
  This will produce a delta function field along positive $ z $-axis.
\section{C,P,T symmetry and Magnetic monopoles}
 Experiments show that the electric current transforms
  as a vector under parity and time reversal
\begin{eqnarray}
\label{PTe}
\begin{array}{cc}
{P \rho_e(x) P^{-1} =  \rho_e(x_P)},\qquad{T \rho_e(x) T^{-1} = \rho_e(-x_P)}\\
{P \vec j_e(x) P^{-1} = -\vec j_e(x_P)},\qquad {T \vec j_e(x) T^{-1} = -\vec j_e(-x_P)},
\end{array}
\end{eqnarray}
where $ x = (t, \vec x) $ and $ x_P = (t, -\vec x) $.
The Maxwell equations will be invariant under parity and time reversal if
\begin{eqnarray}
\label{PTEB}
\begin{array}{cc}
{P \vec E(x) P^{-1} = - \vec E(x_P)},\qquad {T \vec E(x) T^{-1} =  \vec E(-x_P)} \\
{P \vec B(x) P^{-1} =  \vec B(x_P)},\qquad {T \vec B(x) T^{-1} =  -\vec B(-x_P)}.
\end{array}
\end{eqnarray}
 From these field transformations 
 we can deduce the parity and time reversal 
 properties of the magnetic current,
\begin{eqnarray}
\label{PTm}
\begin{array}{cc}
{P \rho_m(x) P^{-1} = -\rho_m(x_P)},\qquad {T \rho_m(x) T^{-1} = -\rho_m(-x_P)}\\
{P \vec j_m(x) P^{-1} = \vec j_m(x_P)},\qquad {T \vec j_m(x) T^{-1} = \vec j_m(-x_P)}.
\end{array}
\end{eqnarray}
 If we look at the time reversal property of monopole current,
 we see that it does not change its sign like ordinary current.
  This looks peculiar, because any physical particle moving in a 
  trajectory will reverse its motion when time reversal is applied.
   However, the confusion ends when we look at the  magnetic 
   charge density, which changes sign under time reversal. 
  So under time reversal, a particle reverses not only the direction
    of its path but also the sign of its magnetic charge, such that the magnetic currents 
    does not change its direction.

  The discussion on charge conjugation (C), parity (P) and time reversal (T) properties of monopole currents is important,
    because violation of $C\,,P$ and $T$ has not 
   been found experimentally in electromagnetism.

 Now we will check whether any physical currents have the P and T properties 
    written in  Eq. (\ref{PTm}).  There are some beautiful theory of 
magnetic monopoles taking monopole currents to be   classical point 
particle current  
{\cite{Dirac : 1948,Rohrlich:1966,zwanziger,Balachandran :
 1975,1979:zwanziger,Goddard :1978,Calucci:1982,Kleinert: 1990}.
  The classical point particle currents for monopole can be written as,
\be
\label{clcurnt}
 j^{\m}(x) = \sum_{i} g_i\int \delta^4(x-x_i) u_i^{\m}d\tau_i.
\ee
Here the particle trajectories are specified by 
$x_i^{\m} = x_i^{\m}(\tau_i)$, $ u_i^{\m} = \dot x^\m = dx_i^{\m}/d\tau_i$, 
and $\tau_i$. The time reversal and parity properties of the above currents are
\begin{eqnarray}
\label{PTmcurnt}
\begin{array}{cc}
{j^0(x) {\rightarrow{\kern-0.4cm}^{P}} \phantom{x}   j^0(x_P)},
\qquad {j^0(x)\buildrel T\over \rightarrow \phantom{x}   j^0(-x_P)}\\
{j^i(x) {\rightarrow{\kern-0.4cm}^{P}} \phantom{x}  -j^i(x_P)},
\qquad {j^i(x) {\rightarrow{\kern-0.4cm}^{T}} \phantom{x}  -  j^i(-x_P)}.
\end{array}
\end{eqnarray}

 However, these parity and time reversal properties do not match with Eq. (\ref{PTm}). 
 These are exactly like those of the electric current.
So this current  cannot be treated as that of a magnetic monopole. To match its behavior with that of a 
monopole current,  one has to define a number $g$ which must change its sign under 
the action of P and T.
However,  numbers do not change under Lorentz transformations. So in order to
 use this current  as a monopole current
we have to redefine time reversal and parity  by changing  the sign of
$ g_i $ along with the usual time reversal operation.

Let us consider CPT in a quantum field theory of magnetic monopoles.
For a complex  scalar field $\Phi$ the particle  current is,
\begin{equation}
\label{scfcurnt}
 j^{\m}(x)= ig(\Phi\d_\mu \Phi^* - \Phi^*\d_\mu \Phi).
\end{equation}
 The above current (\ref{scfcurnt}) can represent a monopole if
 the field $ \Phi $ transforms under parity and time reversal as
\begin{eqnarray}
\label{scalarpt}
\Phi(x)  \buildrel T\over \rightarrow \phantom{x}\Phi^*(-x_P), \qquad
\Phi(x) \buildrel P\over \rightarrow \phantom{x} \Phi^*(x_P).
\end{eqnarray}

  For ordinary conventional  unitary representation of the Lorentz group 
  a quantum scalar field transforms under T and P as
\begin{eqnarray}
\Phi(x)  \buildrel T\over \rightarrow \phantom{x}\Phi(-x_P), \qquad
\Phi(x) \buildrel P\over \rightarrow \phantom{x} \Phi(x_P).
\end{eqnarray}
   However, there is an 
unconventional representation \cite{Weinberg:book,Anselmino:2002} of time reversal 
which converts from  a particle state to an anti-particle 
state with reversed time, which is appropriate for a monopole. The anti-unitary 
property of the time reversal operator is responsible for this conversion. 
The time reversal matrix is block-diagonalizable under anti-unitary 
transformations in the particle state space. The block diagonal part 
of the time reversal matrix converts a particle state to an anti-particle state. However, 
  there is no corresponding  representation for parity because it is a
  unitary operator.  So this unconventional representation is not valid for 
 magnetic monopoles as scalar field.

 For fermionic monopole currents we can write the current {\cite{1979:zwanziger,Kleinert: 1990}}, 
\begin{equation}
\label{ferveccurnt}
j^{\m}(x)= g\bar{\psi}\gamma^{\m}\psi.
\end{equation}
The parity and time reversal properties of this current are
\begin{eqnarray}
\label{PTcurnt}
\begin{array}{cc}
{j^0(x)\buildrel P\over \rightarrow \phantom{x} j^0(x_P)},
\qquad {j^0(x)\buildrel T\over \rightarrow \phantom{x} j^0(-x_P)}\\
{j^i(x) \buildrel P\over \rightarrow \phantom{x}  - j^i(x_P)},
\qquad {j^i(x)\buildrel T\over \rightarrow \phantom{x}  - j^i(-x_P)}.
\end{array}
\end{eqnarray}

However, the above parity and time reversal  properties do not match with the P and T 
properties of monopole current (\ref{PTm}).
So this current cannot be treated as a monopole current under the usual 
definition of parity and time reversal.

      Monopole currents are axial with respect to parity and it is possible 
to write down an axial current {\cite{Abdus Salam:1966,
j.g.taylor:1967,Shu-Yuan Chu : 1973}} with the fermionic fields as
$\bar{\psi}\gamma_5 \gamma^{\m}\psi$. However, the axial fermionic current 
has the time reversal  property of the usual vectors. So it may be possible to
get time reversal property like monopole current as written in Eq. ({\ref{PTm}}) 
if we use unconventional  representation of time reversal.
However,  the axial current does not change its sign under charge conjugation,.
   
   We have seen that all the currents written in equations (\ref{clcurnt}), (\ref{scfcurnt})
   ,(\ref{ferveccurnt}) and also the axial current have some problems if they are to be thought of as
 monopole currents. However, if we can redefine parity and time reversal by also changing
 the sign of $ g $, the magnetic coupling constant,  with  the usual parity and 
 time reversal operation, then we can take 
 some of the above currents (\ref{clcurnt},\ref{scfcurnt},\ref{ferveccurnt})
  as monopole currents. 
  We can generalize this idea by giving a  prescription of redefinition
  of parity and time reversal \cite{N.F.Ramsey:1957}. 
    In a theory  which includes the effect of magnetic monopoles, the TCP theorem 
 would be replaced by a TMCP theorem where T represents simple 
 time reversal, M magnetic monopole conjugation, C electric 
 charge conjugation and P simple inversion of space coordinate.
  It is of course possible to express the theorem in various ways, such as
\begin{equation}
\label{ramsey}
          (TM)\phantom{x}(CM)\phantom{x}(MP) = T' C' P'.
\end{equation}
Here $T'$ indicates an extended time reversal whose definition 
includes magnetic pole conjugation as well, $C'$ represents 
conjugation of both electric and magnetic charges, and $P'$ 
represents a parity transformation includes magnetic pole conjugation as well.

In this prescription a magnetically charged scalar field transforms as
\begin{eqnarray}
&& \Phi(x)  \buildrel T'\over \rightarrow\Phi^*(-x_P)\\
&& \Phi(x)  \buildrel P'\over \rightarrow\Phi^*(x_P).
\end{eqnarray}
    Here we can see that the transformation property of the scalar
     fields under $P'$ and $T'$ are the same as Eq. (\ref{scalarpt}).
$P'$ and $T'$ also act  as ordinary parity and time reversal
on the other fields  which have no magnetic charge. 
 Now it is possible to construct 
 monopole currents using ordinary  vector currents,
  (see \cite{zwanziger}) mentioned in equations
  (\ref{scfcurnt},\ref{ferveccurnt}) .
   
  A magnetic  current $j_{\m}$  
  behaves under $P'$ and $T'$ as
\begin{eqnarray}
\label{vectT'P'}
\begin{array}{cc}
{j_{0}(x)  \buildrel P'\over \rightarrow\ -j_{0}(x_P)}, \qquad
{j_{0}(x)  \buildrel T'\over \rightarrow\ -j_{0}(-x_P)}\\
{j_{i}(x)  \buildrel P'\over \rightarrow\ j_{i}(x_P)},\qquad
 {j_{i}(x)  \buildrel T'\over \rightarrow\ j_{i}(-x_P)}.
\end{array}
\end{eqnarray}
  The transformation (\ref{vectT'P'}) are the same as Eq.(\ref{PTm})
   except $T$and $P$ are replaced by $T'$ and $P'$.

      By defining parity and time reversal  in this way,
      we have removed the possibility of having both electric and magnetic charge 
        of a field (dyon). In this formalism the current 
        constituted from a field transformation either under  $ P' $ and $ T' $
       as in Eq. (\ref{vectT'P'}) or under $ P $ and $ T $ as in Eq. (\ref{PTe}), but not both.

       We can split Fock space into two spaces for all particles one for only magnetic
 monopoles \,($\cal{M} $) and the other for all other particles 
 which do not have any magnetic charge ($\cal{C}$). 
Then we can  define parity, time reversal and charge conjugation
for both the spaces and then take the direct sum of operators 
of the two spaces. We define $P$, $T$ and $C$ for $\cal{C}$ 
and $PM$, $TM$ and $M$ for $\cal{M}$ (\ref{ramsey}) 
where $P$, $T$ and $C$ are ordinary parity,  charge 
conjugation and time reversal and $M$ is the monopole 
conjugation. Now if we take  the direct sum of operators 
of the two spaces we get  $C \oplus M$, $P \oplus PM$, $T \oplus TM$. 
These are the charge conjugation, parity and time reversal 
operators for the whole space.

       So far we have discussed  particle currents for 
magnetic monopoles, but there is another possible way of 
constructing a monopole current.  One can think of a monopole  
as a topological defect and the corresponding monopole currents will be topological currents.
 A beautiful example of such a monopole 
 is the 't Hooft-Polyakov  monopole solution in non-Abelian gauge theory \cite{'tHooft:1974qc, Polyakov:1974ek}
   We will discuss it in detail in chapter \ref{t'hooft_monopole_chapter}.  In this case the monopole current is
\begin{equation}
\label{hooft}
j_m^{\m} = \e^{\m\n\r\l}\e_{abc}\d_{\n}\phi^a\d_{\r}\phi^b\d_{\l}\phi^c. 
\end{equation}
Here $ \phi^i $ is the $ i $-th component  of an SU(2) adjoint scalar field.
    This is the current which has the P and T properties 
    exactly like a magnetic monopole current (\ref{PTm}) 
    in Maxwell equation. But here the fields are in the 
    adjoint representation of a non-Abelian gauge group 
    and the action of time reversal operator on these fields 
    are different from Abelian fields because the generators 
    of the Lie-algebra contains the factor '$ i $', that changes 
    under the time reversal.

     In general  a magnetic monopole current  
     can be treated as a topological current and can be written as
\begin{equation}
j^\m = \e^{\m\n\r\l}\d_{\n}B_{\r\l},
\end{equation}
where $B_{\m\n}$ is an antisymmetric tensor field, which may be constructed
out of field already present in the theory.
For example $B^{\m\n} =  \e^{\m\n\r\l}\e_{abc}\phi^a\d_{\r}\phi^b\d_{\l}\phi^c $ for the current 
written in Eq. (\ref{hooft}).
 In fact $ B_{\m\n} $ can also be constructed by fermionic field as,
\begin{equation}
\label{pauli}
j^{\m}(x)= g\e^{\m\n\r\lambda}\d_{\n}\bar{\psi}\sigma_{\m\n}\psi.
\end{equation}
However, the above (\ref{pauli}) current {\cite{Abdus Salam:1966}} has  mass
 dimension four which means that it is non-renormalizable.

\section{Charge quantization and the Proca model}
In this section we will discuss the classical static monopole 
 in Proca massive Abelian vector field theories in three space 
and one time dimensions. We solve the equations of 
motion of massive electromagnetic field for a static magnetic monopole and an static
electrically charged particle. We
calculate the angular momentum as well for this system to see whether 
its quantization leads to the quantization of charge.

\subsection{Equations of motion and their solution }
The Lagrangian density for electromagnetism with a mass term is
\begin{eqnarray}
\label{procalagrangian}
\M{ L} = - {1\over4}F_{\mu\nu}F^{\mu\nu} + m^2 A^{\mu}A_{\mu}.
\end{eqnarray}

          The field equations of motion with  
a point electric charge $ q $ 
at the origin and a point magnetic charge $ g $ at $ {\vec r} = {\vec r}_g$ are  
\begin{eqnarray}
\label{proca1}
\nabla \cdot{\vec E} &=&- m^2 A^0 + q\delta^3(\vec r),\\
\label{proca2}
{\nabla{\times {\vec B}}} - {\frac{\partial {\vec E}}{\partial t}}
 + m^2 {\vec A}&=& 0,\\ 
 \label{proca3}
 \nabla \cdot{\vec B} &=& 
 g \delta^3 ({\vec r} - {\vec r}_g),\\
 \label{proca4}
  \nabla{\times{\vec E}} + 
 {\frac{\partial{\vec B}}{\partial t}} &=& 0 .
\end{eqnarray}
 We have added the magnetic monopole source to the right hand side
 of Eq.(\ref{proca3}), as we had done for the Maxwell theory.
    We have chosen the
vector potential $ A_\m $ to be time independent because 
the sources are static. 
          The static solution of Eq. (\ref{proca1}) is the Yukawa potential          
\begin{eqnarray}
A^0 &=& q{e^{-mr}\over r}  ,\\ \label{massiveelectricfield}
\vec E &=& q {{e^{-mr}(1+mr)}\over r^2}\hat{r}.
\end{eqnarray}
The electric field falls off exponentially rapidly at distances 
larger than $ m^{-1}$,  i.e. electric fields are screened.
  

   Let us write the magnetic field as
\begin{eqnarray}
\label{procamagneticfield}
 {\vec B} = {\frac g{4\pi}}{\frac {({\vec r}-{\vec r}_g)}{|{\vec r} -{\vec  r}_g|^3}}
  + \nabla \times \vec A,
\end{eqnarray}
 where  $\vec A $ is non singular.  We can insert the expression of
 magnetic field from   Eq. (\ref{procamagneticfield}) into 
  Eq. (\ref{proca2})  to find
\begin{eqnarray}
\label{procaeq1}
\nabla \times \left(\nabla \times \vec A\right) &=& - m^2 \vec A ,
\end{eqnarray}
 because the electric field {$\vec E$} is time independent. This equation simplifies to
\begin{eqnarray}
\label{procaeq2}
 {\nabla}(\nabla \cdot \vec A) - {\nabla}^2 \vec A &=& - m^2 \vec A,
\end{eqnarray}
and the  divergence of Eq.  (\ref{procaeq2}) gives 
\begin{equation}
\label{procacoulombgauge}
\nabla \cdot \vec A = 0. 
\end{equation}
  It follows that Eq. (\ref{procaeq2}) can be written as
\begin{eqnarray}
\label{Proca3}
 {\nabla}^2 {\vec A} - m^2 {\vec A} &=& 0 .
\end{eqnarray}
This is actually three equations for three Cartesian components.
Taking the scalar product of {$\vec A$} with the above Eq. (\ref{Proca3}) 
and integrating over the whole volume we get
\begin{eqnarray}
\label{Proca4}
\int [{\vec A}\cdot {\nabla}^2 {\vec A} - m^2 {\vec A}\cdot{\vec A}] \: d\tau &=& 0.
\end{eqnarray}
Integrating by parts we get
\begin{eqnarray}
\label{proca6}
\int [\nabla \cdot ({\vec A}\cdot \nabla {\vec A}) -
 (\nabla {\vec A})^2 - (m {\vec A})^2 ]\:d\tau &=& 0 .
\end{eqnarray}
Here we assume that  $ \vec A $ goes to zero on the boundary.  
So the equation (\ref{proca6}) becomes
\begin{eqnarray}
\label{proca7}
\int[(\nabla {\vec A})^2 + m^2{\vec A}^2]\:d\tau &=& 0 .
\end{eqnarray}
In the above expression the integrand is positive definite. So the integrand has
 to vanish but it is a sum of squares. So the only solution of the equation 
 (\ref{proca7}) is $ \vec A = 0 $. Then  the only non singular $ \vec A $ in Eq. (\ref{procamagneticfield})
  is  $ \vec A = 0 $ without any electric current. In fact even in the absence of a monopole this says that
  static $ \vec A = 0 $ in the absence of an electric current.

 \subsection{The massive Biot-Savart law }

      If there is a steady current density $ {\vec J}$ in the system then Eq. (\ref{proca2}) becomes
\begin{eqnarray}
\nabla\times{\vec B} + m^2 {\vec A}&=& {\vec J}.
\end{eqnarray}
 Using the definition of vector potential and Eq.
 (\ref{procacoulombgauge}) we can write  the
 above equation as
\begin{eqnarray}
{\nabla}^2{\vec A}- m^2 {\vec A} = - {\vec J},
\end{eqnarray}
because $\nabla\cdot{\vec J} = 0$.
These are nothing but  three  Helmholtz equations,  
one for each Cartesian component. 
 Assuming that ${\vec J}$ goes to zero (outside a compact region), we can read off the solution
\begin{eqnarray}
\label{massivevectorpotential1}
{\vec A(r)} = \int e^{-m|\vec r - \vec r'|} {{\vec J(r')}\over |\vec r - \vec r'|}\: d\tau', 
\end{eqnarray}
where $ d\tau'$ is the infinitesimal volume element.

     This formula gives the vector potential at a point ${\vec r}=(x,y,z)$
      in terms of an integral over the current distribution 
      ${\vec J}(x',y',z')$.  The divergence and the curl are 
      to be taken with respect to the unprimed coordinates.
\begin{eqnarray}
\label{massiveB1}
\nabla\times{\vec A(r)} &=& \int \left[\left(\nabla
{e^{-m|\vec r-\vec r'|}\over |\vec r-\vec r'|}\right) 
\times {\vec J(r')}\right] d\tau', 
\end{eqnarray}
since $\nabla \times{\vec J(r')}=0\,$  (because ${\vec J(r')}$
 does not depend on unprimed coordinates).  so according to Eq. (\ref{procamagneticfield}),
  the magnetic field is
\begin{eqnarray}
\label{massivemagneticfield}
{\vec B}({\vec r}) = {g\over 4{\pi}}{({\vec r}-{\vec r}_g) 
\over {|{\vec r} -{\vec r}_g|^3}} + \int (1+m|\vec r-\vec r'|)
e^{-m|\vec r-\vec r'|}{{{\vec J(r')}\times \hat{r}}\over {|\vec r-\vec r'|^2}}d\tau'.
\end{eqnarray}
If there is no magnetic monopole, the magnetic field becomes
\begin{eqnarray}
\label{massivebiosavart}
{\vec B}({\vec r}) = \int (1+m|\vec r-\vec r'|)e^{-m|\vec r-\vec r'|}
{{{\vec J(r')}\times \hat{r}}\over {|\vec r-\vec r'|^2}}d\tau'.
\end{eqnarray}
This is the Biot-Savart law in massive electromagnetism.
\subsection{Calculation of angular momentum and quantization condition}
The energy momentum tensor for the Lagrangian of Eq. (\ref{procalagrangian}) is
\begin{eqnarray}
\label{energytensor}
T^{\mu\nu} =  - {F^{\mu}} _{\rho}F^{\nu\rho} + 
{1\over4} \eta^{\mu\nu}F^{\alpha\beta}F_{\alpha\beta}
 + m^2(A^\mu A^\nu - {1\over2} \eta^{\mu\nu}A^\alpha A_\alpha).
\end{eqnarray}
The Hamiltonian (energy) and  momentum density are  $ T^{00}$ and $ T^{0i}$, respectively,
which we can write using  Eq. (\ref{energytensor}) as 
 \begin{eqnarray}
 \mbox{Energy} = T^{00} &=& {1\over2}({\vec E}^2 + {\vec B}^2) + {m^2\over2} [({A^0})^2 + (A^i)^2],\\
{\wp}^i= T^{0i} &=& ({\vec E}\times{\vec B})^i + m^2 A^0 A^i,\\
\int {\vec\wp}\: d\tau = {\vec P},
\end{eqnarray}
 where ${\vec P}$ is the total momentum.
 For the  system we are considering, i.e. a static electric charge
 at the origin and a static monopole at $ \vec r_g $, the electric current
 vanishes  $ \vec J(\vec r) = 0 $. So following the equations (\ref{proca7}) and (\ref{massivevectorpotential1})
 we can write ${\vec A} = 0 $.
     Then a standard calculation {\cite{Jackson}} gives ${\vec P} = 0$.
     
The angular momentum for the electromagnetic field is
\begin{eqnarray}
\label{angularmomentum1}
\vec L &=& \int \vec r \times \vec{\wp} \: d\tau . 
\end{eqnarray}
Since the total momentum $ \vec{P} $ vanishes for our system, we can 
calculate this about any origin. Then using Eq. (\ref{massiveelectricfield}) and Eq. (\ref{massivemagneticfield})
   with $ \vec J(\vec r) = 0 $ (the fact that $ \vec{A} = 0 $ for $ \vec J = 0$), we can write
\begin{eqnarray}
\vec L &=& \int {\vec r} \times \left({q \over 4{\pi}} 
{{{ e}^{-mr}(1+mr)}\over r^2}\hat{r} \times 
{g\over 4{\pi}}{{{\vec r}-{\vec r}_g} \over |{{\vec r} - {\vec r}_g|^3}}\right)\\
 &=& \int {f(r) \over r}({\hat {r}} \times ({\hat {r}} \times {\vec B}))\: d\tau,\\
  &=& - \int {f(r) \over r}({\vec B} 
-({\vec B}\cdot \hat {r}) \hat {r})\: d\tau,
\end{eqnarray}
where $ f(r) = {\displaystyle{{q \over 4{\pi}}} 
{{ e}^{-mr}(1+mr)}}$, and the integration is over all space.
  Using the identity $\displaystyle{{\vec B}\over r} =
 \displaystyle{{\hat{r}(\hat{r}\cdot {\vec B})}\over r} + 
 ({\vec B}\cdot \nabla)\hat{r}$ we get
\begin{eqnarray}
 {\vec L} &=& - \int f(r)({\vec B} \cdot \nabla) \hat {r} \:d\tau,\\
 \label{massiveangularmomentum1}
 &=& \int \hat{ r} \nabla \cdot (f(r) {\vec B})
\:d\tau - \int_S  \hat {r}  (f(r) {\vec B} \cdot {\vec ds} ).
\end{eqnarray}
The second term of Eq. (\ref{massiveangularmomentum1})
is zero because it is a angular average of the vector $ \hat r $ over the surface $ S $
and  $f(r)$ also vanishes at infinity. 
So we can write Eq. (\ref{massiveangularmomentum1}) as
\begin{eqnarray}
 {\vec L} &=& \int {\hat {r}} (\nabla f \cdot {\vec B}) 
 \:d\tau + \int \hat {r} f(r) \nabla \cdot {\vec B} \:d\tau.
\end{eqnarray}
Using Eq. (\ref{proca3}) the above equation becomes,
\begin{eqnarray}
{\vec L} &=& \int \hat {r} \nabla f \cdot {\vec B}
 \:d\tau + \int \hat {r} f(r) g \delta^3({\vec r}- {\vec r}_g)\:d\tau\\
&=& \int \hat {r} \nabla f \cdot {\vec B} 
\:d\tau + g f({\vec r}_g){\hat {r}_g} \\ 
&=& \int \hat {r} \nabla f \cdot {\vec B} 
\:d\tau + {qg \over {4 \pi}}e^{-m r_g}(1+mr_g){\hat {r}_g}\\
&=& - {q \over {4 \pi}} \int {\vec r} m^2  
e^{- m r}(\hat {r} \cdot {\vec B})\:d\tau +
 {qg \over {4 \pi}}e^{-m r_g}(1+mr_g){\hat{r}_g}.
\end{eqnarray}
As mentioned earlier, the total linear momentum of the system is zero,    
so the angular momentum is independent of the 
origin of the coordinate system. For simplicity we can choose the
$ z $- axis along $\vec{r}_g $ and write 
     $ {\vec r}_g = a \hat {k}$ and the angular momentum becomes
\begin{eqnarray}
\label{l1}
{\vec L} &=& 
 - {q \over {4 \pi}} \int \hat {r} m^2 r e^{- m r}
 \left(\hat {r} \cdot {g \over {4 \pi}}{{{\vec r} 
 - a \hat {k} } \over {{|{\vec r}- a \hat {k}}|}^3}\right)\:d\tau + {qg \over {4 \pi}}(1+ma)e^{-m a}\hat{k} ,\\
 \label{l2}
&=&  - {{m^2qg}\over{16{\pi}^2}}\int {e^{-mr}({r^2 - ar \cos\theta})
\over{(r^2+a^2 - 2ar {\cos\theta})^{3\over2}}}\hat { r}\: d\tau + {qg \over {4 \pi}}(1+ma)e^{-m a}\hat{k} ,
\end{eqnarray}
where $\cos\theta = \hat{r}\cdot \hat{k}$. To calculate the
integral of Eq. (\ref{l2}) let
us define
\begin{eqnarray}
\label{u}
 U(r) &=& (r^2 + a^2 - 2ar \cos \theta)^{-{1\over2}}, \\
 \label{u'}
 {dU\over dr} &=& -{(r - a \cos\theta)\over(r^2 + a^2 - 2ar \cos\theta)^{3\over2}},
\end{eqnarray}
so that 
\begin{eqnarray}
{\vec L}= {{m^2qg}\over{16{\pi}^2}} \int e^{-mr} r^3 {{dU(r)}\over{dr}}
\:dr \sin\theta\:d\theta\:d\phi \hat{r} + {qg \over {4 \pi}}(1+ma)e^{-m a}\,\hat{k}.
\end{eqnarray}
Integrating by parts over $r$ we get
\begin{eqnarray}
{\vec L}&=& -{{{m^2qg}\over{16{\pi}^2}}\int {{(3r^2 -mr^3)e^{-mr}\sin\theta}
\over{(r^2+a^2 -2ar \cos\theta)^{1\over2}}} \hat{r}\:d\theta \:d\phi 
\:dr + {qg \over {4 \pi}}(1+ma)e^{-ma}\,\hat{k}},\nonumber \\
&=& - {{m^2qg}\over{8\pi}}\int {{(3r^2 -mr^3)e^{- mr}\sin\theta}
\over{(r^2+a^2 -2ar \cos\theta)^{1\over2}}} \cos\theta \,\hat{k}\:d\theta  \:dr \nonumber \\
&& \phantom{xxxxxxxxxxxxxxxxxxxxxxxxxxxxxxx} + {qg \over {4 \pi}}(1+ma)e^{-m a}\,\hat{k}.
 \label{massiveangularmomentum2}
\end{eqnarray}
Using the identity
\begin{eqnarray}
\label{lagendreseries}
{1\over(r^2+a^2 - 2ar \cos\theta)^{1\over2}} = {2\over (2l+1)}\,
 \sum_{l =0}^{l = \infty}{{r_<^l}\over{r_>^{l+1}}}\,\, P_l(\cos\theta),
\end{eqnarray}
where  $P_l(\cos\theta)$  is the Legendre polynomial of order $ l$, we can calculate from 
Eq. (\ref{massiveangularmomentum2})  that,
\begin{eqnarray}
{\vec L} &=& - {{m^2qg}\over{8\pi}}\int_{0}^{\infty} 
(3r^2 -mr^3)e^{- mr}\left[\int_{-1}^{1} P_1(\zeta)
{2\over (2l+1)} \sum_{l =0}^{l = \infty}
{{r_<^l}\over{r_>^{l+1}}}P_l(\zeta)\:d\zeta\right]\hat k \nonumber \\
 &&\phantom{xxxxxxxxxxxxxxxxxxxxxxxxxxxxxx}+ {qg \over {4 \pi}}(1+ma)e^{-m a}\,\hat{ k},
 \label{massiveangularmomentum3}\\
&=& - {qg \over {4 \pi}} \left[2 {e^{-m a} \over {ma}}+ 2{{e^{-m a} - 1} 
\over m^2a^2}\right]\hat{k},  
 \label{massiveangularmomentum4}
\end{eqnarray}
where $\zeta = \cos\theta$.
The exponential can be expanded  and the result becomes
\begin{eqnarray}
\label{massiveangularmomentum}
 {\vec L} =  {{qg} \over {2\pi}} \sum_{n = 0}^{n = \infty} 
 {(-1)^n \over (n+2)}{{(ma)}^n \over {n!}}\hat{k}.
\end{eqnarray}

 The above solution gives the angular momentum of the
   massive electromagnetic field for a charge -magnetic monopole pair. 
The angular momentum depends on two
continuous parameters,  the position $ a $
of the monopole relative to the electric charge and
 the mass $ m $ of the photon and also on the product $ qg $. 
   Quantizing angular momentum does not lead to any simple quantization rule for the
   electric and magnetic charges.
      For $ { m \rightarrow 0}$ we  recover the angular
       momentum calculated by J.J.Thomson,   i.e
        $ |{\vec L}| = \displaystyle{{gq}\over{4\pi}}$ and we can
        quantize electric charge by quantizing the angular momentum.
          
          So, we can conclude that to quantize electric charge
        by quantizing angular momentum, the mass of the photon must be zero.
        Alternatively we can also say that magnetic monopoles 
        in massive electrodynamics does not lead to quantization of electric charge.
        
\chapter{Flux tubes and Confinement of monopoles in Abelian theory}

Experiments show that the angular momentum of every hadron varies linearly
with the square of its mass. By some indirect experimental result
and computer simulation, it is also known that the static potential varies linearly
with the distance between two quarks. These two facts can be simultaneously
satisfied if field lines between two quarks are confined inside a 
thin flux tube such as a Nielsen-Olesen string. In this section we discuss  
 magnetic flux tube configurations in the Abelian Higgs model and
show how we can get a taste of confinement by confining
magnetic monopoles with these Abelian magnetic flux tubes. 
\section{Vortices in Abelian Higgs model}
We start with the Abelian Higgs model,
\begin{eqnarray}
\label{abelianhiggslag}
{\mathcal L} = - \frac{1}{4} F^{\m\n}F_{\m\n} 
+ \half |D_\mu\Phi|^2 - \lambda (|\Phi|^2 - v^2)^2.
\end{eqnarray}
Here $ \Phi $ is a complex scalar field which we decompose as
\begin{equation}
\Phi = \rho e^{i\chi}.
\end{equation} 
The equations of motion are
\begin{eqnarray}
\d_\m F^{\m\n} + e\rho^2 (\d^\n \chi + e A^\n)= 0,\\
\Box \rho - \rho(e A_\m + \d_\m \chi)^2 + \lambda (\rho^2 - v^2)\rho = 0,\\
\d_\m[\rho^2(\d^\m \chi + e A^\m)] = 0.
\end{eqnarray}
The Hamiltonian density can be written as
\begin{eqnarray}
{\mathcal H} = \half {\vec E}^2 + \half {\vec B}^2 +
+\half (\d_0 \rho)^2 + \half(\nabla \rho )^2 + 
 \frac{\rho^2}{2} (e\vec A_0 + \d_0 \chi)^2 +  \frac{\rho^2}{2}(e\vec A
 - \nabla \chi)^2 \nonumber\\ \phantom{xxxxxxxxxxxxxxxxxxxxxx}+ \frac{\lambda}{4}(\rho^2 - v^2)^2.
\end{eqnarray}
We can write  the vacuum configuration as
\begin{eqnarray}
\label{vacuum1}
\rho &=& v\\
A_\m &=& -\frac{1}{e}\d_\m \chi .
\end{eqnarray}
Instead of taking this as vacuum configuration, it is possible to choose a new gauge
 where we can gauge away the $ \d_\m \chi $.
The behavior of the small fluctuations around the vacuum
(\ref{vacuum1}) can be written as
 \begin{eqnarray}
\rho &=& v +  \tilde{\rho}\\
eA_\m &=& - \d_\m\chi +  ea_\m .
\end{eqnarray}
The Lagrangian can effectively be written as,
\begin{eqnarray}
\mathcal{L} = - \frac{1}{4} F^{\m\n}F_{\m\n} 
+ \half e^2v^2 a_\m a^\m + \half \d_\m \tilde{\rho}\d^\m \tilde{\rho}
+ \l v^2 \tilde{\rho}^2 + \mbox{higher order terms},
\end{eqnarray}
and  the equations of motion upto
first order are
\begin{eqnarray}
 \Box  a^\m + e^2 v^2 a^\m &=& 0 \\
 \Box  \tilde{\rho} + 2\l v^2 \tilde{\rho} &=& 0.
\end{eqnarray}
These are the equations of the massive electromagnetic field and a real scalar field.
So the excitations $ a_\m$ and $ \tilde{\rho} $ generate particles
with masses $ ev$ and  $ v \sqrt{2\l} $, respectively. 

 Here we have not given any vacuum solution of the field $ \chi(x) $. This is because
 we have assumed that we can gauge away all configurations of
the field $ \chi(x) $ at the vacuum so that it does not contribute to any physical quantity.
 However  this is not always
true. Let us consider a static solution in cylindrical coordinates
where $ \chi $ and $ |\Phi| $ are independent of the coordinate $ z $. 
It can be written as $ \chi(x,y) = \chi(r, \varphi) $.
 Let us draw a circular loop around the $ z $-axis and demand that
\begin{equation}
\label{theta}
\Phi(r,\varphi) = \Phi(r, \varphi + 2 \pi).
\end{equation}
It follows from Eq. (\ref{theta})  that
\begin{eqnarray}
|\Phi(r, \varphi , z)| &=& |\Phi(r, \varphi + 2\pi, z)|,\\
\label{chiperiod}
\chi(r, \varphi + 2\pi) &=& \chi(r, \varphi) + 2 n\pi.
\end{eqnarray} 
In other words there are  solutions for which $ \chi $ is not a single valued
function. $ \chi $ varies by $ 2\pi n $
($n$ = integer) when we make a complete turn around a closed loop 
and $ \chi $ is undefined along the  $ z $-axis. 
We can calculate the magnetic flux along the $ z $-axis by integrating
 the vector potential $ \vec{A}$ along a large loop  around the $ z $-axis.
On the large loop the vector potential $ \vec{A}  =  \frac{1}{e}\nabla\chi$, because
at large distances $ \vec{a} $ should vanish as it is a solution of the Helmholtz for the 
static situation.
\begin{equation}
\label{loopint-abelianhiggs}
\tilde\Phi = \oint \vec A \cdot \vec dl = \frac{1}{e} \oint \nabla\chi\cdot \vec dl
= \frac{2\pi n}{e},
\end{equation}
\begin{equation}
\mbox{i.e.,}\qquad \tilde\Phi = n \Phi_0,\,\, \Phi_0 =  \frac{2\pi}{e}.
\end{equation} 
The  non-trivial result of the line integral  shows a singular magnetic field along $ z $-axis
and we will call it a flux tube or flux string.
So in general  we can decompose the field
in two parts $ \chi = \chi_r + \chi_s $, where the $ \chi_r $ is the regular part that
can be gauged out by the vector potential and goes to zero at very large distances,
whereas the other part $ \chi_s $ is the multivalued part and this part does not go
to zero at large distances.

The configuration $ \chi_s $ is one of a class of vacuum field configurations
 that are important for topological defects
in field theory. These configurations do not go to zero  at infinity.
 In the Abelian Higgs model these field configurations generate a singular
  magnetic field along the $ z $-axis.
 We can regularize the stress-energy tensor by taking
$ |\Phi| = 0 $ along the $ z $-axis. However, since we need a finite energy configuration
as the starting point of perturbation theory, we must have $ |\Phi| = v$ at spatial 
infinity. 
We can try to get a solution for which $ |\Phi| $ is constant
  far away from the $ z $-axis and goes  smoothly  to zero on  the $ z $-axis.
Far away from the $ z $-axis, $ |\Phi| $ reaches its vacuum value and the vector
potential becomes 
\begin{eqnarray}
\vec A =  \frac{1}{e}\nabla\chi,
\end{eqnarray}
and we can decompose the field $ \chi $
in two parts $ \chi = \chi_r + \chi_s $ very far away from the $ z $-axis.

Geometrically we can say that at large distances the field $ \Phi = ve^{i\chi} $ defines a mapping of a circle
  in space to a circle of radius $ v $ at the plane of complex $ \Phi $.
  The mapping of one circle to another circle can be represented by   $ \pi_1(S^1)$, the fundamental group 
  of the circle in the plane of complex field. We know that   $ \pi_1(S^1) = Z $,
  so the mapping is characterized by an   integer $n = 0,1,2, \cdots,$ the winding
number.  The winding number determines the amount of flux through the flux tube and for any flux tube
configuration $ n \neq 0 $.

To see the exact static behavior of the fields $ \vec A(x)$ and $ \rho(x) $ for flux tube, we have
to solve  the equations of motion for the static case. However, till date
an exact solution has not been found for a flux tube configuration. We can solve
the equations in different regions of space and then we can try to match
the solutions to see the whole picture. Let us write down the equations of
motion in cylindrical coordinates and make all the fields independent of the
$ z $ coordinate. From symmetry and using  the asymptotic solution, we can write 
\begin{equation}
\vec A = A_\varphi \hat\varphi.
\end{equation}
Using Eq.(\ref{chiperiod}) we can write 
\begin{equation}
\chi(r, \varphi) = \varphi
\end{equation} 
for $ n=1 $.
So the equations of motion become
\begin{eqnarray}
\label{eqnaphi}
- \frac{\d}{\d r} \frac{1}{r}\frac{\d}{\d r}\left(r A_\varphi \right)
+ e^2 \rho^2 \left(A_\varphi - \frac{1}{r}\right) = 0, \\
-\frac{1}{\rho} \frac{\d}{\d r}\left(r \frac{\d}{\d r} \rho\right)
+ \left(eA_\varphi - \frac{1}{r}\right)^2\rho + \lambda\left(\rho^2 - v^2\right) = 0.
\label{eqna}
\end{eqnarray}
An exact analytic solution of the above equations is not known.
We can get a solution of  Eq. (\ref{eqnaphi}) at larger $ r $ if we write
\begin{equation}
\rho \simeq v\,
\end{equation}
which is valid for large $ r $. Then a solution for $ A_\varphi $ is \cite{Nielsen:1973cs}
\begin{eqnarray}
\label{Aabelianhiggs}
A_\varphi &=& \frac{1}{e r} + \frac{c}{e}K_1(ev r) 
\stackrel{r\rightarrow \infty}\rightarrow 
\frac{1}{er} + \frac{c}{e} \sqrt{\frac{\pi}{2ev r}} e^{- ev r} + \mbox{higher order terms},
\end{eqnarray}
where $c$ is a constant of integration.
The magnetic field has only its $ z $-component and can be written as
 \begin{eqnarray}
 \label{Babelianhiggs}
 B_z = cv K_0(evr)\stackrel{r\rightarrow \infty}\rightarrow 
 \frac{c}{e} \sqrt{\frac{v\pi}{2e r}} e^{- ev r} + \mbox{higher order}.
\end{eqnarray}
We  define a characteristic length $ \tilde\lambda $ as,
\begin{equation}
\tilde\lambda = \frac{1}{e v},
\end{equation}
which is similar to, in fact corresponds to,  the penetration length in superconductivity.
$ \tilde\lambda $ thus measures  
the region over which the field $ B_z $ (\ref{Babelianhiggs}) is appreciably
different from zero.
To see the configuration of $ \rho $ around the vacuum value, we write
\begin{eqnarray}
\label{rho}
\rho = v + \tilde{\rho}.
\end{eqnarray}
After substitution of Eq.(\ref{rho})  and Eq. (\ref{Aabelianhiggs}) 
into Eq. (\ref{eqna}) and neglecting the higher order, 
 we see that the solution behaves as
\begin{eqnarray}
\label{roabelianhiggs}
\tilde{\rho} \sim e^{-v\sqrt{2\lambda}r}.
\end{eqnarray}
Here we can define a new characteristic length $ \xi $
\begin{eqnarray}
\xi = \frac{1}{v\sqrt{2\lambda}}.
\end{eqnarray}
Thus  $ \xi $ measures the radial distance it  takes for 
the field $ \rho $ to reach its vacuum
value $ v $.

 From the above discussion
 we see that for
flux tube solutions $ \tilde{\lambda}$ must be greater than or equal to $ \xi  $ because the magnetic field
always has a decaying tail after $ \rho $ reaches its vacuum value $ v $. At a length scale less than $\xi $,
there is a region where photon is massless. So we can  say
$ \xi $ is roughly the radius of the core of flux tube. Flux tubes
can be treated as strings if the core is very small. At a length scale  greater 
than $\xi  $  but less than $ \tilde{\lambda}$  fields are getting their masses, and 
above the length scale $ \tilde{\lambda} $, there is almost nothing.
However, the field $ \chi $ is still non zero very far away from the string core.
The integration of Eq. (\ref{loopint-abelianhiggs}) showing that
$ \chi $ can  inform us about the existence of flux tube in a region, where practically 
there exists no physical field.

In chapter $ 1 $ we mentioned that the angular momentum of  a string is proportional to its
 energy squared, i.e. $ J = \sigma' E^2  $, where  $ \sigma' = \frac{1}{2\pi\sigma} 
$ (universal slope of Regge trajectory) and $ \sigma $ is the static mass per unit length of a strings.
 So for massless quarks the mass of a hadron  for lower quantum mechanical levels is $ M_H \sim \frac{1}{\sqrt{\sigma'}} $.
 It is  possible to make a classical estimate of the relation of the universal slope
to the three parameters, $ v $, $ \lambda $ and $ e $
by calculating the energy density at rest for the vortex solution.
We can make a crude estimate of the energy per unit length 
for the flux tube constructed form the Abelian Higgs model,
\begin{eqnarray}
\label{alphav2}
\frac{1}{\sigma'} = \frac{\mbox{Energy}}{\mbox{length of the string}}   
 \sim v^2.
\end{eqnarray}  
The exact ratio $ v^2 \sigma' $ can be computed numerically by solving the differential
equations. What is important, however, is that it is of order unity.

We have seen that the two particles from the
fluctuation of the fields $ \rho $ and $ A_\m $ got 
their masses $ M_s  = v\sqrt{2 \lambda} $  and  $ M_v = ev $ by Higgs mechanism.
The strong coupling limit in the Abelian Higgs model is defined by
setting the coupling constant to be large, i.e.
\begin{eqnarray}
e\gg 1,\qquad \lambda \gg 1.
\end{eqnarray}
In the strong coupling limit we can write using Eq. (\ref{alphav2}) that
\begin{eqnarray}
ev\frac{1}{v}\gg 1 \Rightarrow M_v \gg \frac{1}{\sqrt{\sigma'}},\\
\sqrt{2\lambda} v\frac{1}{v} \gg 1 \Rightarrow M_s \gg \frac{1}{\sqrt{\sigma'}}.
\end{eqnarray}
It follows that $ M_v, M_s \gg M_H $
for strong coupling, 
 which means that the particles corresponding to the local fields $ \rho, $ and $ A_\m $ have masses
$M_v$ and $M_s$ much larger than the typical hadron masses. Thus in this limit low energy
phenomena (low energy meaning energies of the order of $ \frac{1}{\sqrt{\sigma'}} $)
should be dominated by hadrons, i.e. dual strings. 
\section{Flux tubes and effective strings}
\label{effectivestrings}
In this section we will try to describe flux tubes as effective strings.
In the last section we have discussed that there are two length scales 
$ \xi $ and $ \tilde{\lambda} $. The scalar field reaches its vacuum value
 after a distance $ \xi $ from the axis. The magnetic  field
vanishes  after a distance $ \tilde{\lambda} $ from the axis.  
We also discussed   flux tube solutions and there we saw 
that after the scalar field reaches its vacuum value there is a
tail of magnetic field  and we are interested in constructing an effective
theory in the region where  $ \xi < \tilde{\lambda} $. In the strong coupling limit
we can consider the core radius or  $ \xi $ o be  very small and the scalar field reaches its 
vacuum value $ v $  very quickly. Thus we can consider the flux tube as a singular line or a string.
 So we can  construct an effective theory in  the strong coupling limit by taking
the fields $ \Re\Phi, \Im\Phi^* $ to be zero  along the string core. 
The two equations
\begin{eqnarray}
\Re\Phi(\tilde x) = 0,\qquad \Im \Phi(\tilde x) = 0,
\end{eqnarray}
define a two dimensional worldsheet of the string.
The string coordinates are the coordinates of the worldsheet on which the scalar field vanishes. 
We can introduce this in the generating functional by  inserting 
the identity \cite{Baker:1999xn}
\begin{eqnarray}
1 = J^{-1}(\Phi, \tilde x) \int D\tilde{x}\, \delta(Re\Phi(\tilde x))
\delta(Im \Phi(\tilde x)), 
\end{eqnarray}
 and then we can integrate over the fields $ \Phi $, $ \Phi^* $. 
 By doing so we expect that the remaining theory will be a theory of interacting strings.
 
There is another way to construct  the effective theory of strings and this is more geometrical
than the previous one. We have seen that the flux tube solution
is characterized by $ \chi = \chi_s $  being an angular coordinate around
the flux tube. This means that we are defining a map from a spatial circle to 
a circle in field space. This mapping is done by taking $ \chi $ as 
 a function of spatial azimuthal angle $ \varphi $ such that when $ \varphi $ 
 changes from $ 0$  to $2\pi$, $  \chi $ changes from $ 0$ to $2n\pi $.
 The number $ n $ represents the first homotopy class of the mapping. 
This number is the quantum number using which we can identify the flux tubes.
When $ \xi $ is  zero (string core),  we can take $ \chi $ as a
variable to describe the string. Let us consider an example where this is seen easily.
We set $ \chi_s = \varphi  $, where $ \varphi $ is the polar angle
on the $ xy $ plane. Then we can write
\begin{eqnarray}
\tan \varphi &=& \frac{y}{x},\\
\nabla \varphi &=& \frac{1}{x^2+y^2}\left[-y\, \hat i + x \,\hat j\right].
\end{eqnarray}
Now we   calculate the magnetic field due to our asymptotic  vector potential 
$ \vec A = \frac{1}{e}\nabla \varphi $.  Let us 
define a regularized vector potential,
\begin{eqnarray}
\vec A_\e = \frac{1}{e}\frac{1}{x^2+y^2 + \e^2}\left[-y\, \hat i + x \,\hat j\right].
\end{eqnarray}
The regulated magnetic field along the $ z $ axis can be written as  
\begin{eqnarray}
\label{regB}
B^{\e}_z =\frac{1}{e} \frac{2\e^2}{(x^2 + y^2 + \e^2)^2}.
\end{eqnarray}
This function is zero everywhere but singular at $ r = 0 $ in the limit $ \e=0 $ . 
The nature of the singularity can be 
seen if we integrate  $ B^{\e}_z $ over the two dimensional space perpendicular to the string.
\begin{eqnarray}
\int B^{\e}_z \,dxdy &=&\int_0^\infty \frac{1}{e}\frac{2\e^2\,dx\,dy}{(x^2+y^2+\e^2)^2}\\
&=& \int_0^\infty \frac{2\e^2\,2\pi\, r \, dr}{e( r^2+\e^2)^2}\\
&=& \int_0^\infty \frac{4\pi\, t \, dt}{e( t^2+ 1)^2},\, 
\mbox{for $ t = \e r $ }\\
&=& \frac{2\pi}{e}.
\end{eqnarray}
So we can write
\begin{eqnarray}
\lim_{\e\rightarrow 0} \frac{2\e^2}{(x^2 + y^2 + \e^2)^2}
= 2\pi \delta(x)\delta(y).
\end{eqnarray}
The $ z $ component of magnetic field can be written as
\begin{eqnarray}
\label{confinedstaticfield}
B_z = \frac{2\pi}{e} \delta(x)\delta(y).
\end{eqnarray}
We found that the  magnetic field  lines corresponding to the asymptotic vector potential 
are confined only at a point
on the two dimensional plane. This allows us to think of the 
flux tube as a  singular string in three dimensional space. We know from the discussion
of the previous section that we can consider flux tube as a singular string 
 when the distance scale is very large 
or the coupling constants are very large.
So we can think of this singular line as
an effective long distance description of the  flux tube.
 In 3+1 dimensions the confined magnetic field is a worldsheet
and we can define the worldsheet current as,
\begin{eqnarray}
\Sigma^{\m\n} = 2\pi n \int d\tau d\sigma 
\left(\frac{dx^\m}{d \sigma}\frac{dx^\n}{d \tau} 
- \frac{dx^\n}{d \sigma}\frac{dx^\m}{d \tau}\right)\delta^4(x - x(\sigma,\tau))
\end{eqnarray}
where $ \sigma, \tau $ are the parameters of the world sheet.
If we take a parametrization $ \tau = x^0$, $ \sigma = x^3 $, we can calculate
for the world sheet current,
\begin{eqnarray}
\label{sigma30}
\Sigma^{30} = 2 \pi n \delta(x)\delta(y).
\end{eqnarray}
This is nothing but the static magnetic field (\ref{confinedstaticfield}) along the $ z $-axis.
In fact  we can write
\begin{eqnarray}
\Sigma^{\m\n} = 2\pi n \e^{\m\n\r\l}\d_\r\d_\l \chi_s.
\end{eqnarray}
\section{Functional integral and Duality}
\label{dualization1}
In this section we will discuss  the construction of a dual theory
using functional integrals.
Let us first consider the functional integration,\begin{scriptsize}\begin{footnotesize}\end{footnotesize}\end{scriptsize}
\begin{eqnarray}
\label{generatingelectro}
Z = \int \D A_\m \exp \left[-i\int d^4 x \frac{1}{4}(\d_\m A_\n - \d_\n A_\m)
(\d^\m A^\n - \d^\n A^\m)\right],
\end{eqnarray}
which can be rewritten upto normalization constant as
 \begin{eqnarray}
 \label{fa}
 Z = \int \D B_{\m\n} \D A_\m \,\delta\left[B_{\m\n} - (\d_\m A_\n - \d_\n A_\m)\right]
 \exp \left[-i\int d^4 x \frac{1}{4}B_{\m\n}B^{\m\n}\right],
\end{eqnarray}
where we have used the identity of functional
delta\begin{equation}
1 = \int \D \phi \,\,\delta\left[\phi(x) - \tilde{\phi}(x)\right].
\end{equation}
We can exponentiate the delta functional of Eq. (\ref{fa}) by introducing an auxiliary
field $ F_{\m\n} $ and integrate over the field $ A_\m $ and $ B_{\m\n} $,
\begin{eqnarray}
 Z &=& \int \D F_{\m\n}\D B_{\m\n} \D A_\m \, \exp \left[-i\int d^4 x\left\{\frac{1}{4} B_{\m\n}B^{\m\n} + \frac{1}{4}\e^{\m\n\r\l}F_{\m\n}B_{\r\l}
 - \half\e^{\m\n\r\l}F_{\m\n}\d_\r A_\l\right\}\right]\nonumber\\
 &=& \int \D F_{\m\n} \D B_{\m\n} \, \delta(\half\e^{\m\n\r\l}\d_\n F_{\r\l})
 \exp \left[-i\int d^4 x\left\{ \frac{1}{4}B_{\m\n}B^{\m\n} + \frac{1}4{}\e^{\m\n\r\l}F_{\m\n}B_{\r\l}
 \right\}\right]\\
 \label{electroF}
&=& \int \D F_{\m\n} \,\delta(\half\e^{\m\n\r\l}\d_\n F_{\r\l})
 \exp \left[-i\int d^4 x \frac{1}{4}F_{\m\n}F^{\m\n}\right].
 \label{Freps}
\end{eqnarray}

In Eq.\, (\ref{fa}) we are integrating  over  $B_{\m\n} $ 
for which $ B_{\m\n} =  \d_\m A_\n - \d_\n A_\m$ and then
integrating over all possible $ A_\m $. In other words, we are
integrating over all those $ B_{\m\n} $ for which $ \e^{\m\n\r\l}\d_\n B_{\r\l} = 0 $
and that is exactly Eq. (\ref{electroF}). So these two equations are the same up to 
some multiplicative constant. To get Eq. (\ref{generatingelectro}) from Eq. (\ref{Freps}), we can 
solve the delta functional of Eq.  (\ref{Freps}) and write down
Eq. (\ref{fa}) with $ F_{\m\n}$ in place of $B_{\m\n} $ , then integrate over $ F_{\m\n} $.
Here all  equalities between different functional integrations are 
up to some multiplicative constant.

Let us introduce an auxiliary field $ G_{\m\n} $ into 
Eq. (\ref{generatingelectro}) 
to write
\begin{eqnarray}
\label{linearizedF}
Z = \int \D A_\m \D G_{\m\n} \exp\left[i\int d^4x 
\left\lbrace -\frac{1}{4}
G^{\m\n}G_{\m\n} + \half \e^{\m\n\r\l} \d_\m A_\n G_{\r\l}
\right\rbrace\right].
\end{eqnarray} 
The action in the functional integral (\ref{linearizedF})
is 
\begin{eqnarray}
I = \int d^4x 
\left\lbrace -\frac{1}{4}
G^{\m\n}G_{\m\n} + \half \e^{\m\n\r\l} \d_\m A_\n G_{\r\l}
\right\rbrace.
\end{eqnarray}
The equations of motion for the field $ G_{\m\n} $ are
\begin{equation}
G^{\m\n} = \half\e^{\m\n\r\l}F_{\r\l},
\end{equation}
where $ F_{\m\n} = (\d_\m A_\n - \d_\n A_\m) $. So
we can take $ G_{\m\n} $ as a dual field tensor according
to the definition given in Eq. (\ref{maxwell_monopole}).

   In the presence of an external electric source $ j^\m $  
the functional integral (\ref{generatingelectro}) can be written as 
\begin{eqnarray}
\label{dual1}
Z = \int \D A_\m \exp \left[i\int d^4 x \left\{- \frac{1}{4}(\d_\m A_\n - \d_\n A_\m)
(\d^\m A^\n - \d^\n A^\m) - A^\m j_\m\right\rbrace\right].
\end{eqnarray}
We can here introduce a dual field $ G_{\m\n} $ and integrate over $ A_\m $,
\begin{eqnarray}
Z &=& \int \D A_\m \D G_{\m\n} \exp\left[i\int d^4x 
\left\lbrace -\frac{1}{4}
G^{\m\n}G_{\m\n} + \half \e^{\m\n\r\l} \d_\m A_\n G_{\r\l} - A^\m j_\m
\right\rbrace\right]\\
&=& \int \D G_{\m\n} \delta(\half \e^{\m\n\r\l}
 \d_\n G_{\r\l} - j^\m )\exp\left[i\int d^4x 
\left\lbrace -\frac{1}{4}
G^{\m\n}G_{\m\n} 
\right\rbrace\right].
\label{dual2}
\end{eqnarray}
Eq. (\ref{dual2}) can be written as a dual
theory of Eq. (\ref{dual1}) and Eq.
(\ref{dual2}) looks the same as Eq. (\ref{Freps}) except
the Bianchi identity, written in terms of a delta functional.
So, in this case the  introduction of
 dual vector potential is not very easy. However, it becomes easy if we write  
 the form of the current as
\begin{equation}
\label{topocurrent}
j^\m = \half\e^{\m\n\r\l}\d_\n R_{\r\l}.
\end{equation}
Using the above expression of the current, Eq. (\ref{dual2}) can be written as
\begin{eqnarray}
Z = \int \D G_{\m\n}\, \delta\left(\half \e^{\m\n\r\l}
 \d_\n (G_{\r\l} - R_{\r\l}  )\right)\, \exp\left[i\int d^4x 
\left\lbrace -\frac{1}{4}
G^{\m\n}G_{\m\n} 
\right\rbrace\right].
\end{eqnarray}
Using the discussion after Eq. (\ref{Freps}) we can 
introduce a dual vector potential $ b_\m $. The generating functional becomes
\begin{eqnarray}
Z = \int \D G_{\m\n} \D b_\m\,\delta\left(G_{\m\n} - \left\lbrace 
\d_\m b_\n - \d_\n b_\m + R_{\m\n}\right\rbrace \right)\, \exp\left[i\int d^4x 
\left\lbrace -\frac{1}{4}
G^{\m\n}G_{\m\n} 
\right\rbrace\right].\nonumber\\
\end{eqnarray}
Now we can integrate over the field $ G_{\m\n} $ and get
\begin{eqnarray}
Z = \int \D b_\m\,\, \exp\left[i\int d^4x 
\left\lbrace -\frac{1}{4} \left( 
\d_\m b_\n - \d_\n b_\m + R_{\m\n}\right)
\left(\d^\m b^\n - \d^\n b^\m + R^{\m\n}\right) 
\right\rbrace\right].
\end{eqnarray}
We started with a generating functional with a Lagrangian
  \begin{eqnarray}
 \label{la}
\mathcal{L^A} = - \frac{1}{4}(\d_\m A_\n - \d_\n A_\m)
(\d^\m A^\n - \d^\n A^\m) - A^\m j_\m
\end{eqnarray} 
and we ended up with a different Lagrangian 
for the same generating functional (upto a multiplicative constant) , but with a
 dual vector potential,
\begin{eqnarray}
\label{lb}
\mathcal{L}^b = -\frac{1}{4} \left( 
\d_\m b_\n - \d_\n b_\m + R_{\m\n}\right)
\left(\d^\m b^\n - \d^\n b^\m + R^{\m\n}\right). 
\end{eqnarray} 
 In the first Lagrangian ($ \M{L}^A$) the current
(\ref{topocurrent}) couples with the gauge field minimally, whereas  in the second Lagrangian ($ \M{L}^b $)
the current  can only be detected from the Bianchi identity. We 
 can thus conclude that currents become topological in this dualization process.
We can construct a theory of magnetic particles by dualizing
a  theory where the monopole current is minimally coupled with the dual vector potential.
After dualization we get a Lagrangian like (\ref{lb}) with ordinary vector potential
and we can minimally couple electric current with this vector potential. The Lagrangian then 
can be written as
\begin{eqnarray}
\label{lm}
\mathcal{L}^m = -\frac{1}{4} \left( 
\d_\m A_\n - \d_\n A_\m + M_{\m\n}\right)
\left(\d^\m A^\n - \d^\n A^\m + M^{\m\n}\right) - A^\m j_\m
\end{eqnarray}
where the monopole current is
\begin{equation}
\label{topomonocurrent}
j^\m_m = \half\e^{\m\n\r\l}\d_\n M_{\r\l}.
\end{equation}
and $j^\m  $ is the electric current.
So  we can take the magnetic current
as topological and electric current as Noether current to construct
a theory of magnetic monopoles. We will discuss this in \S \ref{monopole_attachment}
 and chapter \ref{t'hooft_monopole_chapter} in more detail.

\section{Dualization in  flux tube  configurations}
In the last three sections we have discussed flux tube configurations
in the Abelian Higgs model and dualization techniques for the electromagnetic field 
using functional integrals. In this section we  discuss the
dualization of the Abelian Higgs model with flux tube configurations. 
We start with the functional integral for the Abelian Higgs model
in $3+1$ dimensions, coupled to an Abelian gauge field $A^e_\mu$.
The partition function is given by
\begin{eqnarray}
Z = \int \D A^e_\mu\D\Phi\D\Phi^* \exp iS\,,
\end{eqnarray}
with the action
\begin{eqnarray}
\label{actionabelianhiggs}
S = \int d^4 x \left(- \frac14 F^e_{\mu\nu}F^{e\mu\nu} +
|D_\mu\Phi|^2 - \frac{\lambda}{4} (|\Phi|^2 - v^2)^2 \right)\,,
\label{flux.Higgs}
\end{eqnarray}
where $D_\mu = \partial_\mu +i e A^e_\mu\,,$ and $F^e_{\mu\nu}
= \partial_\mu A^e_\nu - \partial_\nu A^e_\mu$ is the Maxwell
field strength.
    
We change variables from $\Phi\,,$ $\Phi^*\,$ to the radial
Higgs field $\rho$ and the angular field $\chi\,,$ defined by  
$\Phi = \frac 1{\sqrt 2}\rho\exp(i\chi)\,.$ Then the measure
becomes, in these variables, 
\begin{eqnarray}
\int \D\Phi\D\Phi^*\cdots = \int \D\rho^2 \D\chi\cdots\,.,
\label{flux.measure}
\end{eqnarray}
where the dots represent the measure for any other fields and the
integrand. 

Remembering  the discussions on the flux tube configuration 
in the sections (3.1) and (3.2), we have to handle the functional
integral over $ \chi $ carefully,
since $ \chi $ is not defined on the points where
\begin{eqnarray}
\mbox{Re}\Phi = \mbox{Im}\Phi = 0.
\end{eqnarray}

As mentioned earlier these two equations define the two-dimensional manifold
in  space-time and we should
integrate over all functions that are regular everywhere
except for these two-dimensional manifolds. These two dimensional
singularities are  the Abelian Nielsen-Olesen (ANO) string
world sheets, since the Higgs field is zero at the center of
the ANO string. 

Our intention is to construct an effective  theory of interacting strings
from the Abelian Higgs model.  This is possible 
because as discussed in  \S 2.1,  
 at large distances  strongly coupled theory 
behaves like a string theory. So we can construct a long distance 
effective theory in which the mass of $ \rho $ is very large,
so that it is constant everywhere except
on  the thin flux tubes
in the effective theory. In terms of the coupling constant
we can say that we are interested in the large
 $\lambda\rightarrow \infty$ regime. So
$ \xi $ is almost zero and $ \rho \simeq v $. For the calculation
we will consider the theory  with the radial part of the field held
fixed, i.e. we will ignore the $\rho$-dependent part of the
measure, and set $\rho=v$ (constant) in the action. 
The string part will be taken care of by the singularity
of the field $ \chi $, as we discussed in section (\ref{effectivestrings}).  
The integration over the field 
$ \chi $ must be  handled carefully in the presence of flux tubes.
 This is because  the  theory 
has a topological winding number and we have seen 
that  this winding number is associated with
the field $ \chi $. 
The theory has  a gauge invariance as well
 and gauge invariance is also related to the field $ \chi $.
 The gauge invariance comes in as the redefinition of the field
$ \chi $ every time we do a gauge transformation.

The topological winding number arises from the large gauge transformations at large distances
where the fields get their vacuum configuration, i.e.
the gauge transformation for which  $ \chi $ does  not go to   zero at large distances, 
whereas to maintain  gauge invariance (small)  $ \chi $ must go
 to zero at large distances. 
So at least for large distances $ \chi $ behaves like 
sum of two fields. Here we are interested to do
large distance effective theory or theory with very
strong coupling. As discussed earlier  we can decompose the angular field
$\chi$ into a regular and a singular part\cite{Akhmedov:1995mw},
$\chi = \chi^r + \chi^s\,,$ where $\chi^s$ corresponds to a
given magnetic flux tube configuration, and $\chi^r$ describes single valued
fluctuations around this configuration. The singular part of the
phase of the Higgs field is related to the world sheet $\Sigma$ of
the magnetic ANO string according to the
equation
\begin{eqnarray}
\epsilon^{\mu\nu\rho\lambda}\partial_{\rho}
\partial_{\lambda}\chi^s &=& \Sigma^{\mu\nu}\,,  \\ 
\Sigma^{\mu\nu} &=&
2\pi n\int_{\Sigma}d\sigma^{\mu\nu}(x(\xi))\,\delta^4(x-x(\xi))\,,   
\label{flux.sigma}
\end{eqnarray}
where $\xi = (\xi^1, \xi^2)$ are the coordinates on the world-sheet
of the flux-tube, and $d\sigma^{\mu\nu}(x(\xi)) =
\epsilon^{ab}\partial_a x^\mu \partial_b x^\nu\,.$ In the above
equation $n$ is the winding number~\cite{Marino:2006mk}. Then the partition
function in  the presence of flux tube  reads 
\begin{equation}
Z = \int \D A^{e}_{\mu}\D\chi^s \D\chi^r \exp\left[i\int d^4 x
\left(- \frac14 F^e_{\mu\nu}F^{e\mu\nu} + \frac{v^2}2
(\partial_\mu\chi + e A^e_{\mu})^2 \right)\right]\,.
\label{flux.redZ}
\end{equation}
We will dualize the action using the techniques discussed in the previous
chapter. We begin by linearizing the term $\displaystyle\frac{v^2}{2}
(\partial_{\mu}\chi + e A^e_{\mu})^2$ by introducing an auxiliary
field $C_\mu$ to get
\begin{eqnarray}
&& \int \D\chi^r \exp[i\int d^4 x \frac{v^2}{2}
(\partial_\mu\chi^r + \partial_\mu\chi^s + e A^e_{\mu})^2]\,
\nonumber \\ &&\qquad = \int \D\chi^r \D C_\mu \exp\left[-i\int d^4 x
\left\{ \frac 1{2v^2}\,C^2_\mu + C^\mu (\partial_\mu \chi^r + 
\partial_\mu \chi^s + e A^e_{\mu})\right\}\right]\, \nonumber \\
&&\qquad = \int \D C_\mu \, \delta[\partial_\mu C^\mu]
\exp\left[-i\int d^4x \left\{ \frac 1{2v^2}\,C^2_\mu + C^\mu
(\partial_\mu\chi^s + e A^e_\mu)\right\}\right]\,. 
\label{flux.Cmu}
\end{eqnarray}
As discussed earlier we can resolve the constraint $\partial_\mu C^\mu = 0$ by
introducing an antisymmetric tensor field $B_{\mu\nu}$ and writing
$C_\mu$ in the form $C^\mu = \half
\epsilon^{\mu\nu\rho\lambda}\partial_\nu B_{\rho\lambda}$.
Integrating over the field $C_\mu\,,$ we get
\begin{eqnarray}
Z &=& \int \D A^e_{\mu} \D x_\mu(\xi)  \D B_{\mu\nu}  \exp\left[i\int
d^4 x\left\{ -\frac14 F^e_{\mu\nu}F^{e\mu\nu}  \right.\right.\,
\nonumber \\ &&
\qquad\qquad   \left.\left. + \frac{1}{12 v^2} H_{\mu\nu\rho}
H^{\mu\nu\rho} - \half \Sigma_{\mu\nu}B^{\mu\nu} - \frac e 2 
\epsilon^{\mu\nu\rho\lambda} A^e_\mu
\partial_{\nu}B_{\rho\lambda}\right\}\right]\,.
\label{flux.Cmuout1}
\end{eqnarray}
Here we have written $H_{\mu\nu\rho} = \partial_\mu B_{\nu\rho} +
\partial_\nu B_{\rho\mu} + \partial_\rho B_{\mu\nu}$, and
replaced the integration over $\D\chi^s$ by an integration over
$\D x_\mu(\xi)$ which represents a sum over all configurations of
the worldsheet of the flux tube. Here $x_{\mu}(\xi)$ parametrizes
the surface on which the field $\chi$ is singular. The Jacobian
for this change of variables gives the action for the string on the
background space time~\cite{Akhmedov:1995mw}. The string has a
dynamics given by the Nambu-Goto action, plus higher order
operators~\cite{Polchinski:1991ax}, which can be obtained from the
Jacobian. We will not write the Jacobian explicitly in what
follows, but of course it is necessary to include it if we want to
study the dynamics of the flux tube.

Let us now integrate over the field $A^e_{\mu}$. To do this we 
linearize $F^e_{\mu\nu}F^{e\mu\nu}$ by introducing another
auxiliary field $\chi_{\mu\nu}\,,$
\begin{eqnarray}
&& \int \D A^e_{\mu} \exp\left[i\int d^4x \left\{- \frac14
F^e_{\mu\nu}F^{e\mu\nu}-   
\frac e2 \epsilon^{\mu\nu\rho\lambda}A^e_\mu
\partial_{\nu}B_{\rho\lambda}\right\}\right]\,  \nonumber\\ 
&&\qquad = \int \D A^e_{\mu} \D \chi_{\mu\nu} \exp\left[i\int d^4x 
\left\{- \frac14 \chi_{\mu\nu}\chi^{\mu\nu} + \half
\epsilon^{\mu\nu\rho\lambda} \chi_{\mu\nu}
\partial_\rho A^e_\lambda - \frac e2 \epsilon^{\mu\nu\rho\lambda}
B_{\mu\nu}\partial_\rho A^e_\lambda \right\}\right]\, 
\nonumber\\\label{flux.Alinear0} \\  
&&\qquad = \int \D
\chi_{\mu\nu}\,\delta\Big[\epsilon^{\mu\nu\rho\lambda} 
\partial_{\nu}(\chi_{\rho\lambda}-e B_{\rho\lambda})\Big]\, 
\exp[i\int d^4x\{ -\frac 14 \chi_{\mu\nu}\chi^{\mu\nu}\}]\,.
\label{flux.Alinear1}
\end{eqnarray}
We can integrate over $\chi_{\mu\nu}$ by introducing a vector field
$ A_\m^m $ to solve the
$\delta$-functional, 
\begin{eqnarray}
\chi_{\mu\nu} - e B_{\mu\nu} = \partial_\mu A_\nu^m - \partial_\nu A_\mu^m\,. 
\label{flux.Adual1}
\end{eqnarray}

 $A^m_{\mu}$ can be thought of as a `dual photon' because   $A^m_{\mu}$ 
 appears through the  dualization of the vector potential $A^e_{\mu}$ as discussed in section 
 \S\ref{dualization1}. By looking at the action appearing in the path integral
 of Eq. (\ref{flux.Alinear0}), we see that the fields $ \chi_{\m\n} $ and $ B_{\m\n} $ transform like
 the tensor $ {*F}_{\m\n} = \half \e_{\m\n\r\l}F^{\r\l} $ under  parity 
 and time reversal. 
 The parity and time reversal properties of the field
 $ A^m_\m $ can be checked from Eq. (\ref{flux.Adual1}).  
 \begin{eqnarray}
\label{vectT'P'1}
\begin{array}{cc}
{A^m_{0}(x)  \buildrel P'\over \rightarrow\ -A^m_{0}(x_P)}, \qquad
{A^m_{0}(x)  \buildrel T'\over \rightarrow\ -A^m_{0}(-x_P)}\\
{A^m_{i}(x)  \buildrel P'\over \rightarrow\ A^m_{i}(x_P)},\qquad
 {A^m_{i}(x)  \buildrel T'\over \rightarrow\ A^m_{i}(-x_P)}.
\end{array}
\end{eqnarray}

  The result of the integration is then inserted into
Eq.~(\ref{flux.Cmuout1}) to give
\begin{equation}
Z = \int \D A^m_{\mu} \D x_{\mu}(\xi) \D B_{\mu\nu}
\exp\left[i\int\left\{- 
\frac 14 (e B_{\mu\nu} + \partial_{[\mu}A^m_{\nu]})^2
+ \frac 1{12 v^2} H_{\mu\nu\rho}H^{\mu\nu\rho} - 
\half \Sigma_{\mu\nu}B^{\mu\nu} \right\}\right] \,. 
\label{flux.functional0}
\end{equation}

The equations of motion for the fields $ A^m_\m $ and $B_{\mu\nu}$ can be calculated
from this,
\begin{eqnarray}
\label{absencemonopole}
\partial_\mu G^{\mu\nu} &=& 0,\\
\partial_\lambda H^{\lambda\mu\nu} &=& -\frac{m^2}e\, G^{\mu\nu} -
m^2 \,\Sigma^{\mu\nu} \,,
\label{flux.beom} 
\end{eqnarray}
where $G_{\mu\nu}= e B_{\mu\nu} + \partial_{\mu}A^m_{\nu} -
\partial_{\nu}A^m_{\mu}\,,$ and $m = ev$.  We can think of $ G_{\mu\nu} $ as the dual field tensor
because we get Eq. (\ref{absencemonopole}) by a variation with respect to the dual
gauge field  $ A^m_\m $ of the action in Eq.(\ref{flux.functional0}).

Eq. (\ref{absencemonopole}) shows that there is no
magnetic monopole current (dual current) present in the action.
This is of course expected, since we found these equations
by dualizing the Abelian Higgs model in the presence of 
flux tubes, but without magnetic monopoles. 

Using Eq. (\ref{absencemonopole}) and (\ref{flux.beom})
we get
\begin{equation}
\partial_\nu \Sigma^{\mu\nu} = 0\,.
\label{flux.closed}
\end{equation}
This equation means that  the vorticity tensor current $\Sigma_{\mu\nu}$ is
conserved. The vector $\partial_\nu \Sigma^{\mu\nu}  $ gives  the current of the
 endpoints of the flux string. We will see later that in the presence
  of magnetic monopoles the right
hand side of Eq. (\ref{absencemonopole}) will have the monopole current.
  Eq. (\ref{flux.closed}) means that due to the conservation of magnetic flux
all the flux tubes in the absence of magnetic monopoles are either closed or infinite. 

\section{Attachment of monopoles to the flux tube for confinement}
\label{monopole_attachment}
We will attach magnetic monopoles at the ends of a flux
tube of finite length. We will take the monopoles to be massless fermions and
minimally couple the monopole current to the magnetic or dual
photon. As discussed in chapter 1, the monopole current behaves like an axial 
current under parity. The time reversal property of an axial current is
not like a monopole current and under charge conjugation axial current also does not change
its sign.
However,  following the parity property  of magnetic monopole
current from Eq. (\ref{PTm}) we have taken the monopole current 
as axial. When these magnetic monopoles are couples are coupled to the magnetic
photon discussed earlier, the resulting theory will be  CPT invariant.
 After coupling, we will dualize the theory a
second time to get back to vector gauge fields, now coupled to flux
tubes.

However, a theory containing axial fermionic currents is anomalous
and if we try to dualize the theory, the presence of the anomaly
gives inconsistent results. We can cancel the anomaly by
introducing another species of fermionic monopoles with axial
charge opposite to the previous one. Let us denote the two species
by $q$ and $q'\,,$ with monopole charges $+g$ and $-g\,,$
respectively. So the monopole current becomes 
\begin{eqnarray}
j^\mu_m &=& g\bar q\gamma_5 \gamma^\mu q -
g\bar q'\gamma_5 \gamma^\mu q'\,.
\label{mono.anomaly}
\end{eqnarray}

The partition function of  Eq.~(\ref{flux.functional1}) is modified
to include the fermionic monopoles, minimally coupled to the
`magnetic photon' $A_\mu^m\,,$ so the Lagrangian reads
\begin{eqnarray}
{\mathcal L} = - \frac 14 (e B_{\mu\nu} +
\partial_{[\mu}A^m_{\nu]})^2 + \frac 1{12 v^2}
H_{\mu\nu\rho}H^{\mu\nu\rho} &-& \half \Sigma_{\mu\nu}B^{\mu\nu}\nonumber\\
&+& i \bar q \slash\partial q + i \bar q' \slash\partial q' - A^m_\mu 
j^\mu_m \,.
\label{mono.lag}
\end{eqnarray}
The field equation for $ G_{\m\n}$, Eq. (\ref{absencemonopole}) is now modefied to
\begin{eqnarray}
\d_\m G^{\m\n} = j^\n_m.
\end{eqnarray}
When we take the divergence of Eq. (\ref{flux.beom}) and use this result, we find that
\begin{eqnarray}
\frac 1e \partial_\mu \Sigma^{\mu\nu}(x) +  j^\nu_m(x)\, = 0.
\label{mono.coneq1}
\end{eqnarray}
The above Eq. (\ref{mono.coneq1}) is showing that the endpoint current
of the flux tube is cancelled by monopole currents at every space-time point. 
So we can say that point particle monopoles are attached at the end of the flux tube.
 This equation can also be derived as a consequence of gauge invariance, like current conservation in
 electromagnetism. To see this, we
take a transformation
\begin{eqnarray}
B_{\mu\nu} &\rightarrow& B_{\mu\nu} + \partial_{\mu}\Lambda_{\nu} -
\partial_{\nu}\Lambda_{\mu} \,, \nonumber \\ 
A^m_{\mu} &\rightarrow& A^m_{\mu} - \frac kg \Lambda_{\mu} \,.
\label{mono.gauge}
\end{eqnarray}
The second term of the Lagrangian of Eq.~(\ref{mono.lag}) is
invariant under the above transformation, while the first term is
made invariant by setting $eg = k\,.$ This is related to the Dirac
quantization condition as we shall see shortly. 

Since the flux due to the monopoles is fully confined in the tube,
the flux inside the tube must match with the  flux of a
 monopole with magnetic charge $ g $. The flux inside a flux tube
is known from the Eq. (\ref{loopint-abelianhiggs}).
\begin{eqnarray}
\mbox{Flux inside the flux tube} &=& \mbox{Flux of a monopole with magnetic charge $ g $}\nonumber\\
\frac{2\pi n}{e} &=& 4\pi g\\
 eg &=& \frac n2.
 \label{quant-abelinhiggs}
\end{eqnarray} 
 
So $\displaystyle k = \frac n2$, and we  have the Dirac
quantization condition $\displaystyle eg = \frac n2\,$. 
\pagebreak 
We can now write the partition function as
\begin{eqnarray}
Z[\Lambda_\mu] &=& \int \D A^m_{\mu} \D x_{\mu}(\xi) \D B_{\mu\nu}
\D q \D\bar q \D q' \D\bar q'\, \exp i\int d^4x \left[- \frac 14
(eB_{\mu\nu} + \partial_{[\mu}A^m_{\nu]})^2 \right. \nonumber\\ &&
\left. + \frac 1{12 v^2} H_{\mu\nu\rho}H^{\mu\nu\rho} - \half
\Sigma_{\mu\nu}B^{\mu\nu} -
\Sigma^{\mu\nu}\partial_{\mu}\Lambda_{\nu} +
e\Lambda_{\mu}j^{\mu}_m + i \bar q \slash\partial q + i \bar q'
\slash\partial q' - A^m_\mu j^\mu_m \right]\,.\; \nonumber \\
\label{mono.transpf}
\end{eqnarray}
Since (\ref{mono.gauge}) is only a change of variables, $Z$ cannot
depend on $\Lambda_\mu$. Thus $\Lambda_{\mu}$ can be integrated out
with no effect other than the introduction of an irrelevant
constant factor in $Z$, which we ignore. After integrating over
$\Lambda_\mu\,,$ we get
\begin{eqnarray}
Z = \int \D A^m_\mu\cdots
\delta\Big[\frac 1e \partial_\mu\Sigma^{\mu\nu} +  j^\nu_m\Big] 
&& \exp i\int d^4x \left[- \frac 14 (e B_{\mu\nu} +
\partial_{[\mu}A^m_{\nu]})^2  + \frac 1{12 v^2} H_{\mu\nu\rho}
H^{\mu\nu\rho} \right. 
\nonumber\\
&& \qquad \left. - \half\Sigma_{\mu\nu}B^{\mu\nu} 
+ i \bar q \slash\partial q + i \bar q'
\slash\partial q' - A^m_\mu j^\mu_m 
\right]\,, 
\end{eqnarray}
where the dots represent the measures for the other fields and
$x^\mu$.  One can see from the $\delta$-functional that the
vorticity current tensor is not conserved, but is cancelled by the
current of the added fermions. So the strings are open strings with
fermions stuck at the ends. Now we dualize the theory a second time
and get back to a vector gauge field which is something like a Maxwell
field, but in the presence of monopoles and flux tubes.
Introducing an auxiliary field $\chi_{\mu\nu}$ to linearize the
first term of the Lagrangian, we get
\begin{eqnarray}
Z &=& \int \D A^m_\mu \cdots  \D \chi_{\mu\nu}\,
\delta\Big[\frac 1e\partial_\mu\Sigma^{\mu\nu} +  
j^\nu_m\Big] \exp\,i\int d^4x \Big[- \frac 14
\chi_{\mu\nu}\chi^{\mu\nu} + \half
\epsilon^{\mu\nu\rho\lambda}\chi_{\mu\nu} 
\partial_\rho A^m_\lambda\, \nonumber\\ 
&&  + \frac 14 \epsilon^{\mu\nu\rho\lambda}
\chi_{\mu\nu}B_{\rho\lambda} +  
\frac 1{12 v^2} H_{\mu\nu\rho}H^{\mu\nu\rho}
- \half \Sigma_{\mu\nu}B^{\mu\nu} 
+ i \bar q \slash\partial q + i \bar q'
\slash\partial q' - A^m_\mu j^\mu_m 
\Big ] \,.\nonumber\\
\end{eqnarray}
We can now integrate out $A^\mu_m$, and the result is
\begin{eqnarray}
Z &=& \int \D \chi_{\mu\nu} \cdots
\delta\Big[\frac 1e\partial_\mu\Sigma^{\mu\nu} +  j^\nu_m\Big]\,
\delta\Big[\half \epsilon^{\mu\nu\rho\lambda}
\partial_\nu\chi_{\rho\lambda} - j^\mu_m\Big]\,
\exp\, i\int d^4x \Big[- \frac14 \chi_{\mu\nu}\chi^{\mu\nu}
\nonumber\\   
&& + 
\frac e4 \epsilon^{\mu\nu\rho\lambda}\chi_{\mu\nu}B_{\rho\lambda} + 
\frac 1{12 v^2} H_{\mu\nu\rho}H^{\mu\nu\rho}
- \half \Sigma_{\mu\nu}B^{\mu\nu}  
+ i \bar q \slash\partial q + i \bar q'\slash\partial q' 
\Big ] \,.
\end{eqnarray}
Both the $\delta$-functionals must be satisfied, which requires
$\displaystyle \frac 1e \partial_\nu\Sigma^{\mu\nu} - \half
\epsilon^{\mu\nu\rho\lambda} \partial_\nu \chi_{\rho\lambda} =
0\,. $ This can be solved by introducing a gauge field $A_\mu$,
which allows the integration over $\chi_{\mu\nu}\,.$ Then the
partition function becomes
\begin{eqnarray}
Z &=& \int \D x_\mu(\xi) \D B_{\mu\nu} \D A_\mu\cdots \delta
\Big[\frac 1e\partial_\mu\Sigma^{\mu\nu} +  j^\nu_m\Big] \exp \,i\int d^4x
\Big[- \frac 14 F_{\mu\nu}F^{\mu\nu}
\nonumber\\ 
&& + \qquad \frac 1{12 v^2} H_{\mu\nu\rho} H^{\mu\nu\rho} + \frac
1{2g} \epsilon^{\mu\nu\rho\lambda} B_{\mu\nu} \partial_\rho
A_\lambda  
+ i \bar q \slash\partial q + i \bar q'\slash\partial q' \Big]\,.
\label{mono.final}
\end{eqnarray}
Here $F_{\mu\nu} = \partial_\mu A_\nu - \partial_\nu A_\mu -
\displaystyle\frac {1}{2e} \epsilon_{\mu\nu\sigma\lambda}
\Sigma^{\sigma\lambda}\,,$ the dots represent the fermion measure
and we continue to suppress the action for the flux tube itself, as
in Eq.~(\ref{flux.Cmuout1}).

The vector potential $ A_\m $ has the same parity and time reversal
properties as the usual gauge potential of electromagnetism.
The theory is now in the form we originally intended, and contains
thin tubes of flux. The new feature is that the ends of the flux
tube are {\em sealed} by fermions, so that no flux escapes, all
flux is confined.  We should not think of this as any more than a
toy model of confinement, because the underlying theory is only the
Abelian Higgs model and not quantum chromodynamics. Even then, some
features are interesting enough to be highlighted.

There is a simple argument to calculate the length of the
string. The flux confined inside the string is $ 4 \pi g$, a
constant. The radius of the string core is of the order of
$1/v\sqrt{\lambda}\,.$ {}From this we can calculate the energy per
unit length of the tube to be $\mu \sim g^2 v^2 \lambda\,,$ also a
constant. Such a string, of finite length, would collapse in order
to minimize the energy unless it was stabilized by its angular
momentum.  For a rotating string of length $l\,,$ energy per unit
length $\mu,$ angular momentum $J\,,$ the energy function is $E =
\mu l + J^2/2\mu l^3\,.$ This has a minimum for the length
$L\sim\sqrt{J/\mu}\,.$ We see that for the stable flux tube with
magnetic monopoles at the ends,
\begin{equation}
\frac{J}{E^2} = constant\,,
\end{equation}
similar to the well-known Regge trajectory for mesons.

The gauge field $A_\mu$ is massive, with mass
$m=v/g$~\cite{Cremmer:1974mg, Allen:1990gb}. It does not couple
directly to the fermionic monopoles at the ends. Those fermions are
coupled only through the $\delta$-functional in
Eq.~(\ref{mono.final}), which guarantees that the monopoles must
seal the ends of the string. However, any other gauge field,
Abelian or not, axial or not, may be coupled to these fermions with
charge assignments independent of their charges under $A_\mu^m\,,$
which has been integrated out of the theory. In particular, if we
suggestively rename $q$ and $q'$ to $u$ and $\bar d\,,$ the allowed
configurations are $u\bar d\,, \bar u d\,,$ and $u\bar u \pm d\bar
d\,,$ which can couple to electroweak gauge fields. Note also that
we could have introduced three species of fermions (with charges
$1, 1, -2,$ for example) in Eq.~(\ref{mono.anomaly}), for the
purpose of anomaly cancellation, and we would again get flux tubes
with ends sealed by fermions. But a single species of fermions
would not produce such configurations.

In order to make the model more realistic, one would need to check
if flux is truly confined in the tube or if it escapes when the
tube has a finite thickness. A similar picture starting with an
axial gauge field and ending with a tube of `electric' rather than
magnetic flux will be interesting as well. Further, the freedom to
have other global symmetries in the theory allows in principle that
the $U(1)$ producing the string here may be embedded in an
$SU(N)_{global}\times U(1)_{local}$ symmetry, as
in~\cite{Vachaspati:1991dz}. 

\chapter{Monopoles and flux tubes  in broken SU(2)}
\label{t'hooft_monopole_chapter}
In this chapter we  discuss some aspects
of magnetic monopoles and flux tubes in SU(2)
gauge theory. 
 Under a gauge transformation,  an  adjoint SU(2) field transforms $ \phi\rightarrow U\phi U^\dagger $, a 
  field in the fundamental representation transforms as $  \psi\rightarrow U\psi$ and the gauge field transforms as
$ A_\m \rightarrow U A_\m U^\dagger -i \d_\m U U^\dagger $, where $\displaystyle{ U = e^{-i\alpha(x)^i\tau^i}} $
are space-time dependent SU(2) matrices. We start with a discussion 
about some geometrical aspects of spontaneous symmetry breaking.

\section{Spontaneous symmetry breaking and Higgs vacuum}
\label{sec:ssb}
We start with the Lagrangian
\begin{eqnarray}
\label{L1}
L = - {\frac 12} \Tr\left[G_{\mu\nu}G^{\mu\nu}\right] +
 \Tr\left[ \M{D}_\mu \phi \M{D}^\mu \phi \right]  - 
 {\frac\lambda 4}\left(|\phi|^2 - \xi_1^2\right)^2, 
\end{eqnarray}
where
\begin{eqnarray}
\label{phi22}
\phi = \phi_i\tau_i&,& \,
{\M{D}_\mu{\phi}} = \partial_\mu \phi - i g [A_{\mu},\phi] \,\\
\label{YMtensor}
A_\m = A^i_\m\tau^i&,& \,
G_{\mu\nu} = \partial_\mu {A}_{\nu}- \partial_\nu {A}_{\mu} 
 - i g [{A}_{\mu},{A}_{\nu}].
\end{eqnarray}

The $\phi^i$ are a  triplet of  scalar fields
and $A_\mu$ is the gauge field.
$ \phi $ transforms in the adjoint representation of SU(2). The Hamiltonian density corresponding to the Lagrangian is
\begin{eqnarray}
\label{hamilL1}
\theta_{00} = \half {\M{E}_i}^2 + \half {\M{B}_i}^2
+ \half (\M{D}_0\phi)^2 + \half (\M{D}_i\phi)^2 
+ {\frac\lambda 4}\left(|\phi|^2 - \xi_1^2\right)^2,
\end{eqnarray}
where
\begin{eqnarray}
G^{i0} = \M{E}^i, \, G^{ij} = - \e^{ijk}\M{B}^k.
\end{eqnarray}
The energy is at minimum for $ \theta_{00} = 0$, i.e. vanishes if and only if
\begin{eqnarray}
\label{vac1}
G^a_{\mu\nu} &=& 0,\\
\label{vac2}
\M{D}_\m\phi &=& 0,\\
\label{vac3}
V(\phi) &=& 0 \Rightarrow
|\phi|^2 = \xi_1^2.
\end{eqnarray}
These three equations define the vacuum and Eq. (\ref{vac3})
gives the classical value of $ |\phi|^2 $ at the vacuum.
So  the vacuum  expectation value (vev) of the field $ \phi $ is
non zero. At the vacuum, $ \phi $ lies on the surface of a sphere of radius $ \xi_1 $
in the Lie algebra. For the moment let us fix the direction of $ \phi $
in the third direction of the Lie algebra, i.e.  we choose  the vacuum field configuration
to be $ \phi_0 = \xi_1 \tau^3 $. If we consider fluctuations around
the vacuum and  also for the moment consider fluctuations only along the third direction,
we can write the field as
\begin{eqnarray}
\phi_0 = (\xi_1 + \rho) \tau^3.
\end{eqnarray} 
The Lagrangian becomes
\begin{eqnarray}
\M{L} = -\frac{1}{4} G_i^{\m\n}G^i_{\m\n} + \half \d_\m\rho\d^\m \rho
+ \frac{ g^2}{2}(\xi_1 + \rho)^2 (A^1_\m {A^1}^\m + A^2_\m {A^2}^\m)\nonumber\\
\qquad\qquad -\frac{\lambda}{4}(4\xi_1^2 + \rho^4 + 4 \xi_1\rho^3). 
\end{eqnarray}
From the above Lagrangian we can see that  the
fields $ \rho $, $ A^1_\m $ and $ A^2_\m $ have become massive. The masses are 
$ \sqrt{\lambda}\xi_1 $ and $ g\xi_1 $. We can take $ \xi_1 $ and
$ \lambda $ to be very large compared to the mass scale that 
we are interested in which would be the momentum scale of an external particle for example\,.
Then we can consider this  scale $ \Lambda $  to be a  vacuum for
the fields  $ \rho $, $ A^1_\m $ and $ A^2_\m $ but in this
vacuum $ A_\m^3 $ can excite particles. So at the scale $ \Lambda $,
an external particle will only see the interaction with $ A_\m^3 $. 
 At this scale the Lagrangian can be written as
 \begin{eqnarray}
 \label{brokenL1}
\M{L} = - \frac{1}{4} (\d_\m A_\n^3 - \d_\n A_\m^3)^2.
\end{eqnarray}

We call this vacuum the `Higgs vacuum'. Formally we can define the
`Higgs vacuum' by the solutions of the 
equations,
\begin{eqnarray}
V(\phi) = 0,\qquad
\M{D}_\m \phi = 0.
\end{eqnarray} 
In terms of the position representation we can say that the 
field configurations in a certain region of space-time are in the Higgs vacuum if equations
(\ref{vac2}) and (\ref{vac3}), but not necessarily Eq. (\ref{vac1}),
are satisfied at the region. In terms of scattering we can say the external particles
will experience `Higgs vacuum' upto a length scale $\displaystyle{ \sim \frac{1}{\Lambda} }$.

Let us discuss some features of the Higgs vacuum. 
Let us define the vacuum manifold as \cite{Goddard :1978}
\begin{eqnarray}
\label{vacuummanifold}
\M{M}_0 = \{\phi : V(\phi) = 0 \}.
\end{eqnarray}
Gauge invariance of a Lagrangian requires 
that $ V(\phi) $ is also invariant 
under the action of the group $ G $. 
It follows that if $ \phi $ satisfies Eq.(\ref{vac3}) then so does $ \M{D}(g)\phi $ 
for all $ g\in G $, where $ \M{D}(g) $ is a norm preserving representation of 
the group $ G $ under which $ \phi $ transforms. 
Thus we can write 
\begin{eqnarray}
V(\phi) = V(\M{D}(g)\phi)= 0.
\end{eqnarray}
    So $ G $ acts on $ \M{M}_0 $, i.e. every $ g $ takes
any point on $ \M{M}_0 $ to another point on $ \M{M}_0 $. This means 
that if $ \phi_0 $ is a point on $ \M{M}_0 $ then we can
go to another point simply by acting with a group element
$ g $ on $ \phi_0 $, i.e.
\begin{eqnarray}
\label{orbit1}
\phi'_0 = \M{D}(g)\phi_0.
\end{eqnarray}  
However, that does not mean that $ G $ can span all $ \M{M}_0 $
by acting on $ \phi $.

Two points $ \phi_1 $, $ \phi_2 $ which can be related by an element
$ g\in G $, 
\begin{eqnarray}
\phi_1 = \M{D}(g)\phi_2,
\end{eqnarray}
are said to be on the same orbit.
 So the orbit of $ \phi_0 $ is given by Eq. (\ref{orbit1}).
  If $ \M{D}(g)\phi_0 $ spans all
the points on $ \M{M}_0 $ as $ g $ varies over $ G $,  
we say that $ \M{M}_0 $ consists of a single orbit of the gauge
group $ G $. Another way of saying it is to say that
$ G $ acts transitively on $ \M{M}_0 $. That means for every $\phi_1, \phi_2$
belonging to $ \M{M}_0 $, there is some $  g_{12}\in G $ such that
\begin{eqnarray}
\phi_2 = \M{D}(g_{12}) \phi_1.
\end{eqnarray} 
Starting from a fixed element in $ \M{M}_0 $ , say $ \phi_0 $, we can always associate a group
element $ g_{i0} $ to another element in $ \M{M}_0 $, say $ \phi_i $, by using the relation
\begin{eqnarray}
\phi_i = \M{D}(g_{i0}) \phi_0.
\end{eqnarray}
If $ g_{i0} $ is unique,   $ \phi_i $ is associated with only one group element
 with respect to  $ \phi_0 $, then we can say $ \M{M}_0 $ is isomorphic to $ G $ as a manifold.
  If $ g_{i0} $ is not unique
then we have a subgroup other than the identity which leaves  $ \phi $ invariant. 
 Suppose $ G $ acts transitively on  $ \M{M}_0 $. Let
 $ \phi \in \M{M}_0 $ and let $ H_\phi \subset G $ be the subgroup leaving 
 $ \phi $ fixed,
 \begin{eqnarray}
H_\phi = \{h_\phi\in G |\, h_\phi \phi = \phi \}.
\end{eqnarray} 
$ H_\phi $ is called the stability, or isotropy, or little group of $ \phi $.
 Here we have written $ h_\phi $ instead of $ \M{D}(h_\phi)$ and we shall
use this convention from now on for any representation of the group.

If $ G $ acts transitively on $ \M{M}_0 $ then any
 $ \phi_1, \phi_2 \in \M{M}_0 $ are related by some $ g_{21}\in G $,
 \begin{eqnarray}
 \label{g12}
\phi_2 = g_{21}\phi_1.
\end{eqnarray}
On the other hand, if   $ h_{\phi_1} \in H_{\phi_1} $ and $ h_{\phi_2} \in H_{\phi_2} $ 
are  elements of the little groups of $ \phi_1$ and $ \phi_2 $ respectively,
\begin{eqnarray}
\label{littlegroup1}
h_{\phi_1}\phi_1 = \phi_1 ,\, h_{\phi_2}\phi_2 = \phi_2.
\end{eqnarray}
Using Eq. (\ref{g12}) we can write
\begin{eqnarray}
h_{\phi_1}\phi_1 &=& g^{-1}_{21}\phi_2 ,\\
g_{21}h_{\phi_1} g^{-1}_{21} g_{21}\phi_1 &=& h_{\phi_2}\phi_2 ,\\
\label{hconjugation1}
g_{21}h_{\phi_1} g^{-1}_{21}\phi_2 &=& h_{\phi_2}\phi_2.
\end{eqnarray}
Since Eq. (\ref{hconjugation1}) is true for any $ h_{\phi_1} $ and $ h_{\phi_2} $,
\begin{eqnarray}
\label{hconjugation}
g_{21}H_{\phi_1} g^{-1}_{21} = H_{\phi_2}.
\end{eqnarray}
Thus $ H_\phi $ varies within $ G $ by conjugation and consequently different
$ H_\phi $ are isomorphic.
Using Eq.s (\ref{g12}) and (\ref{littlegroup1}) 
we can write
\begin{eqnarray}
\phi_2 &=& g_{21} h_{\phi_1} \phi_1,\\
\phi_2 &=& h_{\phi_2} \phi_2 = h_{\phi_2} g_{21} h_{\phi_1} \phi_1.
\end{eqnarray} 
So it is not only the element $ g_{21} $ which takes $ \phi_1 $ to $ \phi_2 $,
but all elements of the set $ H_{\phi_2}g_{21}H_{\phi_1} $ also do this, and using  
Eq. (\ref{hconjugation}) we can write the elements as
\begin{eqnarray}
H_{\phi_2}g_{21}H_{\phi_1} &=& g_{21}H_{\phi_1} g^{-1}_{21} g_{21}H_{\phi_1} \\
&=& g_{21} H_{\phi_1} \in G/H_{\phi_1}
\end{eqnarray}
So it is clear from the above equations that the group elements 
that take  $ \phi_1 $ to $ \phi_2 $ are  elements of the left coset space with respect to the little group of $ \phi_1 $. 
Since all little groups are isomorphic in this case, we can write down the coset spaces as $ G/H $.

 We have defined the vacuum manifold $ \M{M}_0 $ by Eq. (\ref{vacuummanifold}).
However, the Higgs vacuum was defined by the two equations (\ref{vac2}) and (\ref{vac3}),
and it is not necessary
that if  $ \phi $ is a solution of the equation $ V(\phi) = 0 $, it
will always be a solution of the equation $ \M{D}_\m \phi = 0 $.  
      We know that all solutions of $ V(\phi) = 0 $
 must lie on the vacuum manifold $ \M{M}_0 $. If $ \phi_0 $
 is a solution of $ V(\phi) = 0 $ then all the $ \phi $'s that are
 related to $ \phi_0$  by a gauge transformation by the gauge group
 $ G $ must lie on $ \M{M}_0 $ because $ V(\phi) = V(\M{D}(g)\phi ) $.
 However, it is not necessary that all the points on $ \M{M}_0 $ 
 must be related  by a gauge transformations of the group $ G $. 
Then what we have is  a non-transitive action on $ \M{M}_0 $ 
 by the group $ G $. In that case the vacuum manifold can be 
 divided into several orbits, each of which consists of 
 points that are related by gauge transformations among themselves.
 
  Each orbit is a homogeneous space of the group $ G $. So each orbit 
  must be isomorphic to $ G $ or a subspace of $ G $.  As we have discussed earlier
  that in this case there is a subgroup $ H $ of $ G $ that makes $ \phi $ invariant
  and the orbit is isomorphic to the coset space $ G/H $.   
  If $ \M{M}_0 $ is isomorphic to $ G $ then $ H = e $, the identity element.

     When the system is at the vacuum, i.e. $ V(\phi) = 0 $,  we can write locally
   $ G \simeq H \times \M{M}_0 $. 
  However, this may not  always  be  the global structure of $ G $. 
   We can say that vacuum manifold breaks the global structure
 of the symmetry group $ G $. For example, for SU(2) adjoint scalars
 we can write $ V(\phi) = (\phi^i\phi^i - 1)^2 = 0 $. So the vacuum manifold
 $ \M{M}_0  \sim  S^2 $. Here we can write the
 group manifold  $ S^3 $ as $ S^2 \times S^1 $ but this is  not possible
 globally. The fundamental representation of SU(3), at the vacuum 
 $ V(\psi) = (\psi^\dagger\psi - 1)^2 = 0 $, breaks the 
 group manifold to $ S^3 \times S^5 $ 
  which cannot be written
 globally. So we can say that the equation $ V(\phi) = 0 $
 breaks the gauge group $ G $ to some local product form.

If  $\M{M}_0$ is not  a single orbit of $ \phi $, then 
we have to choose any one of the orbits of $ \phi $ 
where we can fix our vacuum.  For example if we choose the $ i $-th orbit, say $ \M{O}^i$,
in which $ \phi $ lies at the vacuum, then we can write
the group manifold $ G \simeq H^i \times \M{O}^i $. Here $ H^i $ is the isotropy group
of $ \phi^i \in O^i $ and the coset space is $ O^i $.
 There may be some bigger symmetry group 
$ \tilde{G} $ that makes $ V(\phi) $ invariant and
   acts transitively on $ \M{M}_0 $.
 The  gauge group $ G $ is a subgroup of the group
 $ \tilde{G} $ ($ G \subset \tilde{G} $). 
 So the total symmetry space can be written as
 a local product form $ \M{O}^i\times H^i \times \tilde{G}/G $.
 If $ \tilde{G}/G $ is a group then this group will be the 
  remaining global symmetry
  of the theory.

  Let us illustrate this with an example. In a SU(2) gauge theory we can write
  SU(2) fundamental representation as,
\begin{eqnarray}
\Psi = {\psi_1 \choose \psi_2}.
\end{eqnarray}
 We consider the potential
 \begin{eqnarray}
V(\Psi) &=& \frac{\lambda}{4}(|\Psi|^2 - v^2)^2\\
&=& \frac{\lambda}{4}\left((\Re\psi_1)^2 + (\Im\psi_2)^2 +
 (\Re\psi_2)^2 + (\Im\psi_2)^2 -v^2\right)^2,
\end{eqnarray}
$V(\Psi)  $ has a global $ \tilde{G} $ = SO(4) invariance since we can think of
$ \left(\Re \psi_1,  \Im \psi_2, 
 \Re \psi_2,  \Im \psi_2 \right) $ as a four vector.
 On the other hand
 \begin{eqnarray}
(\Re\psi_1)^2 + (\Im\psi_2)^2 +  (\Re\psi_2)^2 + (\Im\psi_2)^2 - v^2 = 0
\end{eqnarray}
is the equation of a 3-sphere, so the vacuum manifold  $ \M{M}_0 \sim S^3 $.
  The group SO(4) can be written  in a local
product form $SU_{l}(2) \times  SU_{g}(2)$.   We can identify 
the gauge group $ G =  SU_{l}(2)$ as the local gauge group.
Then the quotient group $ SU_{g}(2)\sim SO(4)/SU_l(2) $ is the global symmetry
which survives after the full $SU_l(2)$ symmetry group is broken \cite{hooft:1999}.

Let us discuss the significance of the equation
  $ \M{D}_\m \phi  = 0 $ in symmetry breaking.  
  We can write  $ \phi = |\phi|\hat \phi $, where $ \hat \phi $ is a 
 unit vector in some representation of the gauge group $ G $. The 
 equation $ V(\phi) = 0 $ generally gives $ |\phi| = \xi_1 $ where $ \xi_1 $ is some
 predefined constant. 
 $ \M{D}_\m \phi $ can be written as
 \begin{eqnarray}
 \label{dphi}
\M{D}_\m \phi = \d_\m |\phi| \hat \phi + |\phi|\M{D}_\m \hat{\phi}.
\end{eqnarray} 
The first term on the right hand side of Eq. (\ref{dphi})
 is automatically   zero on the vacuum manifold $ \M{M}_0 $
 because  $ |\phi| $ is a constant, but the second term transforms like $ \phi $
 under a gauge transformation. So it is not necessarily zero on the vacuum
 manifold. However, if it is zero for  a $ \hat{\phi} $ in the vacuum then it will be 
  zero for the whole orbit.  $ \M{D}_\m \hat{\phi} $
 can be made zero  by fixing the gauge at the vacuum.
 For example, in Abelian  Higgs model $ \hat{\phi} $ can be written as
 $ e^{-i\chi} $. $ \M{D}_\m e^{-i\chi} $ will be zero at vacuum 
 if we take $ A_\m = -\frac{1}{e}\d_\m\chi $.
  The gauge fixing may be `partial' because  there may be some subgroup 
  $ H $ of $ G $ under which Eq. (\ref{dphi})
 remains invariant. For example, in SU(2) gauge theory with $ \phi $ as an adjoint scalar filed,  $ \M{D}_\m \hat{\phi} $
 is invariant under the action of the group elements $ e^{-i\chi\hat\phi} $ and here
 $ H = U(1) $.
   So  in this process   it is not  possible to
 fix the gauge at the vacuum for the components of  the gauge field that lies in $ L_H $ , the algebra of $ H $.
   Under this gauge fixing procedure the components of the 
  Yang-Mills tensor $ G_{\m\n} $ that lies in the subspace $ L_G  -  L_H $ becomes zero.
  If we look at the fluctuations of the gauge fields along the subspace $ L_G  -  L_H $ around
  $ \M{M}_0 $, we find that they have mass  $ g\xi_1 $, where $ g $ is the
  coupling constant. However, the components of the Yang-Mills tensor which belong
  to $ L_H $ are non-zero on the vacuum manifold, so $ \M{M}_0 $ is not a vacuum 
 for the gauge fields that belong to $ L_H$. The  Lagrangian only has
 the gauge fields that belong to $L_H $ on $ \M{M}_0 $, as in Eq. (\ref{brokenL1})
 and we say that symmetry is broken
 down from $ G $ to $ H $.

 \section{Monopoles in SU(2) scalar gauge theory}
 
 In the last section we have seen that at the `Higgs vacuum' 
 the gauge group $ G$  can be written as  $ G \sim G/H \times H $,
 where $ H $ is the stabilizer (isotropy or little group) 
 and the coset space $ G/H $ is  the vacuum manifold $ \M{M}_0 $. 
  The fields $ |\phi| $ and  $ A_\m \in L_G - L_H $ 
   are massive around this vacuum. 
 The $ L_H $ valued components
 of the gauge field $ A_\m $ are non-zero in this vacuum, so $ \M{M}_0 $
 is not the vacuum for these components of $ A_\m$. 
 It is also possible 
 to have non-zero components of $ A_\m \in L_G - L_H $ in the vacuum  
 but it is  completely determined by the elements of $ G/H $ and   
 the components of Yang-Mills tensor $ F_{\m\n} \in L_G - L_H $
 are always zero at the vacuum manifold $ \M{M}_0 $.
  This is like the case of $ |\phi| $, whose vacuum expectation
 value is non-zero and there is no dynamics of $ |\phi| $ 
 at the vacuum $ \M{M}_0 $.  
 These vacuum configurations of the  $ L_G - L_H $ 
 valued components of $A_\m  $
 often correspond to some configurations which are  not 
  particles but stable against decay to the `` trivial solution."   
 They are truly distinct, and maintain their integrity, 
 even in the face of extremely powerful forces.
 These configurations are called solitons in gauge theory.

Suppose we consider  SU(2)  gauge theory coupled to an adjoint scalar
 in 3+1 dimensions.
 In this case the solitons are the magnetic monopoles. All these soliton solutions 
 can be constructed by giving a large gauge transformation to the 
 gauge field  $ A_\m $ whose $ L_G - L_H $ valued components are zero in the vacuum
 at large distances. 
 This is equivalent to saying  that we have 
 to write down $ A_\m $ for a constant
  $ \phi $ on $ \M{M}_0 $ and then give
 a large gauge transformation in the vacuum at large distances.
 
  We can give an example of this kind of large gauge transformation in
  the construction of quantized magnetic flux lines  of Abelian Higgs model
  which  we have discussed in chapter 2. 
In this case there is a `kernel' \cite{'tHooft:1974qc} 
in the form of a tube outside which all physical
 fields decrease exponentially to their vacuum configurations.
  The vector potential $ A_\mu$ becomes 
 pure gauge outside the kernel . The flux through the tube is an integer 
 times a constant. The integer is called the winding number of the configuration. 
 The winding numbers for flux tubes are  just the $\Pi_1$ (fundamental group)
  of the vacuum manifold, which is $S^1$, as we have seen in chapter 2. 
If we take the Higgs  field to be a constant then the 
vector potential becomes zero outside the kernel 
and there is no flux tube solution, or we could say that there are flux tubes 
with zero winding number.
 To get flux tubes we have to make a large gauge transformation to make
 the gauge field to be a pure gauge solution outside the kernel or
 in the vacuum at large distances. 
 Magnetic monopole solutions in SU(2) gauge theory with a adjoint
  scalar can be found 
 using the same technique. 
The Lagrangian was described at Eq. (\ref{L1}).
From the structure of the potential it is clear
that the scalar field gets a non-zero vacuum expectation value,
\[\langle\phi^i\phi^i\rangle  = \xi_1^2.\] So the symmetry is broken spontaneously. 
The vector fields  acquire a mass $g\xi_1$ and the Higgs has a mass $\sqrt \lambda \xi_1$ near the Higgs vacuum.
The Higgs vacuum is defined as
 \be
\label{1sthiggs1}
|\phi|^2 &=& \xi_1^2, \\
\label{1sthiggs2}
D_\mu {{\phi}} &=& \partial_\mu \phi - i g [A_{\mu},\phi] = 0.
\ee
From Eq.(\ref{1sthiggs1}) we can see that the vacuum manifold 
$\M{M}_0$ is $ S^2 $ and
the little group $ H $ corresponds to rotation around a point on $ S^2 $. So $ H = U(1) $ and
we can write  $\M{M}_0 = SU(2)/U(1) \sim  S^2 $. 
The gauge group $SU(2)$ is transitive on the vacuum manifold $S^2$
and this transitivity makes the theory independent of the
direction of $\phi_1$ at the vacuum as discussed in \S \ref{sec:ssb}.

It is possible to write  a general solution of Eq.(\ref{1sthiggs2}).
Using the  discussion in \S\ref{sec:aa} Eq.
  (\ref{1sthiggs2}) can be written as
\begin{eqnarray}
\d_\m \vec{\phi} + g \vec{A_\m}\times \vec{\phi} = 0.
\end{eqnarray}
Here we have written the gauge field as
\begin{eqnarray}
A_\m  = \vec A_\m \cdot \vec\tau 
\end{eqnarray}
and $ \vec{\tau} $ are the  Pauli matrices multiplied by half.
Taking the cross product of $ \vec{\phi} $ with the above equation we get
\begin{eqnarray}
\vec{\phi}\times \d_\m \vec{\phi} + 
g \vec{\phi}\times \vec{A_\m}\times \vec{\phi} &=& 0 \\
\label{chandra}
\mbox{i.e.,\qquad}\vec{\phi}\times \d_\m \vec{\phi} + 
g |\phi|^2\vec{A_\m} - g \vec{\phi}\vec{A_\m}\cdot \vec{\phi} &=& 0 .
\end{eqnarray}
At the vacuum $ |\phi|^2 = \xi_1^2 $, so we can write Eq.
(\ref{chandra}) as
\begin{eqnarray}
\label{Amonopole1}
\vec{A_\m} = \left(\vec{A_\m}\cdot \hat{\phi}\right) \hat{\phi}
- \frac{1}{g } \hat{\phi}\times \d_\m \hat{\phi},
\end{eqnarray}  
where 
\begin{eqnarray}
\hat{\phi} = \displaystyle{\frac{\vec{\phi}}{\xi_1} }.
\end{eqnarray}

 In terms of matrices, we can write
 \be
\label{A}
 A_\mu = {\frac 1{\xi_1^2}}2\Tr(A_\mu \phi)\phi + {i\over g{\xi_1^2}} [\phi, \partial_\mu\phi].
\ee
 This is the configuration of the gauge field in the vacuum manifold.
 The first term on the right hand side of Eq. (\ref{A}) is the component that lies
along $ \phi $. It is the massless part of the gauge field and 
it is expected to be non-zero on $ \M{M}_0 $. The other components are
zero  on $ \M{M}_0 $, except the part that is fully describable by
the vacuum configurations of the field $ \phi $. The value of the Higgs field 
is known at the Higgs vacuum, so the second part of the right side of the equation
 is fully known at the Higgs vacuum.  We say that the SU(2) symmetry is spontaneously broken to U(1).  We
  can also say that at the Higgs vacuum
the gauge symmetry is partially fixed up to U(1).
The second term on the right hand side of Eq. (\ref{A}) is 
responsible for monopole solutions.
 To  fix the gauge at the vacuum, we fix the field $ \phi  $ along the radial direction
 of the $ S^2 $. So at the vacuum we can write
\begin{eqnarray}
\label{phi_r}
\phi^i = \xi_1{\frac {r^i}{r}}.
\end{eqnarray}
 Using this Eq. (\ref{phi_r}) we can write 
 the second term of the right hand side of the equation
 (\ref{A}) as
\be
A_\mu^i(\mbox{only the second term}) &=& - {\frac 1{g\xi_1^2}} \epsilon^{ijk}\phi^j\d_\mu\phi^k,\\
    &=& - {\frac 1g}\epsilon_{\mu ij} {\frac {r^j}{r^2}}.
\ee

 Now if we define
\be
\label{defF}
F_{\mu\nu} = 2\Tr[\hat{\phi} G_{\mu\nu}] &=& {\frac i{g\xi_1^3}}2\Tr(\phi[\d_\mu\phi, \d_\nu\phi])\\
&=& -{\frac 1g}\epsilon_{\mu\nu i}{\frac {r^i}{r^3}},
\ee
we can write the magnetic field as, 
\begin{eqnarray}
\vec{B} = Q_m \frac{\hat r}{r^2}.
\end{eqnarray}
Here $\displaystyle{ Q_m = \frac{1}{g}} $ and the arrow indicates a vector in the usual 
three dimensional space.

 The flux for the above field is $\displaystyle{{4\pi}\over  g} $ and 
 a quantization condition for $ n=1 $ can be written $Q_m g = 1$. 
However,  this radial gauge fixing procedure gives a singular 
magnetic field at the position of the monopole.
This singularity can be regularized by  choosing a smooth configuration
 of the Higgs field $ |\phi| $  such that
$ |\phi| = 0 $ at the position 
of the monopole.  On the other hand, in the vacuum
manifold $ \M{M}_0 $, we know that $ |\phi| = \xi_1 $. 
 So, some regions of space-time
cannot be at the vacuum configuration of the 
Higgs field.  These regions were called the ``kernel" by 't Hooft \cite{'tHooft:1974qc}.
The existence of the kernel gives a finite size to a magnetic monopole, because $ |\phi| $ takes
some distance to reach its vacuum value from zero. Here we always assume  that 
 a kernel  always exists if we have a Higgs field for non zero winding number.

   In the Abelian Higgs model one can construct the
    vacuum  by choosing the Higgs field as constant and also
    choosing the gauge field to be zero. A flux tube can be constructed
    by a large gauge transformation at the vacuum (outside the kernel). 
   In other words, if we take the Higgs field to be  constant, we get the flux tube 
   solution with zero winding number. Non-zero winding
  number solutions can be found by applying appropriate large gauge transformation
  to the vacuum solutions.
   The story is the  same for monopoles as well. We shall see that a monopole solution
   can be constructed by giving a large gauge transformation  to the gauge field
   solution for a constant value of the Higgs field. 
 The construction is as follows. Let us fix  $\phi$ over  the vacuum  manifold 
 defined by Eqs. [\ref{1sthiggs1}, \ref{1sthiggs2}] by setting
\be
\phi &=& \xi_1\tau^3.
\ee
It follows that
\be
\partial_\mu \phi &=& 0.
\ee
 Then (\ref{A}) gives
\be
A_\mu &=& A_\mu^3 \tau^3,\\
G_{\mu\nu} &=& [\partial_\mu {A}_{\nu}^3- \partial_\nu {A}_{\mu}^3]\tau^3.
\ee

 The above $A_\mu$ and  $G_{\mu\nu}$ do not have any monopole solutions.
So these are zero winding number solutions. 
Now we can make a  gauge transformation by $ U(x)\in G $ 
to get a general gauge field on this vacuum
solution.
\be
\label{A1}
A_\mu &=& A_\mu^3 U\tau^3U^{\dagger} - {i\over g} \partial_\mu U U^{\dagger}.
\ee
If this $ U $ is a large gauge transformation, we will have a solution with a
non-zero winding number. For this solution we have
\be
\label{A2}
A_\mu^3 &=& 2\Tr(A_\mu U\tau^3U^{\dagger}) + {i\over g}2\Tr[\partial_\mu 
U U^{\dagger}(U\tau^3 U^{\dagger})],
\ee
so that using Eq.(\ref{A2}) we can rewrite Eq.(\ref{A1}) as
\be
\label{A3}
A_\mu &=& 2\Tr(A_\mu U\tau^3U^{\dagger}) U\tau^3  U^{\dagger}  \nonumber \\
 && \qquad \qquad - {i\over g}[\partial_\mu U U^{\dagger} - 2\Tr[\partial_\mu U U^{\dagger}(U\tau^3 U^{\dagger})]U\tau^3U^{\dagger}].
\ee
If we define 
\begin{eqnarray}
\hat{\phi}  =  U\tau^3U^{\dagger},
\end{eqnarray}
then Eq. (\ref{A3}) can be written as
\be
\label{A4}
A_\mu &=&   2\Tr(A_\mu \hat{\phi} )\hat{\phi}  - \frac{i}{g}\left[\partial_\mu U U^{\dagger} - 
2\Tr(\partial_\mu U U^{\dagger}\hat{\phi} )\hat{\phi} \right]
\ee
which in turn can be written as
\be
\label{A5}
 A_\mu =  2\Tr(A_\mu \hat{\phi})\hat{\phi} + \frac{i}{g} [\hat{\phi}, \partial_\mu\hat{\phi}].
\ee
Eq. (\ref{A5}) is nothing but Eq. (\ref{A}).

So we  see that the second term of the right hand side of Eq. 
(\ref{A4}) is responsible for the monopole solution.  Let us define 
\begin{eqnarray}
2\Tr[\hat{\phi} A_\mu] = B_\mu\,,
\end{eqnarray}
 then Eq.(\ref{Amonopole1}) becomes,
\be
\label{Amonopole3}
 A_\mu = B_\mu \hat{\phi} - \frac{1}{g } \hat{\phi}\times \d_\m \hat{\phi} .
\ee
This is the field at the vacuum,
and using this $ A_\m $ we can calculate $G_{\mu\nu}$
 \begin{eqnarray}
\vec{G}_{\m\n} &=& \d_\m \vec{A_\n} - \d_\n \vec{A_\m} +
\vec{A_\m} \times \vec{A_\n}\\
&=& \d_\m\left(B_\nu \hat{\phi} - \frac{1}{g } \hat{\phi}\times \d_\n \hat{\phi} \right)
- \d_\n\left(B_\nu \hat{\phi} - \frac{1}{g } \hat{\phi}\times \d_\n \hat{\phi} \right)
\nonumber \\ 
&&+ \,g\left(B_\mu \hat{\phi} - \frac{1}{g } \hat{\phi}\times \d_\m \hat{\phi} \right)
\times
\left(B_\nu \hat{\phi} - \frac{1}{g } \hat{\phi}\times \d_\n \hat{\phi} \right) \\
\label{Gvector}
&=& \left(\d_\m B_\m - \d_\n B_\m + \frac{1}{g } \hat{\phi}\cdot
\d_\m\hat{\phi}\times\d_\n\hat{\phi}\right)\hat{\phi} - \frac{2}{g}
 \d_\m\hat{\phi}\times\d_\n\hat{\phi}\nonumber \\
 &&-\, \frac{1}{g} \hat{\phi}\times [\d_\m, \d_\n]\hat{\phi},
\end{eqnarray}
using the fact that $ \hat{\phi}\cdot\hat\phi = \Tr(\hat\phi\hat\phi) = \half $.
It is easy to show that
\begin{eqnarray}
\hat{\phi} \times \left(\d_\m\hat{\phi}\times\d_\n\hat{\phi} \right) = 0,
\end{eqnarray}
and following this we can write  
\begin{eqnarray}
  \d_\m\hat{\phi}\times\d_\n\hat{\phi} = \left( \hat{\phi}\cdot
\d_\m\hat{\phi}\times\d_\n\hat{\phi}\right)\hat{\phi}.
\end{eqnarray} 
Using this relation   $ \vec{G}_{\m\n} $ can be written as
\begin{eqnarray}
\label{G-unstablestring-mono}
\vec{G}_{\m\n} = \left(\d_\m B_\m - \d_\n B_\m - \frac{1}{g } \hat{\phi}\cdot
\d_\m\hat{\phi}\times\d_\n\hat{\phi}\right)\hat{\phi} - \frac{1}{g} \hat{\phi}\times [\d_\m, \d_\n]\hat{\phi}.
\end{eqnarray}
The last term in $ G_{\m\n} $ written above is a  string term
and we shall see that this is an unstable string  configuration.
The string part can be written as
\begin{eqnarray}
- \frac{1}{g} \hat{\phi}\times [\d_\m, \d_\n]\hat{\phi} = 
- \frac{1}{g} \sin\theta \,\hat{\theta} \,[\d_\m, \d_\n]\varphi, 
\end{eqnarray}
where we have written in the internal three dimensional space
\begin{eqnarray}
\hat{\phi} &=& \cos\theta\tau^3 + \sin\theta\cos\varphi\tau^1 + \sin\theta\sin\varphi
\tau^2,\\
\hat{\theta} &=& -\sin\theta\tau^3 + \cos\theta\cos\varphi\tau^1 + \cos\theta\sin\varphi
\tau^2.
\end{eqnarray}
Here $ \theta $ and $ \phi $ are the parameters of the field orbit which is a sphere.
A flux string can be constructed if there is a map from any spatial loop to
any loop on the orbit. However, we know  that $ \Pi_1(S^2) = 0 $. 
This means that here any loop on the sphere can be shrunk to a point  
by a suitable gauge transformation.  
For a loop at the equator the flux of the string is $ \displaystyle{\frac{2\pi}{g} }$. 
However, we can make the loop disappear  by taking it to the pole where
the flux is zero and this can be done by setting $ \theta = 0 $ using just a gauge transformation.
 So we can ignore the last term of the right hand side of the  equation
(\ref{G-unstablestring-mono}). In terms of matrices , $ G_{\m\n} $ can be written as
\be
G_{\mu\nu} = \left[\d_\mu B_\nu - \d_\nu B_\mu + {\frac {2i}{g}}\Tr\left(\hat{\phi}
[\d_\mu \hat{\phi},\d_\nu \hat{\phi}]\right)\right]\hat{\phi},  
\ee
where $ \hat{\phi} = \displaystyle{\frac{\phi}{\xi_1}}$ .
We can write down an effective Lagrangian at the vacuum as
\be
\label{vaclag}
L &=& -{\frac 12}\Tr[G_{\mu\nu}G^{\mu\nu}]\nonumber\\
&=& -  {\frac 14} \left[F_{\mu\nu}F^{\mu\nu}\right],
\ee
where we have defined
\be
\label{F}
 \phantom{x}F_{\mu\nu} = \d_\mu B_\nu - \d_\nu B_\mu +
  {\frac {2i}{g}}\Tr\left[\hat{\phi}[\d_\mu \hat{\phi},\d_\nu \hat{\phi}]\right].
\ee

    Let us discuss  the form of  the field $ \phi $ when $ \phi $ is an SU(2)  adjoint scalar
   and can be written as $\phi(x) = |\phi(x)|\hat\phi(x)$ where $x\equiv \vec x$ and under 
    gauge transformations
$\hat\phi$ has a trajectory on $ S^2 $. Since $\phi$ is in the
adjoint of SU(2), we can always write $\phi$ as
\begin{equation}
\phi(x) = |\phi(x) | g(x)\tau ^3 g^{-1}(x) = |\phi(x)| \hat\phi(x) 
\,,
\end{equation}
for some $g(x) \in $ SU(2). Then for a given $\phi(x)\,,$ we can
locally decompose $g(x)$ as $g(x) = h(x)U(x)\,,$ with $h(x) = \exp
(- i\xi (x)\hat\phi(x))\,,$ and we can write
\begin{eqnarray}
\phi(x) = |\phi(x)| U(\varphi(x), \theta(x))\tau ^3
U^\dagger(\varphi(x), \theta(x)). 
\label{hU}
\end{eqnarray}
Here $\xi(x), \varphi(x), \theta(x)$ are angles on $S^3 \simeq$
SU(2). The matrix $U$ rotates $\hat\phi(x)$ in the internal space,
and is an element of SU(2)/U(1), where the U(1) is the one
generated by $h\,.$ If $|\phi|$ is zero at the origin and $ |\phi|$
goes smoothly to its vacuum value $\xi_1 $ on the sphere at infinity,
the field $ \phi $ defines a map from space to the vacuum manifold
such that the second homotopy group of the mapping is $Z$, the set of integers.
Equating $ \hat\phi $ with the unit radius vector of a sphere we can
solve for $U(\theta(x),\varphi (x))$,
\begin{eqnarray}
\label{Umonopole1}
U = \left(\begin{tabular}{cc}
$\cos{\theta\over 2} $ & $-\sin{\theta\over 2}e^{-i \varphi}$ \\
$\sin{\theta\over 2}e^{i \varphi}$ & $\cos{\theta\over 2}$\\
\end{tabular}\right)\, . 
\end{eqnarray}

In other words, an 't~Hooft-Polyakov monopole \cite{'tHooft:1974qc, Polyakov:1974ek} (in the point approximation, or as
seen from infinity) at the origin is described by
\begin{eqnarray}
\label{Umonopole2}
 U = \cos{\theta\over 2}\left(\begin{tabular}{cc}
$e^{i \varphi}$ & 0 \\
0 & $e^{-i \varphi}$\\ 
\end{tabular}\right) + \sin{\theta\over 2}
\left( \begin{tabular}{cc}  
$0\quad$ & $i$\\
$i\quad$ & $0$\\ 
\end{tabular}\right)\,, 
\end{eqnarray}
where $0\le\theta(\vec x)\le \pi$ and $0\le\varphi(\vec x)\le2\pi$
are two parameters on the group manifold.
Both choices, Eq. (\ref{Umonopole1}) and Eq. (\ref{Umonopole2}),  lead to the field configuration
\begin{eqnarray}
\vec\phi &=&  \xi_1{\frac {r^i}{r}}\tau_i
\end{eqnarray}
upon using Eq. (\ref{hU}) with $ |\phi| = \xi_1 $.
For this monopole, $Q_m g = 1\,,$ as we mentioned earlier. A monopole
of charge $n/g$ is obtained by making the replacement $\varphi \to
n\varphi $ in Eq.s~(\ref{Umonopole1}) or (\ref{Umonopole2}).
\begin{eqnarray}
\label{UmultiMonopole}
U_n = \cos{\theta\over 2}\left(\begin{tabular}{cc}
$e^{in \varphi}$ & 0 \\
0 & $e^{-in \varphi}$\\ 
\end{tabular}\right) + 
\sin{\theta\over 2} \left(\begin{tabular}{cc} 
$0\quad $ & $i$\\
$i\quad$ & $0$\\
\end{tabular}\right) \,. \\
\mbox{$n$ = $\pm 1, \pm 2, \pm 3$, ....}\nonumber
\end{eqnarray}

The integer $n$ labels the homotopy class, $\pi_2(SU(2)/U(1))
\sim \pi_2(S^2) \sim Z\,,$ of the scalar field configuration.
Other choices of $U(\vec x)$ can give other configurations. For
example, a monopole-antimonopole pair located on the $ z $
axis~\cite{Bais:1976fr} is given by the
choice
\begin{eqnarray}
\label{M-anti-M}
U = \sin{({\theta_1 - \theta_2})\over 2}\left(\begin{tabular}{cc} 
0 &   $- e^{ -i \varphi}$\\
$e^{i \varphi}$     & 0\\ 
\end{tabular}\right) + \cos{({\theta_1 - \theta_2})\over 2}
\left( \begin{tabular}{cc}  
 $1\quad$ & $0$\\
 $0\quad$ & $1$\\ 
 \end{tabular}\right) . 
\end{eqnarray}
For our purposes, we will need to consider a $\phi$-vacuum
configuration with $U(\vec x) \in SU(2)$ corresponding to a
monopole-anti-monopole pair separated from each other by a distance
$\gg 1/\xi_1.$ Then the total magnetic charge vanishes, but each
monopole (or anti-monopole) can be treated as a point particle.


\section{Flux tubes in SU(2) scalar gauge theory}
In the Abelian Higgs model we can construct a flux tube configuration
by  U(1) symmetry breaking. The asymptotic form of the gauge field
is pure gauge, and if we  map an angle of a loop around the flux tube 
to the gauge parameter of the group U(1) we get a  flux tube solution.
 Geometrically we can say that for a stable flux tube solution in gauge theory, 
 there has to be a non-trivial mapping from 
 a spatial loop in space  to the vacuum manifold $ \M{M}_0 = G/H $.
 That means $ \pi_1(G/H)\neq  0 $, where $ H$ is a subgroup   
 of the group $G$  and $ G/H $ is the coset space.
 
We have seen in last section that SU(2) symmetry can be broken down to
U(1) by an adjoint scalar. It is possible to construct 
a flux tube solution by breaking this U(1) symmetry. 
However, unlike the Abelian Higgs model 
there are  two ways by which we can construct flux tubes
from this U(1) theory.
We can break U(1) by using another SU(2) adjoint scalar
or we can use a fundamental SU(2) scalar. 
As we will discuss, the energy scale of this  symmetry
breaking must be different from the one that breaks SU(2) down to U(1).
This is a requirement for stability and we will take the difference
between energy scales to be very high. The scales determine the masses
of the two Higgs particles.

\subsection{Flux tubes with a second adjoint scalar}
\label{sec:adjointstring}
For the theory with two adjoint scalar fields, we can
write down the Lagrangian as
\begin{eqnarray}
\label{L2}
L &=& - {\frac 12} \Tr\left(G_{\mu\nu}G^{\mu\nu}\right) + 
\Tr\left( D_\mu \phi_1 D^\mu \phi_1 \right) + 
\Tr\left( D_\mu \phi_2 D^\mu \phi_2 \right) \nonumber\\
&&\qquad\qquad\qquad\phantom{xxxxxxxx} - \frac{\lambda_1}{4} \left(|\phi_1|^2 - \xi_1^2\right)^2
 - \frac{\lambda_2}{4}\left(|\phi_2|^2 - \xi_2^2\right)^2. 
\end{eqnarray}
Let us suppose that the  SU(2) symmetry is broken to U(1) at
 a scale $\xi_1$ and the U(1) is broken  at a scale $ \xi_2 $.
 Since $ \phi_1 $ is in the adjoint representation of SU(2),
 we can also think of this as SO(3) being broken down to SO(2),
 which is subsequently broken. 
 We assume that $ \xi_1\gg\xi_2 $ for the stability reason.  
 We choose the vacuum $ \phi_1 $ along the third axis, i.e.
 \begin{eqnarray}
\phi_1 = \xi_1 \tau^3.
\end{eqnarray}
 Below this vacuum scale there is only one massless
 gauge field  present  and that is  $ A^3_\m $ as discussed in \S \ref{sec:ssb}.
 The only gauge transformation that  is allowed on the
 field $ \phi_2 $ is then 
 \begin{eqnarray}
\phi'_2 = \displaystyle{e^{- i\chi\tau^3}\phi_2 e^{ i\chi\tau^3}}.
\end{eqnarray}
 The general form of $ \phi_2 $ in the $ \phi_1 $-vacuum is
 \begin{eqnarray}
\phi_2 = |\phi_2|\displaystyle{e^{- i\chi\tau^3}\hat\rho e^{ i\chi\tau^3}},
\end{eqnarray}
where $ \hat\rho $ is some unit vector in the SU(2) Lie algebra. Then the 
covariant derivative becomes 
\begin{eqnarray}
\M{D}_\m\phi_2 &=& \d_\m \phi_2 -i A^3_\m[\tau^3, \phi_2]\\
&=&\displaystyle{e^{- i\chi\tau^3}\left(\d_\m|\phi_2| \hat\rho 
+ |\phi_2| \d_\m \hat{\rho} -i |\phi_2|(g A^3_\m + \d_\m \chi)
\left[\tau^3, \hat{\rho}\right]\right) e^{ i\chi\tau^3}}. 
\end{eqnarray}
So the Lagrangian becomes
\begin{eqnarray}
\M{L} = -\frac{1}{4} (\d_\m A^3_\n - \d_\n A^3_\m)^2 +\half (\d_\m |\phi_2|)^2
+ |\phi_2 |^2 \Tr\left(\d_\m \hat{\rho} - i(g A^3_\m + \d_\m \chi)
\left[\tau^3, \hat{\rho}\right]\right)^2 \nonumber\\
- \frac{\lambda_2}{4}\left(|\phi_2 |^2 - \xi_2^2\right)^2.
\end{eqnarray}
It followes from the above Lagrangian that further symmetry breaking is possible if
\begin{eqnarray}
[\hat{\rho}, \tau^3] \ne 0,
\end{eqnarray}
because if $[\hat{\rho}, \tau^3] = 0  $, there will be no interaction
between $ A^3_\m $ and $ \hat{\rho} $.
If symmetry breaking happens then we can write
\begin{eqnarray}
A^3_\m = -\frac{1}{g}\d_\m \chi,\qquad |\phi_2| = \xi_2
\end{eqnarray}
in the vacuum. A flux tube through the origin can be constructed if
$ \chi $ is at least isomorphic to the angle around the flux tube. 
So the flux of a tube through the origin \,(along the $ z $- axis) can be calculated by taking
the line integral of $ A^3_\m $ around a loop far away 
from the  flux tube. We consider a distant loop
because there is a kernel near the origin which smooths out the  
line singularity and makes a real flux tube. However,
far away from the flux tube core it looks like a singular line,
 and the flux can be calculated by integrating  $ \vec{B}_3 $ 
over a surface $ \vec\Omega $ encircled by a loop $ C $,
\begin{eqnarray}
\mbox{Flux} &=& \int_\Omega \vec{B_3}\cdot \vec{d\Omega}\\ 
&=& \oint_C \vec{A_3} \cdot \vec{dl} \\ &=& \oint_C \frac{1}{g}\nabla\chi \cdot \vec{dl} \\
&=& \frac{2\pi n}{g}. 
\end{eqnarray}
 Here $ n $ is the winding number of the homotopy class of the 
 mapping from the spatial angle of the loop $ C $ to $ \chi $. For $ n = 1 $
 this mapping is an isomorphism. 
 To see the kernel of the flux tube we need the solutions of the 
 equations of motion. No exact solution is known till date but asymptotic solutions  do exist.
 
 \subsection{Flux tubes with one adjoint and one fundamental scalar}
 In  \S \ref{sec:adjointstring} we have seen that we
  can construct a flux tube configuration by using two adjoint scalars.
  However, instead of using an adjoint scalar to break the U(1)
  symmetry, we can have a fundamental scalar which causes that.
  In this case the starting Lagrangian is
  \begin{eqnarray}
\label{L3}
L &=& - {\frac 12} \Tr\left(G_{\mu\nu}G^{\mu\nu}\right) + 
\Tr\left( D_\mu \phi D^\mu \phi \right) + \Tr(\M{D}_\m \psi)^2 
 \nonumber\\
&&\qquad\qquad\qquad\phantom{xxxxxxxx} - \frac{\lambda_1}{4} \left(|\phi_1|^2 - \xi_1^2\right)^2
- \frac{\lambda_2}{4}\left(|\psi |^2 - \zeta_2^2\right)^2.
\end{eqnarray}
Here the field $ \psi(x) $ is in the fundamental representation of SU(2)} with a
covariant derivative defined by
\begin{eqnarray}
\M{D}_\m\psi = \d_\m\psi -i g A_\m\psi, \qquad\psi(x) = {\psi_1(x) \choose \psi_2(x)} \,.
\end{eqnarray}
As we discussed in the last section the first symmetry is broken by
the vacuum expectation value of the field $ \phi $ from SU(2) to U(1).
As in the  last section, here also we fix the vacuum as
$ \phi = \xi_1\tau^3 $, and call this the $ \phi $-vacuum. As explained
in the previous section, two of the gauge fields become massive at the scale 
$ \xi_1 $ , which we take to be large compared to the other scale $ \zeta_2 $,
$ \xi_1 \gg \zeta_2 $. We will look at the theory below the scale $ \zeta_2 $,
 and ignore the fields which have masses of the order $ \xi_1 $. Then the Lagrangian for the remaining
 fields is given by 
\begin{eqnarray}
\label{fundaeffective}
\M{L} = -\frac{1}{4} (\d_\m A^3_\n - \d_\n A^3_\m)^2 +
 \Tr(\M{D}_\m \psi)^2 - \frac{\lambda_2}{4}\left(|\psi |^2 - \zeta_2^2\right)^2.
\end{eqnarray} 
The equation
\begin{eqnarray}
\label{vacpsi}
\re(\psi_1)^2 + \im(\psi_1)^2 + \re(\psi_2)^2 + \im(\psi_2)^2 = \zeta_2^2
\end{eqnarray}
defines a three sphere ($ S^3 $).  To fix any point on this $ S^3 $ as vacuum, we have to
 fix all the three parameters of the group SU(2). 
So any $ \psi(x) $ that satisfies Eq. (\ref{vacpsi})
can break the residual U(1) of the theory because  two parameters
are fixed already by the vacuum expectation value of the field $ \phi $.
If $ \phi $ is fixed along $ \tau^3 $, then the only allowed
gauge transformations  are represented by the element $ \displaystyle{e^{-i\xi(x) \tau^3}} $.
 So the gauge transformation on the field $ \psi $ in the $ \phi $-vacuum can be written as
\begin{eqnarray}
\psi \rightarrow \psi'(x) = \displaystyle{e^{-i\zeta(x) \tau^3}}\psi(x).
\end{eqnarray}

We choose a form 
\begin{eqnarray}
\label{psi-vac}
\psi(x) = |\psi(x)|\displaystyle{e^{-\xi(x) \tau^3}}{1\choose 0}.
\end{eqnarray} 
Any $ \psi $ can be written like this for some $ \xi $.
 Then we can write the Lagrangian as
\begin{eqnarray}
\M{L} = -\frac{1}{4} (\d_\m A^3_\n - \d_\n A^3_\m)^2 + \half(\d_\m|\psi|)^2
 + \half (g A^3_\m + \d_\m \xi)^2 - \frac{\lambda}{4}\left(|\psi |^2 - \zeta_2^2\right)^2.
\end{eqnarray} 
At the vacuum we can write 
\begin{eqnarray}
|\psi| = \zeta_2,  \qquad A^3_\m = - \frac{1}{g} \d_\m \xi(x).
\end{eqnarray}
We can construct flux tubes in the same way 
as we did for two adjoint scalars, by mapping $ \xi $ onto a spatial circle.
 However, there is a problem of
uniqueness here. To construct a flux tube we have to 
write down a mapping from a spatial angular variable,
say $ \chi $, to the variable $ \xi $ and the homotopy
class of this mapping must be non trivial.
The form of $ \psi $ we have written in above Eq. (\ref{psi-vac}) is
periodic in $ \xi $ with periodicity $ 4\pi $,
\begin{eqnarray}
\label{xiperiodicity}
\psi(\xi) = \psi(\xi + 4\pi).
\end{eqnarray}  
 Since $ \psi(x) $ is a physical scalar  field on space time,  it must be
single valued on the space time points,
\begin{eqnarray}
\label{chiperiodicity}
\psi(\xi(\chi)) = \psi(\xi(\chi + 2\pi)).
\end{eqnarray} 
For flux tube solutions we have seen that there is a mapping
from $ \chi $ to $ \xi $.
The minimum nontrivial flux tube configuration
was constructed for an isomorphism $ \chi \rightarrow \xi $. However, 
when $ \psi $ is in the
fundamental representation,  the two equations
(\ref{xiperiodicity}) and (\ref{chiperiodicity}) are
inconsistent if we take $ \chi = \xi $. So for the consistency of the two equations
(\ref{xiperiodicity}) and (\ref{chiperiodicity}) we have to set 
\begin{eqnarray}
2\chi = \xi(x).
\end{eqnarray}
The  flux of this flux tube will be
\begin{eqnarray}
\mbox{Flux} &=& \int \vec{B_3}\cdot d\vec{\Omega}\\ 
&=& \oint \vec{A_3} \cdot \vec dl \\ &=& \frac{1}{g} \oint \nabla\xi \cdot \vec dl \\
&=& \frac{2}{g} \oint \nabla\chi \cdot \vec dl\\
&=& \frac{4\pi n}{g}. 
\end{eqnarray} 
The minimum non-zero flux for the flux tube constructed by 
one adjoint scalar and one fundamental scalar 
is thus twice the flux of the flux tube constructed 
by two adjoint scalars. 
\chapter{Confinement of monopoles in broken SU(2)}
 In this chapter  we will construct an effective Lagrangian
 from the original SU(2) Lagrangian with two  scalar fields.
We should clarify  the meaning of the word ``effective".
The theory that we are going to discuss is 
a theory of two stage symmetry breaking. In other words,
 the theory has two different energy scales at which the symmetry is broken.
  We shall consider a very high energy scale  $\xi_1$ and 
a low energy scale $\xi_2$.
We will assume that  $ \xi_1\gg \xi_2 $, specific values of $ \xi_1 $, $ \xi_2 $
can be determined if this theory is embedded in a larger theory but we 
will not do so.
Above the high energy  scale there is no isolated 
monopole and below the low energy scale we find confinement.
As  discussed in the previous chapter, spontaneous symmetry breaking
at the scale $ \xi_1 $, creates magnetic monopoles and breaks the SU(2)
down to U(1). This U(1)is then broken at the scale 
$ \xi_2 $ and strings (flux tubes) are produced. We will show
that below the scale $ \xi_2 $ there can be flux tubes of finite
length (open strings) with magnetic monopoles and anti-monopoles
attached to their ends.
 However, to show the attachment of monopoles at the end of confining strings
 and the interactions of confining strings,
  we should consider the theory near the confining 
  scale $ \xi_2 $. 
At  this scale many of the original degrees of freedom are frozen out. The remaining degrees of freedom,
 which are the propagating degrees at this scale, have a description
 in terms of dual variables. These make it convenient to see that the magnetic monopoles are indeed attached
 at the ends  of the string, as was in the case of the Abelian Higgs model with external monopole in chapter 2.
    Here we will  construct a Lagrangian to describe, in terms of these dual variables,  only the physics
  near the length scale $ \xi_2^{-1} $.   This Lagrangian is the one we call `` effective ''.   
  
  \section{Long distance Effective action  by two
  adjoint scalars } 
    
We start from a theory with SU(2) symmetry and a pair of adjoint
scalars, as discussed in \S 3.2 . The non-zero vacuum expectation value of the field
$\phi_1$ breaks the symmetry to U(1) at a scale $\xi_1$.  Below $\xi_1$
the theory is effectively an Abelian theory with magnetic
monopoles.

The Lagrangian for this system is
\begin{eqnarray}
\label{L4}
L = - {\frac 12} \Tr\left(G_{\mu\nu}G^{\mu\nu}\right) + 
\Tr\left( D_\mu \phi_1 D^\mu \phi_1 \right) + 
\Tr\left( D_\mu \phi_2 D^\mu \phi_2 \right) \nonumber\\
 - \frac{\lambda_1}{4} \left(|\phi_1|^2 - \xi_1^2\right)^2
 - \frac{\lambda_2}{4}\left(|\phi_2|^2 - \xi_2^2\right)^2. 
\end{eqnarray}
 The vacuum can be chosen according to the
eqs.~(\ref{1sthiggs1}, \ref{1sthiggs2}).  
Here our plan will be to construct an effective Lagrangian
 from the Lagrangian (\ref{L4}). The effective Abelian Lagrangian with monopole was written
 in Eq. (\ref{vaclag}). For two adjoint scalars
 the Lagrangian becomes 
 \begin{eqnarray}
\M{L} = -  {\frac 14}F_{\mu\nu}F^{\mu\nu} + \half D_\mu
\vec\phi_2 \cdot D^\mu \vec\phi_2 
 -  {\frac{\lambda_2} 4}(|\phi_2|^2 - \xi_2^2)^2,
\label{vaclagadj}
\end{eqnarray}
where, similar to Eq. (\ref{F}) and Eq.(\ref{Amonopole3}), 
\begin{eqnarray}
\label{F00}
F_{\mu\n} &=& \d_{\mu}B_{\n} - \d_\nu B_\mu - 
{\frac 1g}\hat{\phi}_1\cdot\d_\mu \hat{\phi}_1\times
\d_\nu\hat{\phi}_1, \\
\label{B00}
2\Tr[\hat{\phi}_1 A_\mu] &=& B_\mu\, , \qquad
\label{Amonopole22}
 A_\mu = B_\mu \hat{\phi}_1 - \frac{1}{g } \hat{\phi}_1\times \d_\m \hat{\phi}_1 .
\end{eqnarray}

One important difference with the general discussion on spontaneous symmetry breaking
 earlier is that here we did not fix the field $\phi_1$ at a constant internal direction everywhere.  
Since
the gauge group $SU(2)$ is transitive on the vacuum manifold $S^2$,
  this transitivity makes the symmetry breaking independent of the
direction of $\phi_1$.  There is a little group U(1) in the theory
which leaves $\phi_1$ invariant on the vacuum and this little group
becomes the remaining symmetry of the theory.  This  little
group is defined at every point on the vacuum. For the symmetry 
breaking from SU(2) to U(1), 
 the little group action is the rotation around a point on
the vacuum manifold $S^2$. That is why the little group is the same
for every point on $S^2$ and the little group is U(1).  
 However,   $\phi_2$  is also in the adjoint representation of SU(2).
 It has three real scalar components, out  of which three, one
component can be chosen along the direction of the field $\phi_1.$
Then the other two will rotate on a two dimensional plane normal to
$\phi_1$ under the action of the little group U(1).  Since the little group of $ \phi_1 $ is a subgroup
of the gauge group, we can say that  a gauge transformation rotates $\phi_2$ around $\phi_1$.
Flux tubes will be produced when this $U(1)$ symmetry is
spontaneously broken down to $Z_2$.  It is natural to take the
$U(1)$ breaking scale $\xi_2$ to be very small compared to the
$SU(2)$ symmetry breaking scale, $\xi_2\ll \xi_1\,.$

In order to find string configurations, we write
the covariant derivative of $\phi_2$ using
Eq.~(\ref{Amonopole22}),
\begin{eqnarray}
D_\mu\vec\phi_2 &=& \d_\mu\vec\phi_2 + g \vec
A_{\mu}\times\vec\phi_2,\nonumber \\ 
&=& \d_\mu \vec\phi_2 + g \left[B_{\mu}{\hat\phi_1 } - 
{\frac 1g}
\hat\phi_1\times{\d_{\mu}{\hat\phi_1}}\right]\times\vec\phi_2 
,\nonumber\\ 
\label{dmuphi}
&=&{\d_\mu \vec\phi_2} + g B_{\mu}{\hat\phi_1}\times\vec\phi_2 +
\left[\hat\phi_1\left(\d_\mu \hat\phi_1\cdot\vec\phi_2\right) -
  \d_\mu \hat\phi_1  \left(\hat\phi_1\cdot\vec\phi_2\right)
\right] .
\end{eqnarray}
This is of course in the $\phi_1$ vacuum.

For string configurations, $\phi_2$ has to approach its vacuum
value  far away from the string.  The  $\phi_2$ vacuum is defined by 
\begin{eqnarray}
\label{2ndhiggsad1}
|\vec{\phi_2}|^2 = \xi_2^2\,,\\
\label{2ndhiggsad2}
D_\mu \vec{\phi_2} = 0.
\end{eqnarray}
These equations are taken in the $\phi_1$ vacuum, so in particular
we use Eq.~(\ref{dmuphi}) in the left hand side of Eq.~(\ref{2ndhiggsad2}). If
we now dot Eq.~(\ref{2ndhiggsad2}) with $ \hat\phi_1$, 
we get
%
\begin{eqnarray}
\hat\phi_1\cdot \d_\m \phi_2  + \phi_2\cdot \d_\m \hat\phi_1 = \d_\mu(\vec\phi_1\cdot\vec\phi_2) = 0.
\label{phidotphi}
\end{eqnarray}
So in the $\phi_2$ vacuum (which by definition is embedded in the
$\phi_1$ vacuum), the component of $\vec\phi_2$ along $\vec\phi_1$
remains constant.  

As mentioned above, we can decompose $\phi_2$ (not necessarily in
the $\phi_2$ vacuum) into a component along $\phi_1$ and another
component normal to $\phi_1$ in the internal space,
\begin{eqnarray}
\vec\phi_2 = (\hat\phi_1\cdot\vec\phi_2)\hat\phi_1 + \vec K\,, 
\label{defK}
\end{eqnarray}
with
\begin{eqnarray}
\vec{K}\cdot \hat{\phi} &=& 0.
\end{eqnarray}
Then
\begin{eqnarray}
\d_\m\vec K \cdot \hat \phi_1 &=& - \vec{K}\cdot \d_\m\hat{\phi_1}.
\end{eqnarray} 
The form of $ \vec K $ will be important to write 
string degrees of freedom and it will be discussed in next section.
Now we can calculate $ \M{D}_\m\phi_2 $ using the 
above expressions.
\begin{eqnarray}
\M{D}_\m\phi_2 &=& \d_\m \phi_2 + \vec{A}\times \phi_2\\
&=& \d_\m \left((\hat\phi_1\cdot\vec\phi_2)\hat\phi_1 + \vec K\right) \nonumber\\
&& \qquad\qquad\qquad \,+ \left( B_\mu \hat{\phi} - \frac{1}{g } \hat{\phi}\times 
\d_\m \hat{\phi}\right)\times \left((\hat\phi_1\cdot\vec\phi_2)\hat\phi_1 + \vec K\right)\\
&=&\d_\m (\hat\phi_1\cdot\vec\phi_2)\hat\phi_1 
+ (\hat\phi_1\cdot\vec\phi_2)\d_\m\hat\phi_1 + 
\d_\m \vec{K} + g B_\m \hat{\phi_1}\times \vec{K} \nonumber\\
 && \qquad\qquad - (\hat\phi_1\cdot\vec\phi_2)(\phi_1 \times \d_\m\phi_1)\times \phi_1 
 - (\phi_1 \times \d_\m\phi_1)\times \vec{K}\\
 &=& \hat\phi_1\d_\mu(\hat\phi_1\cdot\vec\phi_2) +
\d_\mu\vec K + \hat\phi_1(\vec K\cdot\d_\mu\hat\phi_1) +
gB_\mu\hat\phi_1\times\vec K\,, \\
 &=& \hat\phi_1\d_\mu(\hat\phi_1\cdot\vec\phi_2) +
\d_\mu\vec K - \hat\phi_1(\hat\phi_1\cdot\d_\mu\vec K) +
gB_\mu\hat\phi_1\times\vec K\,. 
\end{eqnarray}

To write down the Lagrangian let us calculate %
\begin{eqnarray}
(D_\mu\vec\phi_2)^2 &=& \left[\d_\mu(\hat\phi_1\cdot\vec\phi_2)\right]^2
+ \left[\left(\d_\mu\vec K - \hat\phi_1(\hat\phi_1\cdot\d_\mu\vec K)\right) +
gB_\mu \hat\phi_1\times\vec K\,\right]^2,\\ 
&=& \left[\d_\mu(\hat\phi_1\cdot\vec\phi_2)\right]^2\nonumber\\
&&+ \left[\left(\d_\mu\vec K\right)^2  - \left(\d_\mu\vec K \cdot \hat\phi_1\right)^2
+ 2 g B^\m\d_\mu\vec K \cdot \hat\phi_1\times\vec K  + g^2 B^\m B_\m |\vec K|^2\right]
\nonumber \\ \\*
&=&\left[\d_\mu(\hat\phi_1\cdot\vec\phi_2)\right]^2 + \left(\d_\mu |\vec K|\right)^2\nonumber\\
&&+ |\vec K|^2\ \left[\left(\d_\mu\hat k\right)^2  - \left(\d_\mu\hat k \cdot \hat\phi_1\right)^2
+ 2 g B^\m\d_\mu\hat k \cdot \hat\phi_1\times\hat k  + g^2 B^\m B_\m \right]\\
&=&\left[\d_\mu(\hat\phi_1\cdot\vec\phi_2)\right]^2 + \left(\d_\mu |\vec K|\right)^2\nonumber\\
&& + |\vec K|^2 \left[\left(\d_\mu\hat k  \times \hat \phi_1\right)^2
+ 2 g B^\m\d_\mu\hat k \cdot \hat\phi_1\times\hat k  + g^2 B^\m B_\m \right],
\end{eqnarray}
where we have defined $ \displaystyle{\hat{k} = \frac{\vec{K}}{|\vec{K}|}} $.
The above expression of $ \left(\M{D}_\m\phi_2\right)^2 $ can be simplified if we use the identity
\begin{eqnarray}
&&\left(\d_\m\hat k \times \phi_1 \cdot \hat k \right)
\left(\d_\m\hat k \times \phi_1 \cdot \hat k \right)
- \left(\d_\m\hat k \times \phi_1 \right) \cdot 
\left(\d_\m\hat k \times \phi_1 \right)\nonumber\\
&=& \left(\d_\m\hat k \times \phi_1 \cdot \hat k \right)
\left(\d_\m\hat k \times \phi_1 \cdot \hat k \right)
- \left(\d_\m\hat k \times \phi_1 \right) \cdot 
\left(\d_\m\hat k \times \phi_1 \right) \hat{k}\cdot \hat{k}\nonumber\\
&=& \left([\d_\m\hat k \times \phi_1] \times \hat k \right)
\cdot \left([\d_\m\hat k \times \phi_1] \times \hat k \right)\nonumber\\
&=&[(\hat k \cdot \d_\m \hat k)\hat{\phi_1}
 - \d_\m\hat k (\hat\phi_1\cdot \hat k)]^2 = 0,
 \label{identity1}
\end{eqnarray}
which holds because  
\begin{eqnarray}
\hat{k}\cdot\d_\m\hat{k} = 0 = \hat{\phi_1} \cdot \hat{k}
\end{eqnarray}
by the definition of $ \hat k $.
Using the identity   of Eq. (\ref{identity1}) we can write
\begin{eqnarray}
(D_\mu\vec\phi_2)^2 &=& \left[\d_\mu(\hat\phi_1\cdot\vec\phi_2)\right]^2 + \left(\d_\mu |\vec K|\right)^2\nonumber\\
&&+ |\vec K|^2 \left[\left(\hat k \cdot \d_\mu\hat k \times \hat\phi_1\right)^2
+ 2 g B^\m \,\hat k \cdot \d_\mu\hat k \times \hat\phi_1 + g^2 B^\m B_\m \right]\nonumber\\
&=& \left[\d_\mu(\hat\phi_1\cdot\vec\phi_2)\right]^2 + \left(\d_\mu |\vec K|\right)^2
+ |\vec K|^2 \left[\hat k \cdot \d_\mu\hat k \times \hat\phi_1
 + g  B_\m \right]^2.
 \label{dphi2sq}
\end{eqnarray}

We put this
expression into the Lagrangian of  Eq.~(\ref{vaclagadj}). Then in order to
extract the string variables, we note that at large distances
away from the string, $\phi_2$ approaches its vacuum value
$|\phi_2| \to \xi_2$. Further, according to Eq.~(\ref{phidotphi}),
$\hat\phi_1\cdot\vec\phi_2$ also approaches a constant, so using
Eq.~(\ref{defK}) we see that $|\vec K|$ should also approach a constant. 
Then the first two terms of Eq.~(\ref{dphi2sq}) disappear at
infinity, as does the last term of Eq.~(\ref{vaclagadj}), and the
Lagrangian at infinity behaves as 
\begin{eqnarray}
{\mathcal L} = {\frac {|\vec K|^2} 2}  \left[\hat k \cdot \d_\mu\hat k \times \hat\phi_1
 + g  B_\m \right]^2 - {\frac
  14} F^{\mu\n}F_{\mu\n},
\label{vaclag2}
\end{eqnarray}
where now $|\vec K|$ is a constant. 

Since $|\vec\phi_2|$,  and the component of $\vec\phi_2$ along
$\hat\phi_1$,  both approach constant values at infinity, and so does
$|\vec K|,$ the only degree of freedom remaining in
$\vec\phi_2$ at infinity is an angle $\chi$ which parametrizes the
rotation of $\vec\phi_2$ around $\hat\phi_1.$ The first term inside
the brackets in Eq.~(\ref{vaclag2}) provides the $\partial_\mu\chi$
as we will see below.  This is the angle which is mapped onto a
circle at infinity to produce a flux string. Further, the system is
in the $\phi_1$-vacuum, i.e. $\vec\phi_1$ is in a vacuum
configuration given by Eqs.~(\ref{1sthiggs1}) and
(\ref{Amonopole1}). So in particular we can choose this
vacuum to contain 't~Hooft-Polyakov monopoles as discussed after
Eq.~(\ref{defF}).

\section{Monopoles and Strings from the effective action}
With the above in mind, let us parametrize the $\phi_1$-vacuum as
discussed in Eq. (\ref{hU}). 
\begin{eqnarray}
\label{Uphi}
\vec\phi_1 = \xi_1 U(\vec x)\tau_3U(\vec x)^{\dagger}\,,\qquad {\rm
  with\,} U(\vec x) \in  
SU(2)/U(1). 
\end{eqnarray}
Appropriate choices of $ U(\vec x) $ provide different monopole configurations,
some examples were given in Eqs.
(\ref{Umonopole1}), (\ref{Umonopole2},
(\ref{UmultiMonopole}), (\ref{M-anti-M}).  

For our purposes, we will need to consider a $\phi_1$-vacuum
configuration with $U(\vec x) \in SU(2)/U(1)$ corresponding to a
monopole-anti-monopole pair separated from each other by a distance
$\gg 1/\xi_1$ \cite{Bais:1976fr}. Then the total magnetic charge vanishes, but
the monopole and anti-monopole can be treated as  point particles.

We also need to choose the form of the vector $\vec K$ as in
Eq.~(\ref{defK}), so that it is orthogonal to $\hat \phi_1 =
\vec\phi_1/\xi_1$ in the internal space and rotates around $\hat
\phi_1$. Let us write $ \hat{k} $ in terms of matrices as
\begin{eqnarray}
\label{hatk}
\hat k \equiv \hat k(\vec x)^i\tau^i =
e^{-i\chi(\vec x)\hat\phi_1(\vec x)}U(\vec x) 
\tau_2U^{\dagger}(\vec x)e^{i\chi(\vec x)\hat\phi_1(\vec x)}\,.
\end{eqnarray}
We have used $\tau_2$ to write $\hat k$ here but 
we can substitute any constant vector orthogonal to $\tau_3$
without affecting the results below. The $\displaystyle{\hat\phi_1(\vec x) = \frac{\vec \phi_1}{\xi_1}}$ used
here is constructed according to Eq.~(\ref{Uphi}) with $U(\vec x)$
as described above. Then $\chi(\vec x)$ is the angle by which the
vector $U(\vec x)\tau_2U^{\dagger}(\vec x)$ is rotated in the group.

To get string and monopole terms in the Lagrangian we have to 
calculate the term $ \hat k \cdot \d_\m \hat{k} \times \hat \phi_1$.
Let us first calculate $ \d_\m \hat{k} $.
Using Eq. (\ref{hatk}) we can write
\begin{eqnarray}
\d_\m \hat k = \d_\m (g\tau^2 g^\dagger ),
\end{eqnarray}
where $ g = hU $ and 
\begin{eqnarray}
\label{h1}
h(x) = \displaystyle{  e^{-i\chi(\vec x)\hat\phi_1(\vec x)}}.
\end{eqnarray}
Then
\begin{eqnarray}
\d_\m \hat k &=& \left[\d_\m g g^\dagger , \hat k \right]\\
\label{rr}
&=& \left[\d_\m h h^\dagger + h \d_\m U U^\dagger h^\dagger , \hat k \right].
\end{eqnarray}
 For the sake of convenience, we write Eq. (\ref{rr}) as a vector equation, 
\begin{eqnarray}
\d_\m \hat k &=& i \vec{R}_\m \times \hat k,
\end{eqnarray}
where $ \vec{R}_\m = \d_\m h h^\dagger + h \d_\m U U^\dagger h^\dagger.
$
It follows from this that
\begin{eqnarray}
\hat k \cdot \d_\m \hat{k} \times \hat \phi_1 &\equiv & 
\hat k \times \d_\m \hat{k} \cdot\hat \phi_1\nonumber\\
&=& i \hat k \times \left(\vec{R}_\m \times \hat k\right) \cdot\hat \phi_1\nonumber\\
&=& i \left(\vec{R}_\m  - \hat k (\hat k \cdot \vec{R}_\m)\right)
\cdot \hat \phi_1\nonumber \\
&=& i \vec{R}_\m\cdot \hat \phi_1,
\end{eqnarray}
where we have used the fact  that $ \hat \phi_1 \cdot \hat k  = 0$.
Using Eq. (\ref{h1}) and the expression of 
$ R_\m $ written above, we can write
\begin{eqnarray}
\hat k \cdot \d_\m \hat{k} \times \hat \phi_1  &=& i
2\Tr\left[\d_\m h h^\dagger \hat{\phi_1}\right] 
+ i  2\Tr\left[\d_\m U U^\dagger  \hat{\phi_1}\right] \\
&=& i 2\Tr\left[\d_\m (U \displaystyle{e^{-i \chi \tau^3}}U^\dagger)
U \displaystyle{e^{i \chi \tau^3}}U^\dagger\hat{\phi_1}\right] 
+ i  2\Tr\left[\d_\m U U^\dagger  \hat{\phi_1}\right] \\
&=& i 2\Tr\left[\d_\m (U \displaystyle{e^{-i \chi \tau^3}}U^\dagger)
U \displaystyle{e^{i \chi \tau^3}}\tau^3U^\dagger\right] 
+ i  2\Tr\left[\d_\m U U^\dagger  \hat{\phi_1}\right]\\
&=& i 2\Tr\left[U^\dagger\d_\m U \tau^3 
+ \d_\m U^\dagger U \tau^3 \right] + \d_\m \chi 
+ i  2\Tr\left[\d_\m U U^\dagger  \hat{\phi_1}\right]\\
&=& \d_\m \chi + i  2\Tr\left[\d_\m U U^\dagger  \hat{\phi_1}\right]\\
\label{thetaN}
&=& \d_\mu \chi(\vec x) + g N_\mu(\vec x),
\end{eqnarray}  
where the vector $N_\mu$ is given by 
\begin{eqnarray}
\label{Nvector}
N_\mu = \frac{2i}{g}  \Tr\left[ \d_\mu U U^{\dagger}\hat \phi_1\right].
\end{eqnarray}
$\chi$ is the angle which is mapped onto a circle in space to
exhibit the flux tube. As we will see now, $N_\mu$ is the (Abelian)
field corresponding to magnetic monopoles.

Let us calculate the field strength tensor for $N_\mu$,
\begin{eqnarray}
\label{magfield}
\d_\mu N_\nu - \d_\nu N_\mu =  -\frac{1}{g} \hat\phi_1\cdot\d_\mu
\hat\phi_1\times \d_\nu\hat\phi_1 + \frac{2i}{g} \Tr[(\d_{[\mu}\d_{\nu]}U)
U^{\dagger}\hat\phi_1]\,.
\end{eqnarray}
If we use the $U(\vec x)$ of Eq.~(\ref{Umonopole2}), the first term
on the right hand side of this equation is the field strength of a
magnetic monopole at the origin, while the second term is a gauge
dependent line singularity, commonly known as a Dirac string. In
this case,
\begin{eqnarray}
\label{nmu}
N_\mu = -\frac{1}{g}(1+\cos\theta)\d_\mu\psi\,.
\end{eqnarray}
If $\theta$ and $\psi$ are mapped to the polar and the azimuthal
angles, $N_\mu$ is the familiar 4-potential of a magnetic monopole
with a Dirac string~\cite{Dirac:1931kp}. For the $U(\vec x)$ of the
monopole-anti-monopole pair of Eq.~(\ref{M-anti-M}), the first term
of Eq.~(\ref{magfield}) gives the Abelian magnetic field of a
monopole-anti-monopole pair, while the second term again contains a
Dirac string.

The Dirac string is a red herring, and we are going to ignore it,
for the following reason. The singular Dirac string appears because
we have used a $U(\vec x)$ which is appropriate for a point
monopole. If we look at the system from far away, the monopoles
will look like point objects, and it would seem that we should find
Dirac strings attached to each of them. However, we know that the
't~Hooft-Polyakov monopoles are actually not point objects, and
their near magnetic field is not describable by an Abelian
four-potential $N_\mu,$ so if we could do our calculations without
the far field approximation, we would not find a Dirac string. 

There is another way of confirming that the Dirac string will not
appear in any calculations. In the far field approximation, we have
written the Lagrangian of Eq.~(\ref{vaclag}) as
Eq.~(\ref{vaclag2}), which we can rewrite using Eq.~(\ref{thetaN})
as
\begin{eqnarray}
\M{L} = - {\frac 14}\left(\d_{\mu}B_{\nu} - \d_\nu B_\mu +
  M_{\mu\nu}\right)^2 + {\frac {|\vec K|^2}2} \left(gB_\mu
  +\d_\mu\chi + gN_\mu\right)^2\,,
\label{effective1}
\end{eqnarray}
where $|\vec K|$ is a constant as mentioned earlier, and $M_{\mu\nu}$ is
the monopole field,
\begin{eqnarray}
M_{\mu\nu} =  - \frac{1}{g} \hat{\phi}_1
\cdot\d_\mu \hat{\phi}_1\times \d_\nu\hat{\phi}_1\,.
\end{eqnarray}
The second term of the Lagrangian~(\ref{effective1}) is the one
which exhibits a flux tube or a `physical' string,  as opposed to
the unphysical Dirac string, which is an artifact of the far field
approximation and can be relocated by a gauge transformation. An
exactly analogous term appears in the Abelian Higgs model, where
instead of $|\vec K|$ we get the physical Higgs field. This 
model also exhibits a flux string, and just like in the Abelian
Higgs model, we know that the flux string here will appear along
the zeroes of $|\vec K|$, even though Eq.~(\ref{effective1}) is written in
the far field approximation, where $|\vec K|$ is a constant. The Dirac
string is also an artifact of the far field approximation, and we
can get rid of it by choosing $U(\vec x)$ such that the Dirac
string lies along the zeroes of $|\vec K|$, i.e., along the core of the
flux string. Then the troublesome line singularity, which appears
in the second term of Eq.~(\ref{effective1}), is always multiplied
by zero, and we can ignore it for the rest of the calculations.



\section{Low energy effective action  
 with one adjoint and one fundamental scalars}
In this section we will first construct an effective Lagrangian
 from the  SU(2) Lagrangian with one adjoint scalar 
 and one fundamental scalar field.  
Like the last section here we shall also
consider two mass scales  $\xi_1$, $\xi_2$ with $ \xi_1\gg \xi_2 $. 
At the length scale shorter than  $\xi_1^{-1}$ there is no isolated 
monopole. At scales larger than the long distance scale $\xi_2^{-1}$ we can
 see confinement.  Like the last two sections here also we will 
  construct a Lagrangian which only describes the physics
  near the length scale $ \xi_2^{-1} $ .
 We consider an SU(2) gauge theory coupled to an
adjoint scalar field as well as a fundamental scalar field. The two
fields break the symmetry at two scales. At the higher energy scale the
adjoint scalar breaks the symmetry down to U(1) and produces 't
Hooft-Polyakov magnetic monopoles~\cite{'tHooft:1974qc,
  Polyakov:1974ek, Prasad:1975kr}. The fundamental scalar breaks
the remaining U(1) symmetry at a lower energy scale and produces a flux
string.

We start with the Lagrangian
\begin{eqnarray}
  \label{lagrangian_1} 
  {\M{ L}} = - {\frac
    12}\Tr\left(G_{\mu\nu}G^{\mu\nu}\right) + \Tr\left(D_\mu \phi
    D^\mu     \phi \right) + \half(D_\mu \psi)^{\dagger}(D^\mu
  \psi)\nonumber\\ \phantom{xxxxxxxxxxxxxxxxxxxxxxxxxx}- {\frac{\lambda_1} 4}(|\phi|^2 - \xi_1^2)^2  
-  {\frac{\lambda_2} 4}(\psi^\dagger\psi - \xi_2^2)^2. 
\end{eqnarray} 
Here $\phi$ is in the adjoint representation of $SU(2), \,\phi =
\phi^i\tau^i$ with real $\phi^i\,$ and $\psi$ is a fundamental scalar complex
doublet of $SU(2)$. The $SU(2)$ generators $\tau^i$ satisfy
$\Tr(\tau^i\tau^j) = \half\delta^{ij}$. The covariant derivatives
$D_\mu$ and the Yang-Mills field strength tensor $G_{\mu\nu}$ are
defined as %
\begin{eqnarray}
  \left(D_\mu\phi\right)^i &=& \partial_\mu \phi^i + g
  \epsilon^{ijk}A_{\mu}^j\phi^k\,,\qquad (D_\mu \psi)_\alpha = \d_\mu \psi_\alpha -
  igA_\mu^i\tau^i_{\alpha\beta} \psi_\beta\,, \hfil\\ 
  G_{\mu\nu}^i &=& \partial_\mu {A^i}_{\nu}- \partial_\nu {A^i}_{\mu}
  + g \epsilon^{ijk}{A^j}_{\mu}{A^k}_{\nu}\,.\hfil.
  \end{eqnarray}

The adjoint scalar $\phi$ acquires a vacuum expectation value 
$\vec \xi_1$ which is a vector in internal space, and breaks the
symmetry group down to U(1). The 't Hooft-Polyakov monopoles are
associated with this breaking.  
As in the Eqs.~(\ref{1sthiggs1}) and (\ref{1sthiggs2}),  the vacuum is defined by
\begin{eqnarray}
|\vec\phi|^2 = \xi_1^2, \qquad D_\m\vec\phi = 0.
\end{eqnarray}

Like the equation (\ref{fundaeffective}) here also we can 
write the Lagrangian in the $\phi$-vacuum 
\begin{eqnarray}
\M{L} = -  {\frac 14}F_{\mu\nu}F^{\mu\nu} 
+ (D_\mu \psi^{\dagger})(D^\mu \psi)
 -  {\frac{\lambda_2} 4}(\psi^\dagger \psi - \xi_2^2)^2 ,
\end{eqnarray}
 where
\begin{eqnarray}
\label{F12}
F_{\mu\n} &=& \d_{\mu}B_{\n} - \d_\nu B_\mu - 
{\frac 1g}\hat{\phi}\cdot\d_\mu \hat{\phi}\times
\d_\nu\hat{\phi}, \\
\label{B}
2\Tr[\hat{\phi} A_\mu] &=& B_\mu ,\,\qquad
\label{Amonopole2}
 A_\mu = B_\mu \hat{\phi} - \frac{1}{g } \hat{\phi}\times \d_\m \hat{\phi} .
\end{eqnarray}
The last term of Eq.~(\ref{F12}) is the `monopole term' as discussed in \S 3.2.  

 After  the original
SU(2) is broken down to U(1) in the $\phi$-vacuum, the only remaining gauge symmetry
of the SU(2) doublet $\psi$ is a transformation by the little group
U(1). We will find flux tubes when this U(1) symmetry is
spontaneously broken down to nothing. The elements of this U(1) are
$h(x) = \exp[-i\xi (x)\hat\phi(x)]\,,$ rotations by an angle
$\xi(x)$ around the direction of $\phi(x)$ at any point in
space.This U(1) will be broken by the vacuum configuration of
$\psi\,$. 
 Like the quation (\ref{Uphi}) here also we write
\begin{eqnarray}
\label{Uphi0s}
\vec\phi  = \xi_1 U(\vec x)\tau_3U(\vec x)^{\dagger}\,,\qquad {\rm
  with\,} U(\vec x) \in  
SU(2)/U(1). 
\end{eqnarray}
Although in principle this process is the same for the fundamental
scalar as it was for an adjoint scalar, there are some important differences in the construction,
as we will see.

Let us then define the $\psi$-vacuum by
\begin{eqnarray}
 \label{2ndhiggs1}
 \psi^{*i}\psi^i = \xi_2^2\\
 \label{2ndhiggs2}
 D_\mu \psi = 0,
\end{eqnarray}
where $ D_\mu $ is defined using $ A_\mu $ in the $\phi$-vacuum, as
in Eq.~(\ref{Amonopole2}).  Multiplying Eq.~(\ref{2ndhiggs2}) by
$\psi^\dagger\hat\phi$ from the left, its adjoint by
$\hat\phi\psi$ from the right, and adding the results, we get
\begin{eqnarray}
0 & =& \psi^\dagger\hat\phi D_\mu\psi  +
(D_\mu\psi )^\dagger  \hat\phi\psi\, \nonumber\\
 &= & \psi^\dagger\hat\phi\d_\mu \psi +
 (\d_\mu\psi^\dagger)\hat\phi\psi  -ig\psi^\dagger
 \left[A_\m , \hat\phi \right]\psi\,  \\
 &=&  \psi^\dagger\hat\phi\d_\mu \psi +
 (\d_\mu\psi^\dagger)\hat\phi\psi + \psi^\dagger\d_\mu \hat\phi \psi\\
&= & \d_\mu\left[\psi^\dagger\hat\phi\psi\right]\,,
\end{eqnarray}
from which it follows that 
\begin{eqnarray}
\label{phi}
\psi^\dagger\hat\phi\psi = \mathrm{constant}\,,
\end{eqnarray}
or explicitly in terms of the components,
\begin{eqnarray}
\label{psi1}
\psi_i^\dagger \tau^\alpha_{ij}\psi_j\hat\phi^\alpha = \Tr\left[{\psi^\dagger}_i\sigma^\alpha_{ij}\psi_j
  \tau_\alpha\hat\phi\right] = \mathrm{constant}\,.
\end{eqnarray} 
Since $ \psi^\dagger\psi =  $ constant,  it follows that the components of the adjoint vector $ {\psi^\dagger}_i\sigma^\alpha_{ij}\psi_j
  \tau_\alpha $ parallel and orthogonal to $\phi$
are both constants.  Then we can decompose
\begin{eqnarray}
\label{psi2}
{\psi^\dagger}_i\sigma^\alpha_{ij}\psi_j \tau_\alpha =
\xi^2_2\cos\theta_c \hat\phi +  \xi^2_2 \sin\theta_c \hat k\,,
\end{eqnarray}
where $\hat k$ is a vector in the adjoint, orthogonal to
$\hat\phi\,.$ Following the equation (\ref{hatk})we can also write  $\hat k$ as
\begin{eqnarray}
\label{h}
\hat k = hU\tau^2U^\dagger h^\dagger\,,
\end{eqnarray}
where $h$ and $U$ are as defined  in \S 3.2 .

Using the identity $ {\sigma^\alpha}_{ij} {\sigma^\alpha}_{kl} =
\delta_{il}\delta_{kj} - \half \delta_{ij}\delta_{kl}$, we find
that $\psi$ is an eigenvector of the expression on the left hand
side of Eq.~(\ref{psi2}) (see Appendix A.2). Then writing the right hand side of that
equation in terms of $h$ and $U$, we find that $\psi$ can be
written as
\begin{eqnarray}
\label{psi11}
\psi = \xi_2hU \left(\begin{tabular}{l}
                                 $\rho_1$\\
                                  $\rho_2$\\
                         \end{tabular}\right)\,,
\end{eqnarray}
%
where   $\rho_1$ and $\rho_2$ are constants.  Keeping $U$ fixed, we
vary $\xi$ and find the periodicity
\begin{eqnarray}
\psi(\xi) = \psi(\xi + 4\pi)\,.
\end{eqnarray}
This $\xi$ is the angle parameter of the residual $U(1)$ gauge
symmetry and in the presence of a string solution, this $\xi$ is
mapped onto a circle around the string. In order to make $\psi$ single
valued around the string, we need $\xi = 2 \chi$, where $ \chi $ is
the angular coordinate for a loop around the string. Next let us
calculate the Lagrangian of the scalar field $\psi$.  We have, writing $ \rho $
for the constant doublet of Eq.(\ref{h}),
\begin{eqnarray}
  D_\mu\psi &=& \d_\mu\psi -igA_\mu\psi\\
&=& \d_\mu (hU\rho) -ig\left[B_\mu \hat\phi +
    ig\left[\hat\phi,
      \d_\mu\hat\phi\right]\right]hU\rho\\ 
&=& \d_\mu (Uh_0\rho) -ig\left[B_\mu \hat\phi +
    ig\left[\hat\phi, 
      \d_\mu\hat\phi\right]\right]Uh_0\rho\\ 
&=& \d_\mu U h_0\rho -i(2\d_\mu\chi  + g B_\mu) U h_0\tau^3 \rho
+ \left[\hat\phi, \left[\d_\mu U U^\dagger, \hat\phi\right]\right]Uh_0\rho\\
&=&   \d_\mu U h_0\rho -i(2\d_\mu\chi  + g B_\mu) U h_0\tau^3 \rho
+ \hat\phi 2\Tr\left(\hat\phi\d_\mu U U^\dagger\right)Uh_0\rho - \d_\mu Uh_0\rho
\nonumber\\  \\
&=& - iUh_0\tau^3\rho \left[2\d_\mu\chi  + g\left(B_\mu
      +N_\mu\right)\right]\,,
\end{eqnarray}
where $h_0 = e^{- i2\chi\tau_3}\,,\, \rho^i\rho^i = {\xi_2}^2\,,$ and
we have used the identity $ U^\dagger h U = \exp  (- 2i \chi\tau^3)\,.$
We have also introduced the Abelian `monopole field' $ N_\m $ as defined in
Eq. (\ref{Nvector}).
%
%
The  term  $ N_\m $ reproduces the magentic field of the monopole
configuration with the Dirac string. As in the earlier construction,  this singular string is a red
herring, and we are going to ignore it.  We have discussed the reason for this at the end of \S 4.2. 
  Then we can write  our effective Lagrangian 
\begin{eqnarray}
\label{effective2}
\M{L} &=& -{\frac 14}F^{\mu\nu}F_{\mu\nu}   
+ {\frac {{\xi_2}^2}2}\left(\d_\mu\chi + e\left(B_\mu +
    N_\mu\right)\right)^2.
\end{eqnarray}
Here we have defined the electric charge as $e = \frac g2$ and written
the magnetic charge as $Q_m = \frac 1{2e}$ .  Then $ Q_m e = \half $.

\section{Dual theory and confinement }
Let us now dualize the low energy effective action in order to
express the theory in terms of the macroscopic string variables.
The functional integration $Z$ can be written using 
Eq. (\ref{effective1}) or Eq.( \ref{effective2}). The functional integration for the
Lagrangian  of Eq. ( \ref{effective2})  can be recovered  if we replace the terms 
$ |\vec K| $ and $ g $ of the first equation into $ \xi_2 $ and $ e $ 
of the second equation.  So we have decided to use Eq.(\ref{effective2}) for dualization. 
\begin{eqnarray}
Z &=& \int \D B_\mu\D\chi\exp i
\int d^4 x \left[- {\frac 14}F_{\mu\nu}F^{\mu\nu} + 
{\frac {\xi_2^2}2}\left(eB_\mu +\d_\mu\chi +
eN_\mu\right)^2\right]\,.
\label{flux.Higgs11}
\end{eqnarray}
We know that the field $ \chi $ cannot be treated as  a regular field that reaches to zero 
very fast at large distances because in the presence of topological defects
 it has non zero values at very large distances.  Following the discussion in \S 2.4, in the presence of flux tubes here also
 we  decompose the angle $\chi$ 
into a part $\chi^s$ which measures flux in the tube,  and a part
$\chi^r$ describing single valued fluctuations around this
configuration, $\chi = \chi^r + \chi^s\,.$ 

Now we have integrations over both $\chi^r$ and $\chi^s\,,$ and the
second term in the action can be linearized by introducing an auxiliary field $C_\mu\,,$
\begin{eqnarray}
&& \int \D\chi^r \exp\left[i\int d^4x {\frac {\xi_2^2}2}
\left( eB_\mu +\d_\mu\chi_s +\d_\mu\chi_r
+ eN_\mu \right)^2
\right] \,
\nonumber \\ 
\label{cmu}
&& = \int \D\chi^r \D C_\mu \exp\left[-i\int d^4 x
\left\{ \frac 1{2\xi_2^2}\,C^2_\mu + C^\mu (eN_\mu + e B_{\mu} 
+ \partial_\mu \chi^r +
\partial_\mu \chi^s  )\right\}\right].\,\nonumber\\
\end{eqnarray}

The integration over $ \chi_r $ can be replaced by a
functional integration over a regular vector field $f_\m $
by introducing a delta functional in $ Z $. So the integration
over $\chi_r  $ in Eq. (\ref{cmu}),
\begin{eqnarray}
\int \D\chi^r \exp\left[i\int d^4 x C^\mu\partial_\mu \chi^r\right]
&=& \int \D f_\m \delta\left( \d_\m f_\n - \d_\n f_\m \right)
\exp\left[i\int d^4 x C^\mu f_\mu \right]\\
&=& \int \D f_\m \D B_{\m\n} \exp\left[i\int d^4 x C^\mu f_\mu  - 
\displaystyle{\frac {\xi_2}2}
\epsilon^{\mu\nu\rho\lambda}\partial_\mu f_\n B_{\rho\lambda}\right]
\nonumber \\  \\
&=&\int \D f_\m \D B_{\m\n} \exp\left[i\int d^4 x C^\mu f_\mu  - 
\displaystyle{\frac {\xi_2}2}
\epsilon^{\mu\nu\rho\lambda}f_\m \partial_\n B_{\rho\lambda}\right]
\nonumber \\ \\
&=&\int \D B_{\m\n} \delta\left( C^\mu - \displaystyle{\frac {\xi_2}2}
\epsilon^{\mu\nu\rho\lambda}\partial_\nu B_{\rho\lambda}\right),
\end{eqnarray}
 where we have introduced a second rank tensor field $ B_{\m\n} $
 by exponentiating the delta functional.

 Integrating over the field
$C_\mu\,,$ we get the partition function,
\begin{eqnarray}
Z = \int \D B_{\mu}&&\D x_\mu(\xi) \D B_{\mu\nu}\exp\left[i\int
d^4 x\left\{ -\frac14 F_{\mu\nu}F^{\mu\nu}  + \frac{1}{12 }
  H_{\mu\nu\rho} H^{\mu\nu\rho}  \right.\right.\,
\nonumber \\ 
&& \left.\left.- \displaystyle{\frac {\xi_2}2}  \Sigma_{\mu\nu}B^{\mu\nu} 
 - {\frac {e\xi_2 } 4} 
\epsilon^{\mu\nu\rho\lambda} M_{\mu\n}     B_{\rho\lambda} 
- {\frac {e\xi_2 } 2} 
\epsilon^{\mu\nu\rho\lambda} B_\mu
\partial_{\nu}B_{\rho\lambda}\right\}\right]\,,
\label{flux.Cmuout}
\end{eqnarray}
here we have written $F_{\mu\nu}= \partial_\mu B_\nu
- \partial_\nu B_\mu +  M_{\mu\nu}\,,$ defined
$H_{\mu\nu\rho} = \partial_\mu B_{\nu\rho} + \partial_\nu
B_{\rho\mu} + \partial_\rho B_{\mu\nu}\,,$ and also written $M_{\mu\n} = (\d_\mu N_\n -
\d_\n N_\mu)\,.$
$ \Sigma_{\m\n} $ is defined as
\begin{eqnarray}
\epsilon^{\mu\nu\rho\lambda}\partial_{\rho}
\partial_{\lambda}\chi^s  = 
2\pi n\int_{\Sigma}d\sigma^{\mu\nu}(x(\xi))\,\delta^4(x-x(\xi))  
\equiv  \Sigma^{\mu\nu}\,,
\label{def.sigma}
\end{eqnarray}
if $ \chi $ winds
around the tube $n$ times.
Here $\xi = (\xi^1, \xi^2)$ are the coordinates on the worldsheet
and $d\sigma^{\mu\nu}(x(\xi)) = \epsilon^{ab}\partial_a
x^\mu \partial_b x^\nu\,.$

 Let us now integrate over the field $B_{\mu}$ which we can do by
introducing an auxiliary field $ \chi_{\m\n} $, 
\begin{eqnarray}
&& \int \D B_{\mu} \exp \left[ i\int d^4x \left\{- \frac14
F_{\mu\nu}F^{\mu\nu}-   
\frac {e\xi_2}2 \epsilon^{\mu\nu\rho\lambda}B_\mu
\partial_{\nu}B_{\rho\lambda} - {\frac {e\xi_2} 4} 
\epsilon^{\mu\nu\rho\lambda} M_{\mu\n} B_{\rho\lambda} \right\} 
\right] \,
  \nonumber\\ 
&& = \int \D B_{\mu} \D \chi_{\mu\nu} \exp\left[i\int d^4x 
\left\{- \frac14 \chi_{\mu\nu}\chi^{\mu\nu} +  {\frac 1 4}
\epsilon^{\mu\nu\rho\lambda} \chi_{\mu\nu}F_{\rho\lambda}  
\right.\right.\,
\nonumber \\ 
&&  \left.\left.\phantom{xxxxxxxxxxxxxxxxxxxxxx} 
- {\frac {e\xi_2}2}  
\epsilon^{\mu\nu\rho\lambda}
B_{\mu\nu}\partial_\rho B_\lambda - {\frac {e\xi_2} 4} 
\epsilon^{\mu\nu\rho\lambda} M_{\mu\n}     B_{\rho\lambda}
\right\}\right]\,  
\nonumber \\  
&& = \int \D
\chi_{\mu\nu}\,\delta\Big[\epsilon^{\mu\nu\rho\lambda} 
\partial_{\nu}(\chi_{\rho\lambda}-e\xi_2
B_{\rho\lambda})\Big]\,\nonumber\\  
&&\phantom{xxxxxxx}\exp\left[i\int d^4x\left\{ -\frac 14 
\chi_{\mu\nu}\chi^{\mu\nu} + 
 {\frac 1 4}  \epsilon^{\mu\nu\rho\lambda}   (\chi_{\mu\nu} - 
e\xi_2 B_{\mu\n})M_{\rho\lambda}\right\}\right]\,.
\label{flux.Alinear}
\end{eqnarray}
We can integrate over $\chi_{\mu\nu}$ by solving the
$\delta$-functional in the same way as before as
\begin{eqnarray}
\chi_{\mu\nu} = e\xi_2  B_{\mu\nu} + \partial_\mu A_\nu^m - \partial_\nu
A_\mu^m\, , 
\label{flux.Adual}
\end{eqnarray}
and thus dualizing the vector potential $ B_\m $ to a theory
of a magnetic photon $ A^m_\m $.
The result of the integration is then inserted into
Eq.~(\ref{flux.Cmuout}) to give
\begin{eqnarray}
Z =&& \int \D A^m_{\mu} \D x_{\mu}(\xi) \D B_{\mu\nu}\nonumber\\
&&\exp\left[i\int\left\{- 
\frac 14 \left(e\xi_2  B_{\mu\nu} + \partial_{[\mu}A^m_{\nu]}\right)^2
+ \frac 1{12 } H_{\mu\nu\rho}H^{\mu\nu\rho} - 
{\frac {\xi_2} 2} \Sigma_{\mu\nu}B^{\mu\nu} - j_m^{\m}A^{m\mu}
\right\}\right]. \hfil \nonumber\\ 
\label{flux.functional1}
\end{eqnarray}
Here $j_m^{\mu} = - {\frac 1{2}} \epsilon^{\mu\nu\rho\lambda}\d_\n
M_{\rho\lambda}$ is the current of magnetic monopoles.

   The functional integration (\ref{flux.functional1}) can be calculated
from the partition function(\ref{flux.Higgs}).
 in a straightforward way.
  The integration over $\chi$ in Eq. (\ref{flux.Higgs})
   becomes integrations over both
$\chi^r$ and $\chi^s\,$.  However $\chi^r$ is a single-valued
field, so it can be absorbed into the gauge field $ B_\mu $ by a
redefinition, or gauge transformation, $B_\mu \to B_\mu
+ \partial_\mu\chi^r$. We can linearize the action by
introducing auxiliary fields $C_\mu, B_{\mu\nu}$ and $A^{m}_\mu$,
\begin{eqnarray}
Z &=& \int \D B_\mu \D C_\mu \D \chi_s \D B_{\mu\nu} \D
A^m_{\mu}\nonumber \\
&& \exp i \int d^4 x 
\left[ -\frac{1}{4} G^{\mu\nu}G_{\mu\nu} + \frac{1}{4}
  \epsilon^{\mu\nu\rho\lambda} 
G_{\mu\nu}F_{\rho\lambda} - \frac{1}{2\xi_2^2}C_\mu^2 - C^\mu (eB_\mu
+ eN_\mu + \d_\mu \chi_s) \right]\,,\nonumber \\
\end{eqnarray}
where we have written $G_{\mu\nu} = \d_\mu A^m_\nu - \d_\nu A^m_\mu 
+ e\xi_2 B_{\mu\nu}$ and $F_{\mu\nu}= \partial_\mu B_\nu
- \partial_\nu B_\mu + M_{\mu\nu}\,$.  Now we can integrate over
$B_\mu$ easily to get
\begin{eqnarray}
Z = \int \D C_\mu \D \chi_s \D B_{\mu\nu} \D A^m_{\mu} 
\delta\left( C^\mu - \frac{\xi_2}{2}
\epsilon^{\mu\nu\rho\lambda}\d_\nu B_{\rho\lambda}\right)
\exp i \int d^4 x \qquad\qquad \qquad\nonumber\\
\left[ -\frac{1}{4} G^{\mu\nu}G_{\mu\nu} + \frac{e\xi_2}{4}
  \epsilon^{\mu\nu\rho\lambda} 
B_{\mu\nu}M_{\rho\lambda} - A^m_\m j_m^\mu - \frac{1}{2\xi_2^2}C_\mu^2
 - C^\mu ( eN_\mu + \d_\mu \chi_s) \right]\,.\nonumber\\
\end{eqnarray}
Here 
$j_m^{\mu} = - {\frac 1{2}} \epsilon^{\mu\nu\rho\lambda}\d_\nu
M_{\rho\lambda}$ is the magnetic monopole current.  Integrating
over $ C_\mu $ we get
\begin{eqnarray}
Z = \int \D \chi_s \D B_{\mu\nu} \D A^m_{\mu} 
\exp i \int d^4 x 
\left[ -\frac{1}{4} G^{\mu\nu}G_{\mu\nu} + \frac{1}{12}
  H^{\mu\nu\rho}H_{\mu\nu\rho} -  
\frac{\xi_2}{2} \Sigma_{\mu\nu}B^{\mu\nu} - A^m_\m j_m^\mu  \right].\nonumber\\
\end{eqnarray}
Here we have defined
$H_{\mu\nu\rho} = \partial_\mu B_{\nu\rho} + \partial_\nu
B_{\rho\mu} + \partial_\rho B_{\mu\nu}\,,$ used
Eq.~(\ref{def.sigma}) and also used $M_{\mu\nu} = (\d_\mu N_\nu - \d_\nu N_\mu)\,.$

As discussed before,  the integration over $\D\chi^s$ can be replaced  by an
integration over $\D x_\mu(\xi)$, representing a sum over all the
flux tube worldsheets,  where $x_{\mu}(\xi)$ parametrizes the surface
of singularities of $ \chi $. The Jacobian for this change of
variables gives the action for the string on the background space
time~\cite{Akhmedov:1995mw, Orland:1994qt}. The string has a
dynamics given by the Nambu-Goto action, plus higher order
operators~\cite{Polchinski:1991ax}, which can be obtained from the
Jacobian.  Since we are not investigating the dynamics of the string here,
we will simply assume that this has been done.
\begin{eqnarray}
  Z = \int \D x_\mu(\xi) \D B_{\mu\nu} \D A^m_{\mu} 
  \exp i \int d^4 x 
  \left[ -\frac{1}{4} G^{\mu\nu}G_{\mu\nu} + \frac{1}{12}
    H^{\mu\nu\rho}H_{\mu\nu\rho} -  
    \frac{\xi_2}{2} \Sigma_{\mu\nu}B^{\mu\nu} - A^m_\m j_m^\mu  \right],\nonumber\\
\label{flux.functional}
\end{eqnarray}
The equations of motion for the field $B_{\mu\nu}$ and $A^{\mu}$
can be calculated from this to be
\begin{eqnarray}
\label{flux.Beom}
\partial_\lambda H^{\lambda\mu\nu} &=& -m \, G^{\mu\nu} -
\frac{m}{e} \,\Sigma^{\mu\nu} \,,\\
\d_\mu G^{\mu\nu} &=& j_m^\mu,
\label{flux.Aeom} 
\end{eqnarray}
where $G_{\mu\nu}= m  B_{\mu\nu} + \partial_{\mu}A^m_{\nu} -
\partial_{\nu}A^m_{\mu}\,,$ and $m = e \xi_2$. Combining
Eq.~(\ref{flux.Beom}) and Eq.~(\ref{flux.Aeom}) we find that
\begin{equation}
\frac 1e  \partial_\mu \Sigma^{\mu\nu}(x) + j_m^\mu(x) = 0\,.
\label{mono.coneq}
\end{equation}
The same equation that we found in the theory with two adjoint scalars. As in that 
case, it  follows that a vanishing magnetic monopole
current implies $\partial_\mu \Sigma^{\mu\nu}(x) = 0\,,$ or in
other words if there is no monopole in the system, the flux tubes
will be either  closed or or infinite. There is however an important difference between
this and the previous construction. Here
the magnetic flux through the tube is
$\displaystyle{\frac{2n\pi}e}\,,$ while the total magnetic flux of
the monopole is $4m\pi Q_m\,,$ where $n, m$ are
integers.  Since $\displaystyle{e Q_m = \half} $ for the  fundamental SU(2) scalar,
 it follows that we can have a string that confines a monopole and anti-monopole pair
for every integer $ n $. However, this was not true for the construction involving
two adjoint scalars because in that case $ e = g $ and $ gQ_m = 1 $ which makes $ n = 2m $.
This means for confinement of monpoles with $ m = 1 $ will be
possible if $ n = 2 $.

 Although these string configuration could 
be broken by creating a monopole-anti-monopole pair, there is a
hierarchy of energy scales $\xi_1\gg \xi_2\,,$ which are respectively
proportional to the mass of the monopole and the energy scale of
the string. So this hierarchy can be expected to prevent string
breakage by pair creation.

The conservation law of Eq.~(\ref{mono.coneq})
also follows directly from $ Z $ in
Eq.~(\ref{flux.functional}) and this can be shown
by making a gauge transformation as discussed in chapter 2. However, this can also be
derived  by introducing a variable 
$B'_{\mu\nu} = B_{\mu\nu} +
{\frac 1m } (\partial_\mu A^m_\nu - \partial_\nu A^m_\mu)$ and
integrating over the field $ A^m_\mu $. If we do so
we get
\begin{eqnarray}
\label{confinement}
Z =\int \D x_{\mu}(\xi) \D B'_{\mu\nu}&&
\delta\Big[\frac1e\partial_\mu\Sigma^{\mu\nu}(x)  
+  j^\nu_m(x)\Big] \nonumber\\
&&\exp\left[i\int\left\{\frac 1{12} H_{\mu\nu\rho}H^{\mu\nu\rho} 
- \frac 14  m^2 {B'}^2_{\mu\nu}- 
{\frac m{2e}} \Sigma_{\mu\nu}{B'}^{\mu\nu} \right\}\right] \,, \hfil\,\nonumber \\
\end{eqnarray}
with the delta functional showing the conservation law
(\ref{mono.coneq}).  Thus these strings are analogous to the
confining strings in three dimensions~\cite{Polyakov:1996nc}. There
is no $A^m_\mu$, the only gauge field which is present is
$B'_{\mu\nu}$. This $B'_{\mu\nu}$ field mediates the direct
interaction between the confining strings.

\chapter{Conclusion}
In this thesis we have studied the static and dynamical properties of magnetic monopoles and 
flux tubes. Then we have studied configuration which carry  both magnetic monopoles and flux tubes.
For the system with both monopole and flux tube we have showed the confinement of monopoles by flux tubes.
By confinement we mean that when monopoles are attached at the ends of a flux tube, no flux escapes
from the tube to far away. Thus all flux is ``confined". To show this we have calculated three  different cases.
For the Abelian Higgs model first we have constructed a long distance effective
theory of flux tubes and we described the system by dual variables. 
Then we have shown that if we add monopoles in the system  externally then they
  confine themselves  by attaching  to the ends of the flux tubes to seal the flux.  
  The second case that we have considered is an SU(2) gauge theory with two adjoint
  scalar fields and the last case   is an SU(2) gauge theory with one adjoint and one fundamental scalar fields.
 For all the three cases the attachment of monopoles at the end of flux tubes
 have been shown by a delta functional. 
The delta functional 
enforces that at 
every point of space-time, the monopole current cancels the
currents of the end points of flux tube.  So the monopole current
must be non-zero only at the end of the flux tube.
The last  functional integration Eq.~(\ref{confinement}) does not carry Abelian gauge field
$A^m_{\mu}$, only a massive second rank tensor gauge field is present. 
 All these confirm the permanent attachment of monopoles at the end of
the flux tube which does not allow gauge flux to escape out of the
flux tubes.  There are important differences between the results
from the construction involving two adjoint scalar fields  and the one involving one adjoint
and one fundamental scalar. In the first case the mass of the Abelian photon
 will be zero  if the two  vevs are
  aligned in the same direction. However, this can never happen for the second case
  where the scalars are in different representations. Also, in the first case,
   the  flux inside the tube
for $ n = 1 $  is only $ \displaystyle{\frac{2\pi}{g}} $, whereas if the 
second scalar is in the fundamental representation then the flux inside the tube
for $ n = 1 $  will be $ \displaystyle{\frac{4\pi}{g}} $. 
 So when two adjoint scalar fields are used,  monopole confinement is not possible with
a single $ n = 1 $ flux tube with winding number  $ n = 1 $. However, it is possible with a single $ n = 2 $
flux tube construction or with two $ n=1 $ flux tubes attached
with two oppositely charged monopoles. There 
may be a possibility in which two $ n=1 $ flux tubes can attach to
a monopole from opposite directions.
\appendix
\chapter{A review on SU(2) and the rotation group}
Here we shall talk about mainly the rotation group O(3)
 and relation of O(3) with a group of special complex
 2$ \times $2 matrices (SU(2)).
\section{Rotation group}
A general spatial rotation is defined by a transformation 
\begin{eqnarray}
\label{r'}
r'_i = R_{ij}r_j,
\end{eqnarray}
where $ R $ is a rotation matrix. Since rotations preserve distances
from the origin,
\begin{eqnarray}
x'^2 + y'^2 + z'^2 = x^2 + y^2 + z^2.
\end{eqnarray} 
It follows that
\begin{eqnarray}
R^T R = 1,
\end{eqnarray}
and all these $ R $ form a group called $ O(3) $.
If we restrict $ \det R = 1 $ then we call the resulting group $ SO(3) $. 
The matrix for rotation around $ z, x, y $ can be written as
\begin{eqnarray}
\label{R}
 R_z (\theta ) = 
\left(\begin{tabular}{ccc}
$\cos \theta $ & $\sin \theta $ & $0$ \\
$- \sin \theta $ & $\cos \theta $ & $0$ \\
$0$& $0$& $1$
\end{tabular}\right),\,
R_x(\phi ) = 
\left(\begin{tabular}{ccc}
$ 1 $ & $ 0 $ & $ 0 $ \\
$ 0 $ & $\cos \phi $ & $\sin \phi$ \\
$ 0 $ & $ - \sin \phi $ & $ \cos \phi $
\end{tabular}\right)\\
R_y(\psi ) = 
\left(\begin{tabular}{ccc}
$\cos \psi $ & $ 0 $ & $ - \sin \psi $\\
$ 0 $ & $ 1 $ & $ 0 $ \\
$\sin \psi $ & $0$ & $  \cos \psi $
\end{tabular}\right) \qquad \qquad \qquad
\end{eqnarray}
These group elements have generators(angular momentum),
\begin{eqnarray}
J_z 
 = \left(\begin{tabular}{ccc}
$0 $ & $ -i $ & $0 $\\
$ i $ & $ 0 $ & $ 0 $ \\
$0 $ & $0$ & $  0 $
\end{tabular}\right),
J_x =  \left(\begin{tabular}{ccc}
$0 $ & $ 0 $ & $0 $\\
$ 0 $ & $ 0 $ & $-i $ \\
$0 $ & $i$ & $  0 $
\end{tabular}\right),
J_y = \left(\begin{tabular}{ccc}
$0 $ & $ 0 $ & $i $\\
$ 0 $ & $ 0 $ & $ 0 $ \\
$-i $ & $0$ & $  0 $
\end{tabular}\right)
\end{eqnarray}
These generators form a basis for a vector space
with anti symmetric product
\begin{eqnarray}
\label{so3algebra}
\left[J^i, J^j \right] = i\e^{ijk}J^k.
\end{eqnarray}
This vector space is called Lie Algebra of SO(3).

 A general rotation matrix in three dimensions  can be written as
 \begin{eqnarray}
R_{Euler} = R_z(\psi)R_x(\theta)R_z(\phi),
\end{eqnarray}
here $ \psi, \phi, \theta $ are called Euler's angles.
\section{A representation of rotation by Cayley- Klein parameters
or SU(2) group}

Let us consider a vector  
\begin{eqnarray}
\label{psi}
\Psi = \left(\begin{tabular}{c}
$ u $ \\$ v $  \end{tabular}
\right) 
\end{eqnarray}
in a complex two dimensional vector space. A 
transformation  of this vector
\begin{eqnarray}
\label{su2transformation}
u' = \alpha u + \beta v \\
v' = \gamma u + \delta v
\end{eqnarray}
or \begin{eqnarray}
\Psi' = g \,\Psi, \,\mbox{where}\\
\label{g}
g = \left(\begin{tabular}{cc}
$\alpha$ & $\beta$ \\ $\gamma$ & $\delta$ \end{tabular}
\right). 
\end{eqnarray}
will leave the norm $ \psi^\dagger\psi = uu^* + vv^* $ invariant if
\begin{eqnarray}
\label{su2constraint}
g^\dagger g =1, \qquad\det g = 1.
\end{eqnarray}
i.e 
\begin{eqnarray}
\alpha\alpha^* + \beta\beta^* &=& 1,\, \gamma\gamma^* + \delta\delta^* = 1\\
\alpha\gamma^* + \beta\delta^* &=& 0,\, \gamma\alpha^* + \delta\beta^* = 0\\
\alpha\delta - \beta \gamma &=& 1
\end{eqnarray}
Above Eq.(\ref{su2constraint}) suggests that 
\begin{eqnarray}
\label{U2}
\delta = \alpha^*, \gamma = - \beta^*,\\
\label{g2}
g = \left(\begin{tabular}{cc}
$\alpha$ & $ - \gamma^* $ \\
$ \gamma $ & $ \alpha^* $
\end{tabular}
\right)
\end{eqnarray}
These parameters $ \alpha, \gamma $ are called Cayley-Klein parameters.
Using Eq. (\ref{U2}) we can write down Eq.
(\ref{su2transformation}) as 
\begin{eqnarray}
u' =& \alpha u  - \gamma^* v \\
v' =& \gamma u + \alpha^* v,
\end{eqnarray}
 and this can also be written as
 \begin{eqnarray}
- {v^*}' =& \alpha (- v^*)- \gamma^*  u^* \\
{u^*}' =& \gamma (- v^*) + \alpha^* u^*.
\end{eqnarray}
So  we can also define a vector 
\begin{eqnarray}
\label{psic}
\Psi_c = {-v^* \choose u^*} 
 =  \left(\begin{tabular}{cc}
 0 & $-1$\\
 1 & 0
\end{tabular}\right) \Psi^*,
\end{eqnarray}
which transforms the same way as $ \Psi $ and is orthogonal to $ \psi $ or $ \Psi^\dagger_c \Psi = 0$.
These two vectors $ \Psi, \Psi_c $ are called SU(2) spinors. 
 All the  $ g $'s together with the spinor's, represent
 a group of Special Unitary $ 2 \times 2 $ complex matrices (SU(2)).
Now we shall see that the SU(2) group also represents spatial rotations.

The Lie algebra of SU(2) consists of $ 2\times 2 $ Hermitian matrices with trace zero.
One can consider this  a three dimensional real vector space (i.e., it is closed
 under multiplication by real numbers). A basis is given by
the Pauli matrices multiplied by half,
 \begin{eqnarray}
 \tau^i = \half\left[\left(\begin{tabular}{cc}
0 & 1\\
1 & 0
\end{tabular}\right), \left(\begin{tabular}{cc}
0 & $-i$\\
i & 0
\end{tabular}\right), \left( \begin{tabular}{cc}
1 & 0\\
0 & $-1$
\end{tabular}
\right) \right],\, i = 1,2,3.
\end{eqnarray} 
This is an orthogonal basis if we define the inner product 
appropriately,
\begin{eqnarray}
 \tau^i\cdot\tau^j = 2 \Tr(\tau^i\tau^j) = \delta^{ij},
\end{eqnarray}
and the commutation relations can be written as
\begin{eqnarray}
\left[\tau^i, \tau^j\right] = i \e^{ijk}\tau^k .
\end{eqnarray}
The SU(2) Lie algebra is isomorphic to the SO(3) Lie algebra (\ref{so3algebra}) and
 there is a
homomorphism from SU(2) onto SO(3 ).
 These groups are then locally ``the same":
The proof is an application of the Frobenius theorem.
We can exhibit the classical
homomorphism as
\begin{eqnarray}
Ad:SU(2) \rightarrow SO(3),
\end{eqnarray}
here the adjoint action is defined by
$Ad (g)Y =gY g^{ - l}  $ for $ g \in $ SU(2).
The adjoint representation
  of SU(2) on its 3-dimensional Lie algebra  
yields the standard representation of SO(3) on $R^3$.
To show this explicitly,  let us define a map
 \begin{eqnarray}
 \label{mapr3gl}
{}_* &:& R^3 \rightarrow L_G(\mbox{SU(2)}), \qquad X \mapsto X_*\\
\label{X*}
X_* &=& x^i\tau^i = \half \left(\begin{tabular}{cc}
 $ z $ & $x - iy$ \\
 $x + iy$ & $ -z$
\end{tabular}\right)
\end{eqnarray}
where $ L_G $ is the Lie Algebra of SU(2).
This linear transformation maps $ R^3 $ onto the space
 of traceless hermitian matrices and
has inverse given by
\begin{eqnarray}
x = 2\Tr(X_* \tau^1),\qquad y = 2\Tr(X_* \tau^2),\qquad
z = 2\Tr(X_* \tau^3).  
\end{eqnarray}

Every Lie group G acts on its Lie algebra $ L_G $ by the adjoint action
\begin{eqnarray}
Ad: G \rightarrow L_G(g),\qquad Ad(g)X = gXg^{-1}, \qquad g\in G.
\end{eqnarray}
Under this adjoint action we can write down
\begin{eqnarray}
\label{X'}
X_*' = g X_* g^{-1}, \qquad\forall  g \in\mbox{SU(2)},
\end{eqnarray}
here with each $2\times 2$ $ \, g  \in $ SU(2)
 we associate a $3\times 3$ matrix.
  
 We also note that
\begin{eqnarray}
 X_*'\cdot  X_*' = X_*\cdot X_* = x^2 + y^2 + z^2
\end{eqnarray}
So this transformation\,(\ref{X'}) represents rotations 
in $ R^3 $. To see
different rotations we have to
write down different values of the  
parameters $ \alpha, \gamma $
 in Eq. (\ref{g2}) and use Eq. (\ref{X'}).
 
For example a rotation around the $z$-axis is represented by
\begin{eqnarray}
\label{g_phi}
g_\varphi = \left(
\begin{tabular}{cc}
$e^{\frac i2 \varphi}$ & 0\\
0 & $e^{- \frac i2 \varphi}$
\end{tabular}
\right)
\end{eqnarray}
 With this $ g_\varphi $ Eq. (\ref{X'}) can be written as
 \begin{eqnarray}
X_*' &=& x'_i \tau^i \\ &=& g_\varphi X_* 
g_\varphi^\dagger\\
&=& g_\varphi (x \tau_1 + y \tau_2 + z \tau_3) 
g_\varphi^\dagger\\
&=& g_\varphi x^i\tau_i g_\varphi^\dagger \\
 &=& x^i g_\varphi \tau_i g_\varphi^\dagger\\
&=& R_z^{ij}(\varphi)x_j \tau^i , 
\end{eqnarray}
where $ \tau^i = \half \sigma^i \,(\mbox{\,Pauli matrices})$ 
and $ R_z(\varphi) $ is defined by Eq. (\ref{R}).
Since $ \tau^i $ are linearly independent
\begin{eqnarray}
x'_i = R_{ij} x_j,
\end{eqnarray}
and this is same as Eq.(\ref{r'}).  So we have seen 
that  it is possible to represent rotation  
by SU(2) transformation of matrices of the form 
$X_* = x^i \tau_i$.
$X_* $ in Eq.(\ref{X*})  is called an adjoint vector of 
SU(2) and Eq.(\ref{X'}) is the transformation of 
the adjoint vector under SU(2). That means for a given $ g\in SU(2) $ we can
always write down a single $ R\in SO(3) $. However, the converse is not 
true, because to get a single 
rotation matrix $ R $ we could have used  both $ g $ 
and $\tilde g = hg $ in Eq. (\ref{X'}), where
$ [h, X_*'] = 0 $ and $ h\in $ SU(2).
In fact we can check that $ g = \{ 1, -1\} $ map
to $ R = 1 $. So the kernel of this mapping is $ Z_2 $. So the mapping from
 SU(2) to SO(3) is not an isomorphism
but a 2 to 1 homomorphism. 
So we have shown  that rotation group SO(3)
  can be represented
by SU(2) adjoint representations.

Now we shall show that rotations can also be represented
by SU(2) spinors. Let us define two eigenvectors  of
$ \tau^3 $ as
\begin{eqnarray}
\tau^3 |\uparrow\rangle &=& \half |\uparrow\rangle\\
\tau^3 |\downarrow\rangle &=& - \half |\downarrow\rangle
\end{eqnarray}
We have defined two SU(2) spinors in Eq.s (\ref{psi}),( \ref{psic}) as
\begin{eqnarray}
\label{spinor1}
\Psi &=& {u \choose v} = u |\uparrow\rangle  + v |\downarrow\rangle \\
\Psi_c &=& {-v^* \choose u^*} = -v^* |\uparrow\rangle  + u^* |\downarrow\rangle\\
|u|^2 + |v|^2 &=& 1
\end{eqnarray}
Now let us calculate the quantity $\Psi^\dagger \phi \Psi  $ where $ \phi $
is an adjoint vector of SU(2).
\begin{eqnarray}
\Psi^\dagger \phi \Psi &=& \Psi^\dagger \phi^i \tau^i \Psi\\
&=& \phi^i \Psi^\dagger \tau^i \Psi
= \Tr(\phi \Phi),
\label{fundaphi}
\end{eqnarray}
where $ \Phi = \Psi^\dagger \sigma^i \Psi \tau^i $. Using the definitions 
of a spinor from Eq. (\ref{spinor1})we can write 
\begin{eqnarray}
\label{Phi00}
\Phi &=& (|u|^2 - |v|^2)\tau^3 + (v^*u + u^*v)\tau^1 
+ i(v^*u - u^*v)\tau^2\\
&=& U(u,v)\tau^3 U^\dagger(u,v)
\end{eqnarray}
 where \begin{eqnarray}
 \label{Uspinor1}
 U(u,v) = 
\left(\begin{tabular}{cc}
$u$ & $-v^*$\\
$v$& $u^*$
\end{tabular}
\right).
\end{eqnarray}
This is the same as Eq. (\ref{U2}) and the spinors can be written as
\begin{eqnarray}
\label{psi33}
\Psi = U(u, v) |\uparrow\rangle ,\\
\label{psic33}
\Psi_c = U(u, v)|\downarrow\rangle.
\end{eqnarray}
In Eq.(\ref{Phi00}) the components of $ \Phi $ are real
numbers and $ \Phi $ represent  an $ SO(3) $ Lie algebra vector. Now we are going to 
calculate the spinors that represents unit vector in 3D
space. To do that we  first calculate $U(u, v)$ by comparing
the components of $ \Phi$ and components of $ \hat{r} $ and we get
\begin{eqnarray}
|u|^2 - |v|^2 &=&  \cos\theta\\
v^*u + u^*v &=& \sin\theta \cos\varphi\\
i(v^*u - u^*v) &=& \sin\theta \sin\varphi.
\end{eqnarray}
Using the above equations we can write $ \Phi = \hat{r}^i \tau^i $.
 There are many solutions of these equations, We write down two possibilities,
 solutions.
 \begin{eqnarray}
u = \cos\frac \theta 2 \displaystyle{e^{- i \frac \varphi 2}}&,&\qquad
v = \sin\frac \theta 2 \displaystyle{e^{ i \frac \varphi 2}}\\
u = \cos\frac \theta 2 &,& \qquad v = \sin\frac
 \theta 2 \displaystyle{e^{ i \varphi}}.
\end{eqnarray}
The corresponding  SU(2) elements according to Eq.(\ref{Uspinor1}) are then
\begin{eqnarray}
U_1(\theta , \varphi ) &=& \left( \begin{tabular}{cc}
$\cos\frac \theta 2 \displaystyle{e^{- i \frac \varphi 2}}$ &
$- \sin\frac \theta 2 \displaystyle{e^{- i \frac \varphi 2}}$\\
$\sin\frac \theta 2 \displaystyle{e^{ i \frac \varphi 2}}$ &
$\cos\frac \theta 2 \displaystyle{e^{ i \frac \varphi 2}}$
\end{tabular}
\right), \\
U_2( \theta , \varphi ) &=&\left( \begin{tabular}{cc}
$\cos\frac \theta 2 $ &
$- \sin\frac \theta 2 \displaystyle{e^{- i \varphi}}$\\
$\sin\frac \theta 2 \displaystyle{e^{ i \varphi}}$ &
$\cos\frac \theta 2 $
\end{tabular}
\right).
\end{eqnarray}
Using equations (\ref{psi33}) and (\ref{psic33}) we also find the spinor
\begin{eqnarray}
\Psi_1(\theta , \varphi) = 
{\cos\frac \theta 2 \displaystyle{e^{- i \frac \varphi 2}}\choose 
\sin\frac \theta 2 \displaystyle{e^{ i \frac \varphi 2}}}&,&\, 
\Psi_1^c(\theta , \varphi) = 
{- \sin\frac \theta 2 \displaystyle{e^{- i \frac \varphi 2}}\choose 
\cos\frac \theta 2 \displaystyle{e^{ i \frac \varphi 2}}} \\
\Psi_2(\theta, \varphi) = 
{\cos\frac \theta 2 \choose 
\sin\frac \theta 2 \displaystyle{e^{ i \varphi}}}&,&\, 
\Psi_2^c(\theta, \varphi) = 
{- \sin\frac \theta 2 \displaystyle{e^{- i \varphi}}\choose 
\cos\frac \theta 2 }.
\end{eqnarray}
We can write down a group element $ h =
 \displaystyle{exp(-i\varphi \Phi)}$
for which  $ U_1( \theta ,  \varphi ) 
= h(\varphi, \Phi) U_2( \theta , \varphi ) $.
 So we have seen that
there is no unique  way to write down a SU(2) 
element for a SO(3) element.
In general we can write the group elements as 
\begin{eqnarray}
G(\psi, \theta, \varphi) = h(\hat r , \chi) U_1( \theta , \varphi). 
\end{eqnarray}
It suggests that there is a class of spinors for a single unit
SO(3) vector. This class is a circular orbit which maps
to a point on $ S^2 $, whereas the $ S^2 $ is the orbit 
of a unit SO(3) vector. The reason  can be understood
very easily. The orbit of a unit spinor is
$ S^3 $ and locally $ S^3 \simeq S^1\times S^2 $. So
the mapping $ S^3 \rightarrow  S^2$ is possible if we
manage to map $ S^1 $ to a point on $ S^2 $ and 
a unit adjoint vector does the job. This geometry will be useful
when we discuss spontaneous symmetry breaking. So far 
we have shown how to construct an SO(3) vector from SU(2) spinors. 

Now we shall show the converse. 
In Eq. (\ref{fundaphi}) we have seen that an SU(2) adjoint vector $ \Phi $ 
 can be constructed by SU(2) spinors as $ \Phi = \Psi_i^{\dagger} \sigma_{ij}^\alpha \Psi_j \tau^\alpha $.
The squared length of the vector $ \Phi $ can be calculated.
\begin{eqnarray}
|\Phi|^2 &=& 2\Tr(\Phi\Phi)\\
&=&  \Phi^\alpha\Phi^\alpha\\
&=& \Psi_i^{\dagger} \sigma_{ij}^\alpha \Psi_j
\Psi_k^{\dagger} \sigma_{kl}^\alpha \Psi_l\\
&=& (2\delta_{il}\delta_{kj} - \delta_{ij}\delta_{kl})
\Psi_i^{\dagger} \Psi_j
\Psi_k^{\dagger} \Psi_l\\
&=& \Psi_i^{\dagger} \Psi_i \Psi_j^{\dagger} \Psi_j= |\Psi|^4.
\end{eqnarray}
$ \Phi $ is a matrix and we can multiply it from the left to the 
spinors. So we can multiply the adjoint vector $ \Phi $ with the 
spinor $\Psi $ by which it was constructed.
\begin{eqnarray}
\Phi_{ij}\Psi_j &=& \Psi_k^{\dagger} \sigma_{kl}^\alpha \Psi_l
\tau_{ij}^\alpha \Psi_j \\
 &=&  \half \sigma^\alpha_{ij}\sigma^\alpha_{kl}
 \Psi_k^{\dagger} \Psi_l\Psi_j\\
 &=& (\delta_{il}\delta_{kj} - \half \delta_{ij}\delta_{kl})
 \Psi_k^{\dagger} \Psi_l\Psi_j\\
 &=& \half |\Psi|^2 \Psi_i\\
 &=& \half |\Phi| \Psi_i ,
\end{eqnarray}
and 
\begin{eqnarray}
\Phi_{ij}\Psi^c_j &=& \Psi_k^{\dagger} \sigma_{kl}^\alpha \Psi_l
\tau_{ij}^\alpha \Psi^c_j \\
 &=&  \half \sigma^\alpha_{ij}\sigma^\alpha_{kl}
 \Psi_k^{\dagger} \Psi_l\Psi^c_j\\
 &=& (\delta_{il}\delta_{kj} - \half \delta_{ij}\delta_{kl})
 \Psi_k^{\dagger} \Psi_l\Psi^c_j\\
 &=& - \half |\Psi|^2 \Psi^c_i\\
 &=& -  \half |\Phi| \Psi_i,
\end{eqnarray}
here we have used the identity $ \delta_{il}\delta_{kj} = \half \delta_{ij}\delta_{kl}
 + \half \sigma^\alpha_{ij}\sigma^\alpha_{kl}. $
So we  see that  spinors are the eigenvectors
of an  adjoint vector $ \Phi $ with eigenvalues $ \pm \half|\Phi| $.
However, as we have seen before, the spinors are not unique because they can be multiplied by
a matrix that commutes with $ \Phi $. The  non trivial SU(2) element that commute
with $ \Phi $ is $ \exp(-i\chi \hat\Phi) $ for all $ \chi $. So
we have seen that for every  adjoint (SO(3)) vector there are
two classes of SU(2) spinors with eigenvalues $ \pm \half|\Phi| $.

\section{Some useful notation for SU(2) fields} 
\label{sec:aa} 

Any $ 2\times 2 $ complex matrix  $ X $  can be written as
\begin{eqnarray}
\label{X}
X = X_0 {\bf 1} + X_i\sigma_i , 
\end{eqnarray}
where 
\begin{eqnarray}
\label{Xi}
 X_0 = \half \Tr(X) ,\qquad X_i = \half \Tr(X\sigma^i).
\end{eqnarray}
Substitute Eq. (\ref{Xi}) into Eq. (\ref{X})
we get
\begin{eqnarray}
X_{ij} = \half X_{kk}\delta_{ij} + \half X_{lk}\sigma^\alpha_{kl}
\sigma^\alpha_{ij}.
\end{eqnarray}
We compare the coefficients of $X_{lk}$ both sides
and get
\begin{eqnarray}
\label{deltaidentity1}
\delta_{il}\delta_{kj} = \half \delta_{ij}\delta_{kl}
 + \half \sigma^\alpha_{ij}\sigma^\alpha_{kl}.
\end{eqnarray}

Any two  fields $ A $ and $ B $ in the adjoint representation of SU(2) can be written as
\begin{eqnarray}
A = A^i \tau^i,\qquad B = B^i \tau^i.
\end{eqnarray}
Using the orthogonality relation of the algebra
we can calculate the trace.
\begin{eqnarray}
\label{dotproduct}
2\Tr(AB) = A^iB^i = \vec{A}\cdot \vec{B}
\end{eqnarray}
So the trace can be written (upto a factor of $ 2 $) \,as dot product between two vectors
$ \vec{A} $ and $ \vec{B} $.
Let us calculate the commutator of the fields $ A$ and $ B $.
\begin{eqnarray}
\left[A, B\right] &=& A^j B^k \left[\tau^j, \tau^k\right]
= i \e^{ijk}\tau^i A^j B^k\\
&=& i \left(\vec{A}\times \vec{B}\right)^i\tau^i
\end{eqnarray}
We can invert the above equation by using orthogonality relations
and write the cross product between two vectors $ \vec{A} $ and $ \vec{B} $ as,
\begin{eqnarray}
\label{crossproduct}
\left(\vec{A}\times \vec{B}\right)^i &=& -\,2i \Tr\left(\tau^i\left[A, B\right]\right).
\end{eqnarray}
The commutator of three fields can be written as 
\begin{eqnarray}
\left[A,\left[B,C\right]\right] &=& i\e^{klm}B_lC_m\left[A, \tau^k\right]
= - \e^{ijk}\tau^i A_j \e^{klm}B_lC_m\\
&=&-\left(\delta^{il}\delta^{jm} - \delta^{im} \delta^{jl}\right)\tau^i A_jB_lC_m\\
&=& C\, 2\Tr\left(AB \right) - B\, 2\Tr\left(AC \right)
\label{doublecross}
\end{eqnarray}



\pagebreak
\begin{center}
List of Publications 
\end{center}
$ 1)^* $
  C.~Chatterjee and A.~Lahiri,
  Flux dualization in broken SU(2),
  JHEP {\bf 1002}, 033 (2010)
  [arXiv:0912.2168 [hep-th]].

$ 2)^* $
  C.~Chatterjee and A.~Lahiri,
  Monopoles and flux strings from SU(2) adjoint scalars,
  JHEP {\bf 0909}, 010 (2009)
  [arXiv:0906.4961 [hep-th]].

3)
  C.~Chatterjee and S.~Gangopadhyay,
  $\kappa$-Minkowski and Snyder algebra from reparametrisation symmetry,
  Europhys.\ Lett.\  {\bf 83}, 21002 (2008)
  [arXiv:0806.0758 [hep-th]].

4)
  C.~Chatterjee, S.~Gangopadhyay, A.~G.~Hazra and S.~Samanta,
  String non(anti)commutativity for Neveu-Schwarz boundary conditions,
  Int.\ J.\ Theor.\ Phys.\  {\bf 47}, 2372 (2008)
  [arXiv:0801.4189 [hep-th]].

$ 5)^* $
  C.~Chatterjee and A.~Lahiri,
  Monopole confinement by flux tube,
  Europhys.\ Lett.\  {\bf 76}, 1068 (2006)
  [arXiv:hep-ph/0605107].
   \vskip 2cm
   \bf{The papers with $ * $ are included in this thesis.}


\addcontentsline{toc}{chapter}{Bibliography}
\begin{thebibliography}{99}
\label{bb}
\bibitem{Gross:1973ju}
  D.~J.~Gross and F.~Wilczek,
  Phys.\ Rev.\  D {\bf 8}, 3633 (1973).
\bibitem{Politzer:1973fx}
  H.~D.~Politzer,
  Phys.\ Rev.\ Lett.\  {\bf 30}, 1346 (1973).

\bibitem{Bali:2000gf}
  G.~S.~Bali,
  Phys.\ Rept.\  {\bf 343}, 1 (2001)
  [arXiv:hep-ph/0001312].
  
  
\bibitem{hooftstringnotes}
 G.~ 't Hooft,
INTRODUCTION TO STRING THEORY
version 14-05-04



\bibitem{Mandelstam:1974pi}
  S.~Mandelstam,
  Phys.\ Rept.\  {\bf 23} (1976) 245.
Mandelstam S 1976 Phys. Rep. 23 245
\bibitem{Nambu:1975ba}
  Y.~Nambu,
  Phys.\ Rept.\  {\bf 23} (1976) 250.
  \bibitem{hooft:1976}
G.~'t Hooft    , 1976 High Energy Physics ed A Zichichi (Bologna: Editrice Compositori)
   \bibitem{hooft:1999} 
  G.~'t Hooft,
TOPOLOGICAL ASPECTS OF
QUANTUM CHROMODYNAMICS   
arxiv:Hep-th/9812204
\bibitem{'tHooft:1999au}
 G.~'t Hooft,
  ``Monopoles, instantons and confinement,''
  arXiv:hep-th/0010225.
\bibitem{Abrikosov:1956sx}
  A.~A.~Abrikosov,
  Sov.\ Phys.\ JETP {\bf 5}, 1174 (1957)
  [Zh.\ Eksp.\ Teor.\ Fiz.\  {\bf 32}, 1442 (1957)].
\bibitem{Nielsen:1973cs}
  H.~B.~Nielsen and P.~Olesen,
  Nucl.\ Phys.\  B {\bf 61}, 45 (1973).
  
\bibitem{Baker:1984dv}
  M.~Baker, J.~S.~Ball and F.~Zachariasen,
  Phys.\ Lett.\  B {\bf 152}, 351 (1985).
\bibitem{Maedan:1988yi}
  S.~Maedan and T.~Suzuki,
  Prog.\ Theor.\ Phys.\  {\bf 81}, 229 (1989).

\bibitem{Baker:1997bg}
  M.~Baker, J.~S.~Ball and F.~Zachariasen,
  Phys.\ Rev.\  D {\bf 56}, 4400 (1997)
  [arXiv:hep-ph/9705207].
  Phys.\ Rev.\  D {\bf 51}, 1968 (1995).


\bibitem{Nambu:1977ag}
  Y.~Nambu,
  Nucl.\ Phys.\  B {\bf 130} (1977) 505.
\bibitem{Vachaspati:1992fi}
  T.~Vachaspati,
  Phys.\ Rev.\ Lett.\  {\bf 68}, 1977 (1992)
  [Erratum-ibid.\  {\bf 69}, 216 (1992)].
  Nucl.\ Phys.\  B {\bf 397}, 648 (1993).

\bibitem{deVega:1976rt}
  H.~J.~de Vega,
  Phys.\ Rev.\  D {\bf 18}, 2932 (1978).
\bibitem{Auzzi:2003fs}
  R.~Auzzi, S.~Bolognesi, J.~Evslin, K.~Konishi and A.~Yung,
  Nucl.\ Phys.\  B {\bf 673} (2003) 187


\bibitem{Hanany:2004ea}
  A.~Hanany and D.~Tong,
  JHEP {\bf 0404} (2004) 066


\bibitem{Shifman:2002yi}
  M.~Shifman and A.~Yung,
  Phys.\ Rev.\  D {\bf 66}, 045012 (2002)
\bibitem{Kalb:1974yc}
 M.~Kalb and P.~Ramond
  Phys.\ Rev.\ D{9}{1974}{ 2273}.
\bibitem{Davis:1988rw}
  R.~L.~Davis and E.~P.~S.~Shellard,
  Phys.\ Lett.\ B {\bf 214}, 219 (1988).
R.~L.~Davis,
 Phys.\ Rev.\  D {\bf 40}, 4033 (1989)

\bibitem{Lund:1976ze}
  F.~Lund and T.~Regge,
  Phys.\ Rev.\  D {\bf 14}, 1524 (1976).

\bibitem{Mathur:1991ip}
  M.~Mathur and H.~S.~Sharatchandra,
  Phys.\ Rev.\ Lett.\  {\bf 66}, 3097 (1991).

\bibitem{Lee:1993ty}
  K.~M.~Lee,
  Phys.\ Rev.\ D {\bf 48}, 2493 (1993)
  
\bibitem{Baker:1991bc}
  M.~Baker, J.~S.~Ball and F.~Zachariasen,
  Phys.\ Rept.\  {\bf 209}, 73 (1991).

\bibitem{'tHooft:1974qc}
  G.~'t Hooft,
  Nucl.\ Phys.\  B {\bf 79}, 276 (1974).

\bibitem{Polyakov:1974ek}
  A.~M.~Polyakov,
  JETP Lett.\  {\bf 20}, 194 (1974)
  [Pisma Zh.\ Eksp.\ Teor.\ Fiz.\  {\bf 20}, 430 (1974)].
\bibitem{Prasad:1975kr}
 M.~K.~Prasad and C.~M.~Sommerfield,
  Phys.\ Rev.\ Lett.\  {\bf 35}, 760 (1975).
  \bibitem{Akhmedov:1995mw}
  E.~T.~Akhmedov, M.~N.~Chernodub, M.~I.~Polikarpov and
  M.~A.~Zubkov, 
  Phys.\ Rev.\ D {\bf 53}, 2087 (1996)
\bibitem{Chatterjee:2006iq}
  C.~Chatterjee and A.~Lahiri,
  Europhys.\ Lett.\  {\bf 76}, 1068 (2006)
\bibitem{Hindmarsh:1985xc}
  M.~Hindmarsh and T.~W.~B.~Kibble,
  Phys.\ Rev.\ Lett.\  {\bf 55} (1985) 2398.
\bibitem{Shifman:2004dr}
  M.~Shifman and A.~Yung,
  Phys.\ Rev.\  D {\bf 70}, 045004 (2004)
  [arXiv:hep-th/0403149].


\bibitem{Kneipp:2003ue}
  M.~A.~C.~Kneipp,
  Phys.\ Rev.\  D {\bf 69} (2004) 045007


\bibitem{Auzzi:2003em}
  R.~Auzzi, S.~Bolognesi, J.~Evslin and K.~Konishi,
  Nucl.\ Phys.\  B {\bf 686} (2004) 119

\bibitem{Eto:2006dx}
  M.~Eto {\it et al.},
  Nucl.\ Phys.\  B {\bf 780} (2007) 161
  \bibitem{Jackson}
Jackson, John David, 
Classical electrodynamics 3rd ed. 

\bibitem{J.J.Thomson}
J. J. Thomson, Elements of the Mathematical Theory of Electricity and 
Magnetism, Cambridge University Press. 
\bibitem{M.N.Saha}
M.N.Saha Indian.\ J.\ Physics {\bf 10}(1936) 145.

\bibitem{Dirac:1931kp}
  P.~A.~M.~Dirac,
  Proc.\ Roy.\ Soc.\ Lond.\  A {\bf 133} (1931) 60.
\bibitem{Wu:1976qk}
  T.~T.~Wu and C.~N.~Yang,
  Phys.\ Rev.\  D {\bf 14}, 437 (1976).
  Nucl.\ Phys.\  B {\bf 107}, 365 (1976).
  \bibitem{felsagar}
   ~ Bjorn Felsager ,  ``Geometry, Particles and Fields '', 
 - New York: Springer-verlay, 1998.
   

\bibitem{Dirac : 1948}
Paul A.M. Dirac, 
    Phys.Rev.74:817-830,1948.

\bibitem{Rohrlich:1966}
F.~Rohrlich,
Phys.\ Rev. {\bf 150}, 1104 (1966).

\bibitem{zwanziger}
Daniel Zwanziger,
 Phys.Rev.D3:880,1971.

 Phys.Rev.176:1480,1968.
   Phys.Rev.176:1489-1495,1968.


\bibitem{Balachandran : 1975}
A.P. Balachandran, R. Ramachandran, J. Schechter, Kameshwar C. Wali, Heinz Rupertsberger,
Phys.Rev.D13:354,1976.

\bibitem{Goddard :1978}
P.Goddard, David.I.Olive, 
    Rept.Prog.Phys.41:1357,1978.

\bibitem{1979:zwanziger}
Richard A. Brandt, Filippo Neri, Daniel Zwanziger,
    Phys.Rev.D19:1153,1979.

\bibitem{Calucci:1982}
G. Calucci, R. Jengo, 
    Nucl.Phys.B197:93,1982.

\bibitem{Kleinert: 1990} 
H. Kleinert,
    Phys.Lett.B246:127-130,1990.
  
   \bibitem{Weinberg:book}
S. Weinberg In: The Quantum Theory of Fields. Vol. 1: Foundations, Cambridge University Press, Cambridge, UK (1995), p. 100.

\bibitem{Anselmino:2002}
M. Anselmino, V. Barone, A. Drago, F. Murgia,
    Nucl.Phys.Proc.Suppl.105:132-133,2002.[HEP-PH 0111044]

\bibitem{Abdus Salam:1966}
Abdus Salam,
Phys.Lett.22:683-684,1966.

\bibitem{j.g.taylor:1967}
 J.~G.~Taylor,
   Phys.\ Rev.\ Lett.\  {\bf 18}, 713 (1967).
 

\bibitem{Shu-Yuan Chu : 1973}
Shu-Yuan Chu, 
    Phys.\ Rev.\ D 7:853-856,1973.
    
    \bibitem{N.F.Ramsey:1957}
N.F.Ramsey 
Phys.\ Rev.\ {\bf109}, 225 (1957)
\bibitem{Baker:1999xn}
  M.~Baker and R.~Steinke,
  Phys.\ Lett.\  B {\bf 474}, 67 (2000)
  [arXiv:hep-ph/9905375].
 \bibitem{Marino:2006mk}
   E.~C. ~Marino,
 J.\ Phys.\ A {\bf 39}, L277 (2006).
\bibitem{Polchinski:1991ax}
  J.~Polchinski and A.~Strominger,
  Phys.\ Rev.\ Lett.\  {\bf 67}, 1681 (1991).

\bibitem{Vachaspati:1991dz}
  T.~Vachaspati and A.~Achucarro
{  Phys.\ Rev.\ D}{44}{1991}{ 3067}.
qq






\bibitem{Giacomelli :2000}
G. Giacomelli et al., 
DFUB-2000-09 (May 2000) 29p.
MAGNETIC MONOPOLE BIBLIOGRAPHY.
.
 



    


  


 
\bibitem{Cremmer:1974mg}
E.~Cremmer, J.~Scherk
{Nucl.\ Phys.\ B}{72}{1974}{ 117}.

\bibitem{Allen:1990gb}
  T.~J.~Allen, M.~J.~Bowick and A.~Lahiri
{  Mod.\ Phys.\ Lett.\ A}{6}{1991}{ 559}.
 
 
  
 




\bibitem{Nambu:1974zg}
  Y.~Nambu,
  Phys.\ Rev.\  D {\bf 10} (1974) 4262.






  



  








\bibitem{Chatterjee:2009pi}
  C.~Chatterjee and A.~Lahiri,
  JHEP {\bf 0909}, 010 (2009)






















\bibitem{Corrigan:1975hd}
  E.~Corrigan, D.~I.~Olive, D.~B.~Fairlie and J.~Nuyts,
  Nucl.\ Phys.\  B {\bf 106}, 475 (1976).

\bibitem{Bais:1976fr}
  F.~A.~Bais,
  Phys.\ Lett.\  B {\bf 64}, 465 (1976).







\bibitem{Orland:1994qt}
  P.~Orland,
  Nucl.\ Phys.\  B {\bf 428}, 221 (1994)





\bibitem{Polyakov:1996nc}
  A.~M.~Polyakov,
  Nucl.\ Phys.\  B {\bf 486}, 23 (1997)

\bibitem{Goldhaber: 1977}
   Alfred S. Goldhaber, Phys.Rev.D16:1815,1977.

\bibitem{Steven Weinberg:1965}
Steven Weinberg
    Phys.Rev.138:B988-B1002,1965.

\bibitem{Bardakci: 1978}
  K Bardakci, S. Samuel, .
    Phys.Rev.D18:2849,1978. 
 
\bibitem{Suzuki:1988}
Shinji Maedan, Tsuneo Suzuki,Prog.Theor.Phys.81:229-240,1989.

\bibitem{Halpern:1977}   
M.B. Halpern, W. Siegel, 
    Phys.Rev.D16:2486,1977

\bibitem{Antonov :2000}
Dmitri Antonov, Dietmar Ebert, 
    Eur.Phys.J.C12:349-359,2000.
    [HEP-TH 9812112]

\bibitem{C.R. Hagen:1965}
C.R. Hagen, 
    Phys.Rev.140:B804,1965.

\bibitem{Ferreira: 2006}
P. Castelo Ferreira,
    J.Math.Phys.47:072902,2006.
    [HEP-TH 0510063]



\bibitem{Mignani}
R. Mignani, Phys.Rev.D13:2437,1976.

\bibitem{Callan: 1982}
Curtis G. Callan, Jr., Phys.Rev.D26:2058-2068,1982.




\bibitem{Blagojevic:1988}
M. Blagojevic, P. Senjanovic, THE QUANTUM FIELD THEORY
    OF ELECTRIC AND MAGNETIC CHARGE.
    Phys.Rept.157:233,1988.    
    
\end{thebibliography}
\end{document}